\documentclass[11pt]{report}
\usepackage{epsfig}
\begin{document}

\newcommand{\MF}{{\large{\manualMETA}\-{\manualFONT}}} 
\newcommand{\be}{\begin{equation}}   
\newcommand{\ee}{\end{equation}}
\newcommand{\bea}{\begin{eqnarray}}
\newcommand{\eea}{\end{eqnarray}}
\newcommand{\prb}{Physical Review B~}
\newcommand{\prl}{Physical Review Letters~}
\newcommand{\non}{\nonumber}
\newcommand{\la}{\lambda}
\newcommand{\w}{\tilde}
\newcommand{\al}{\alpha}
\newcommand{\g}{{\rm g}}
\newcommand{\La}{\Lambda}
\newcommand{\ga}{\gamma}
\newcommand{\Ga}{\Gamma}
\newcommand{\bk}{{\bf k}}
\newcommand{\bq}{{\bf q}}
\newcommand{\br}{{\bf r}}
\newcommand{\bE}{{\bf E}}
\newcommand{\up}{\uparrow}
\newcommand{\ol}{\overline}
\newcommand{\down}{\downarrow}

\pagestyle{empty}
\begin{center}
{\bf Abstract}\\
Strongly Correlated States in Low Dimensions\\ 
Dmitry Green \\ 
\vspace{-6pt}
2001
\end{center}
This thesis is a theoretical analysis of sample two- and one-dimensional
systems. The two-dimensional examples are the quantum Hall liquid and anomalous
paired states. The most widely accepted effective theory of the quantum Hall
liquid is based on the so-called Chern Simons Lagrangian, but it is not
entirely satisfactory. We obtain the first derivation of an alternative
effective theory from microscopic principles. Our formulation allows for a
first principles derivation of physical quantities such as the effective mass
and compressibility and contains the first analytical observation of the
magnetoroton. The formalism developed along the way is also applied to paired
states in anomalous supeconductors, a topic of much recent interest. The
one-dimenisonal system is the carbon nanotube. Gas uptake in nanotube bundles
is currently attracting a wealth of research with both applied and fundamental
implications. We propose adsorption of gases on the surface of a single tube,
finding strong correlations and symmetries that have not been observed yet. The
properties of these states are directly relevant to other one-dimensional
structures such as spin ladders and stripes and raise interesting and open
questions.
\begin{titlepage}
\begin{center}
\vspace*{.5in}
{\large Strongly Correlated States in Low Dimensions}\\
\vspace*{2.0in}
A Dissertation\\ \vspace{-6pt}
Presented to the Faculty of the Graduate School\\ \vspace{-6pt}
of\\ \vspace{-6pt}
Yale University\\ \vspace{-6pt}
in Candidacy for the Degree of\\ \vspace{-6pt}
Doctor of Philosophy\\
\vspace{1.75in}
by\\ \vspace{-6pt}
Dmitry Green\\
\vspace{0.5in}
Dissertation Director: Prof. Nicholas Read\\
\vspace{0.5in}
December, 2001
\end{center}
\end{titlepage}

\pagestyle{plain}
\setcounter{page}{1}
\tableofcontents
\newpage
%\addtocontents{toc}{\contentsline{chapter}{List of Figures}{4}} 
\listoffigures

\newpage
\section*{Acknowledgements}
I learned a great deal from the faculty at Yale.  My advisor, Nick Read, has been my main guide and teacher, and I have benefitted from courses and discussions with (in alphabetical order) Sean Barrett, Subir Sachdev,  R. Shankar,  Samson Shatashvili, and Doug Stone.  

Jo-Ann Bonnett (Graduate Registrar, Physics) and her predecessor, Jean Belfonti, have made my life at Yale immeasurably easier.   And, I appreciate the various fellowships from the NSF and Yale that have helped me along the way.  

I also want to go back in time a little to thank my undergraduate thesis advisor at the University of Chicago, Prof. Ugo Fano, who passed away recently.  His combination of humor, warmth, and intellectual vitality and integrity are unique, and it was truly exciting to have worked with him.   

Last, but best, thanks to my fianc\'ee, Katie.  
\addcontentsline{toc}{chapter}{Acknowledgements} 

\newpage
\thispagestyle{empty}
\begin{center}
\vspace*{2.5in}
\Large{For my grandmother,}\\
{\it Rakhil Babinskaya}
\end{center}

\renewcommand{\thesubsection}{\thesection.\alph{subsection}}
\renewcommand{\thefootnote}{\fnsymbol{footnote}}

\chapter{Introduction}
\label{chap:Introduction_LLL}

\hyphenation{quantum}
\hyphenation{effect}
%%%%%%%%%%%%%%%%%%%%%%%%%%%%%%%%%%%%%%%%%%%%%%%%%%%%%%%%%%%%%%%%%%%%%%%%%%%%%%
\section{Composite Particles in the Fractional Quantum Hall Effect}
\label{sec:intro_LLL}
%%%%%%%%%%%%%%%%%%%%%%%%%%%%%%%%%%%%%%%%%%%%%%%%%%%%%%%%%%%%%%%%%%%%%%%%%%%%%%
\hyphenation{quan-tum}
\hyphenation{ef-fect}

The technological innovation that made discovery of the quantum Hall effect (QHE) possible is called a MOSFET, or metal oxide semiconductor field effect transistor.  Under suitable conditions, electrons are effectively confined to two dimensions by an inversion layer \cite{GirvinPrange}.  Inversion layers are formed at an interface of a semiconductor and an insulator or between two semiconductors with one of them acting as an insulator.  The original system in which the QHE was discovered was between Si (semiconductor) and ${\rm SiO}_2$ (insulator).  More recently the semiconductor-semiconductor system GaAs-${\rm Al}_x{\rm Ga}_{1-x}{\rm As}$ (with GaAs acting as the semiconductor) has been used.  The parameter is approximately $x\sim 0.2$.  Typically the layers of GaAs and AlGaAs are grown with atomic precision using molecular beam epitaxy.  The necessary donors  that are required for the inversion layer to form are implanted away from the interface allowing very high electron mobility within the inversion layer.   For some samples, an external voltage may be used to control the density of electrons.

The original discovery by von Klitzing, Dorda, and Pepper in 1980 has come to be known as the integer quantum Hall effect (IQHE).  In brief, they found that when a magnetic field is applied perpendicular to the electron layer, the electron current response is purely transverse and quantized.  More precisely, the current density responds to an electric field by $j_i=\sum_j\sigma_{ij}E_j$, where the conductivity tensor is 
\bea
\sigma=\left[\begin{array}[2]{cc}
0 & -\nu e^2/h \\ \nu e^2/h & 0 \end{array}\right]~.
\non
\eea
$h$ is Planck's constant, $e$ is the electron charge, and $\nu$ is a small integer.  Thus the conductivity is quantized in fundamental units, independently of specific material parameters.  The diagonal conductivity vanishes, so the state is dissipationless.  The actual measured quantities are the conductance and/or resistance.  Theoretical understanding of this effect is based on models of independent electrons in the presence of disorder, and is quite developed by now.

In this thesis we will be concerned with an effect that was discovered shortly after the IQHE, which is known as the fractional quantum Hall effect (FQHE).  In 1982, Tsui, St\"{o}rmer, and Gossard found that, in extremely pure samples at very low temperature, the integer $\nu$ above can be replaced by a hierarchy of rational numbers, $\nu=p/q$.  In the quantum limit $\omega_c\tau\gg 1$, where $\omega_c=eB/m$ and $\tau$ is the electronic scattering time, $\nu$ shows plateaus at these fractions as the chemical potential is varied.  The leading fractions were found to have $q=$odd, forming an incompressible quantum liquid.  Since then, $q=$ even states are understood to have their own complementary set of phenomena \cite{HLR}.  In contrast to the odd denominator fillings, they are compressible Fermi liquid-like states and are not characterized by plateaus in the conductivity (at least for $p\leq 3$; $\nu=5/2$ may be an exception \cite{Read52}).  Theoretical understanding of the FQHE has progressed rapidly, but is not yet complete.  For instance, problems which depend in detail on the interaction and disorder, such as transitions between the plateaus, are not well understood.  Questions have also arisen recently on the nature of the effective theory for $\nu<1$ even in pure samples.  This is the issue we will tackle in this part of the thesis.

Let us briefly summarize the salient ingredients of the theory.  The fundamental length scale for electrons in a magnetic field $B$ is 
\bea
\ell_B = \left(\frac{\hbar c}{eB}\right)^{1/2}.
\non
\eea
It is independent of material parameters and is in the range of $50-100$\AA.  The independent particle states are parameterized by Landau levels (LL) of energy $E_n= \hbar\omega_c(n+1/2)$.  Each LL is highly degenerate; the number of states per unit area of one full LL is given by 
\bea
\rho_0=1/2\pi\ell_B^2 = eB/hc
\non
\eea 
The last equality can be rewritten as $\rho_0=B/\Phi_0$, where $\Phi_0$ is the unit flux quantum, so that the degeneracy of one full LL is counted by the number of flux quanta in the external field.  In this simplified model, electrons successively occupy the Landau levels, and the proportion of occupied states is denoted by the filling fraction, $\nu$.  If the electron density is $\rho$ then  
\bea
\nu=\rho/\rho_0~.
\label{eq:nu}  
\eea
The IQHE occurs at integral $\nu$, when an integral number of levels are full.

However, the FQHE occurs when certain rational fractions of LL's are filled.  The theoretical understanding of this phenomenon starts with Laughlin's approach in 1983 \cite{Laughlin83}.  He proposed a variational wavefunction for the ground state at $\nu=1/p$, $p=$odd, which is a fractionally filled lowest Landau level (LLL).  Assuming that $\omega_c$ is large compared to the electron-electron interaction, only the LLL should describe the physics at low energies.  For $N$ electrons with coordinates $z_i=x_i+iy_i$, the Laughlin wavefunction is,
\bea
\psi(z_1,\ldots,z_N)=\prod_{i<j}^N(z_i-z_j)^p\prod_i^N e^{-|z_i|^2/4\ell^2_B}~.
\non  
\eea
This function is composed of single particle states in the LLL and is properly antisymmetric in keeping with fermionic statistics.  The two main features of $\Psi$ are (i) there is a zero on each electron, and (ii) each electron sees other electrons as magnetic flux due to the accumulated phase in dragging one coordinate around another.  The basic excitations are quasiholes and quasielectrons with fractional charge $\pm e/p$.  Clearly this kind of effect is due to strong correlations in the fluid.  The quasiparticles also obey fractional statistics, as articulated by Halperin \cite{Halperin84} and by Arovas et al. \cite{Arovas}.  Experimental data is consistent with fractional charge at $\nu=1/3$ and that the basic excitations at $\nu=1/2$ are neutral \cite{expt,Shimshoni,Goldman}.  Already we see a collective behavior that is very different from Fermi liquids.  In fact, the Laughlin state describes a strongly correlated quantum liquid that cannot be reached perturbatively from the Fermi liquid.  As originally noted by Laughlin, it is accepted that the FQHE liquid is isotropic and incompressible.

Much of the current understanding of quantum liquids relies on effective field theories.  In the case of the FQHE the most successful has been the Chern-Simons (CS) theory \cite{Zhang}.  Fortuitously, CS actions were being formally developed concurrently in a purely field theoretic context  \cite{CS}.  One starts with a transformation that represents each electron as a boson plus an odd number of $\widetilde\phi$ of $\delta$-flux tubes.  Fractional statistics emerges as a Berry phase when particles or vortices are dragged around flux tubes.  After the transformation, the action contains a U(1) Chern-Simons term that couples to the fermion density.  At the fractions $\nu=1/q$ ($q$=odd), the statistical gauge potential ${\bf a}$ is determined by the relation
\bea
\rho=-\frac{\nu}{2\pi}\nabla\times{\bf a}~
\label{eq:CS}
\eea
(where $\hbar=c=1$).  The full Lagrangian includes ${\bf a}$ by minimal coupling and the statistical gauge transformation allows us to replace the fermions by a bosonic field, $\phi$.  The complete gauge invariant Lagrangian is
\bea
{\cal L}_{CS} = \phi^\dagger(i\partial_t-A_0-a_0)\phi -\frac{1}{2m}\left|\left(i\nabla+{\bf A}+{\bf a}\right)\phi\right|^2+\frac{\nu}{4\pi}\epsilon^{\mu\nu\la}a_\mu\partial_\nu a_\la~\non\\
-\frac{1}{2}\int d^2y \rho(x) V(x-y) \rho(y)~,
\label{eq:L_CS}
\eea
where $A_0,{\bf A}$ is the external potential.  The third term is the ${\rm U}(1)$ CS term ($\epsilon$ is the Levi-Civita symbol) and the last term is the interaction.  The defining equation (\ref{eq:CS}) for ${\bf a}$ can be viewed as an equation of motion.  To completely determine ${\bf a}$, one can use conservation of charge, $\partial_t\rho+\partial_i J_i=0$, to obtain the dynamics equation, $J_i=-\frac{\nu}{2\pi}\epsilon^{ij}\partial_t a_j$.  Together with eqn. (\ref{eq:CS}) this completely determines the statistical gauge field ${\bf a}$. 

The net field is the sum of the external field and the $\delta$-flux tubes.  On the average, for a uniform density of particles, the two fields cancel when $\nu=1/\widetilde\phi$ (that is, $\langle{\bf A+a}\rangle=0$) and we are left with bosons in zero net field.  The Laughlin state can thus be interpreted as a bose condensate of the electron-flux tube composite.  In principle, one can substitute bosonic particles for electrons and use an even $\widetilde\phi$.  Some understanding of the states with even denominator, e.g. $\nu=1/2$, has been achieved in this framework \cite{HLR}. In this case, the composite fermions form a Fermi sea and the resultant phase is Fermi-liquid like.  Paired states of fermions or bosons are envisioned as pairing of particle-flux tube composites in zero field, since the external field vanishes at mean field, as discussed in the introductory Section \ref{sec:intro_pairing} and in Chapters \ref{chap:PERM} and \ref{chap:BCS}.  Any analysis beyond mean field must proceed with caution in any case; standard techniques are exact only in the limit $\widetilde\phi\rightarrow 0$, but the FQHE states require $\widetilde\phi\geq 2$.   

Other, perhaps more fundamental, difficulties with the CS approach have been appreciated by several workers since the beginning (see e.g. \cite{Read94}).  At zero temperature, the usual assumption is that the inter-electron interactions $\sim\nu^{1/2}e^2/\varepsilon\ell_B$ are weak compared to the cyclotron frequency $\omega_c$ so the physics should be dominated by LLL states when $\nu<1$.  The kinetic energy is just a constant in any given LL so we are faced with a macroscopically degenerate perturbation theory with a purely dynamical Hamiltonian.  In particular, consider the Fermi liquid-like state at $\nu=1/2$.  In the mean field approximation, the effective mass of excitations close to the Fermi surface is the bare mass $m$, however the low-energy excitations should have an effective mass $m^*$ determined solely by the interactions.  The problem is partially resolved by Fermi liquid fluctuations, which introduce a Landau interaction parameter $F_1$ that renormalizes the bare mass by $m^{-1}=m^{*-1}(1+F_1)$.  The most serious problem is that interactions do not play a role in the compressibility; whether or not interactions are included in the fluctuations, the result is a finite compressibility for the Fermi liquid due to the mass $m$.  On the other hand, a partially filled Landau level of non-interacting particles should have an infinite compressibility.  The same difficulty is present in the odd denominator filling fractions; regardless of the interaction, the compressibility vanishes \cite{Zhang}.  There has been a renewed interest in these puzzles recently, stemming from new developments in composite fermion theory.

It has become clear that a more physical way of looking at the composite particles is as bound states of one particle and $\widetilde\phi$ vortices.  The composites are literally dipoles.  Since the quantum Hall liquid cannot be reached perturbatively from a normal electron fluid, Landau's Fermi liquid theory cannot be applied, and the new quasiparticle must be built up from scratch. This notion developed steadily starting from Laughlin's observation that particles see other particles as flux \cite{Laughlin83}.  Jain used it to obtain the basic sequence of fractional quantum Hall plateaus \cite{Jain}, and Haldane \cite{Haldane83} and Halperin \cite{Halperin84} constructed a complete hierarchy of states.  Their approach relied solely on the analytic properties of ground state wavefunctions.  The corresponding field theoretic implementation is the class of CS models outlined above.  In spite of the successes of these descriptions, neither does justice to the particle-vortex composites as bound states (dipoles) in their own right.  

This thesis will clarify that perhaps this is the essential ingredient required for a self-consistent understanding of the effective mass and compressibility issues.   Several authors have applied the dipole scenario with some success \cite{Read94,SM,Shankar99,Shankar01,ReadHalf,HP,Lee}.  The works of Lee \cite{Lee} and of Shankar and Murthy \cite{SM,Shankar99,Shankar01} reconsider the Chern-Simons action and recover some of the dipole physics.  Here, we will approach the problem from the opposite direction by working in the LLL at the outset without any singular flux attachment.  The language no longer includes $\delta$-flux tubes or $\widetilde\phi$, rather particle-vortex dipoles are the basic building blocks.  Our guide is a formalism introduced  by Haldane and Pasquier \cite{HP} and applied by Read \cite{ReadHalf}.  These authors considered {\it bosons} at $\nu=1$, which is presumably qualitatively identical to fermions at even denominator filling fractions.  Both the effective mass and compressibility puzzles inherent in the original Chern-Simons approach are resolved in this way (at least for bosons at $\nu=1$). 

Our first task, in the following chapter, is to extend the Haldane-Pasquier formalism to arbitrary $\nu$ for either fermions or bosons.   In the case of bosons, Chapter \ref{chap:Fermions}, we will obtain an effective theory microscopically by following Read's analysis at $\nu=1$.  The effective action does lead to a consistent picture of the mass and compressibility, but differs fundamentally from the CS action; there is no {\it a priori} reason the two actions ought to look similar since we project to the LLL from the beginning.   However, it is gratifying that some of our results overlap with those of Shankar and Murthy, who do start with CS.  

In Chapter \ref{chap:Bosons}, we construct a phenomenological Landau-Ginzburg field theory for $\nu=1/p$, i.e. the underlying particles must be fermions for $p=$odd and bosons for $p=$even.  The spectrum in this case contains a magnetoroton dip, which is the first analytical observation of this phenomenon.

A central theme running through both of these chapters is the internal structure of the composite particles.  To compensate for the extra degrees of freedom of the vortices, a set of constraints is introduced and appears thorughout our models.

%The beginning of Chapter \ref{chap:Fermions} lays out the generalized formalism starting with one and two particles in a magnetic field, the building blocks of the bound state (Sections \ref{sec:SingleParticle} and \ref{sec:TwoParticles}).  We then describe the Fock space in Section \ref{sec:Fock_Physical}.  The vortices introduce extra degrees of freedom that must be projected out, which is the content of the constraints that will appear throughout.  The remainder of the chapter is an application of the formalism. 

%%%%%%%%%%%%%%%%%%%%%%%%%%%%%%%%%%%%%%%%%%%%%%%%%%%%%%%%%%%%%%%%%%%%%%%%%%
\section{Pairing in Two Dimensions}
\label{sec:intro_pairing}
%%%%%%%%%%%%%%%%%%%%%%%%%%%%%%%%%%%%%%%%%%%%%%%%%%%%%%%%%%%%%%%%%%%%%%%%%%
 
The standard theory for superconductivity was introduced by Bardeen, Cooper, and Schrieffer (BCS) almost fifty years ago \cite{BCS}.  The original ground state was a paired state of fermions in the s-wave channel, or relative angular momentum $l=0$.  Since then, there have been numerous generalizations to non-zero angular momentum and to other more complicated order parameters.  For a review, of the rich phase diagrams in He$^3$, see the book by Vollhardt and W\"{o}lfle \cite{Helium3}.  We refer to the non-zero angular momentum states as anomalous in the sense that they violate both parity and time reversal symmetry.  Typically, BCS theory is applied to fermionic particles, which can be thought of as pairing into bosons which then condense.  A less widely appreciated body of work treats pairing of bosons \cite{Nozieres}.  We explore instances of both statistics in Chapters \ref{chap:PERM} and \ref{chap:BCS}.   
There is growing evidence of anisotropic superconductivity in the perovskite oxides.  In Sr$_2$RuO$_4$, both experiment \cite{p-wave_expt} and theory \cite{p-wave_thry} support p-wave ($l=-1$) pairing. Similarly, d-wave pairing ($l=-2$) is by now well established in the high temperature superconductors \cite{d-wave_expt}.  p-wave pairing has also been observed in superfluid He$^3$ in the so-called ``A-phase'' \cite{Helium3}~.

In the context of the fractional quantum Hall effect, Halperin \cite{Halperin} proposed that under certain conditions, electrons can form pairs that condense into a Laughlin state of charge-2 bosons.  Since then, various alternatives for p- and d-wave pairing have been explored by several groups \cite{MR,RR,Greiter,Hf,Morf,Bonesteel,ReadGreen}.  An intriguing possibility is that the observed plateau at $\nu=5/2$ is the Pfaffian state of Moore and Read \cite{MR,Morf,nu52}.

Recently it has been proposed that the spin conductivity of the class of p- and d-wave states is transverse and quantized \cite{Senthil99}.  A remarkable series of earlier papers by Volovik \cite{Volovik} contained some of these predictions, as well.  However, the precise proof of this proposal has not been shown.  Section \ref{chap:BCS} contains our derivation of this effect with a conserving approximation.  In the context of the original BCS theory, a conserving approximation is required for a correct description of the collective mode \cite{BCS,anderson58} since it respects charge conservation.  Analogously, in our case, the conserving approximation will respect the conservation of spin current, leading to the correct result. 
 
A less familiar application of BCS theory is to paired states of bosons (see for example \cite{Nozieres}).  Typically, there is a competition between the usual single-particle condensate and a pure paired state, or a phase of coexistence is possible.  

Pure p-wave pairing of spin-$1/2$ bosons is characterized by a BCS wavefunction in the form of a permanent.  We will show that p-wave pairing can be interpreted as a condensate of spin waves.  As parameters in the Hamiltonian are tuned, one reaches another single particle condensate with helical spin order.  The permanent sits right on the transition and contains a single anti-Skyrmion, which is yet another single particle condensate.    
 
In the context of the fractional quantum Hall effect (FQHE), the permanent describes singlet pairs of spin-1/2 composite bosons with filling factor $\nu=1/p$, $p$ odd \cite{MR,RR}.  It is the unique ground state wavefunction of a Hamiltonian which penalizes the closest approach of three spin-1/2 fermions for a fixed number of flux quanta piercing the bulk.  The exact number of flux threading the bulk depends upon the geometry, e.g. on whether the system is on the torus, on the sphere, or on the plane.  Although the permanent itself is difficult to treat analytically, its Hamiltonian contains an infinite set of degenerate zero-energy eigenstates when flux is added, among which is the polarized Laughlin state, and the rest are interpreted in one of two equivalent ways as either quasiholes or as spin wave excitations.  The $p=1$ Laughlin state is most amenable to analysis as it is a Slater determinant of single particle wavefunctions (Section \ref{sec:Wavefunctions}).  Accordingly, it will serve as the prototype for our exact statements.         

We also consider d-wave pairing of spinless bosons, which is known as the Haffnian in FQHE literature \cite{Hf}.  A rich phase diagram of competing single particle and paired condensates emerges, with the Haffnian sitting on a phase boundary.

In the case of pairing in the FQHE, the bosons (fermions) are to be thought of as composite bosons (fermions), as described in Chapter \ref{chap:Fermions}.  We can use CS mean field theory such that the CS gauge field cancels the external field on the average, and we are left with composite particles in zero net field.  Then we can treat the composite particles within a BCS approximation.  The case of composite fermions, which was addressed recently \cite{ReadGreen}, serves as a point of reference for composite boson pairing.  The central theme in ref. \cite{ReadGreen} was the existence of two regimes, weak- and strong-pairing.  The weak-pairing phase can be characterized by a nontrivial topological winding of the BCS order parameter in momentum space; we will see an example in the discussion of the quantum Hall effect for spin in Chapter \ref{chap:BCS}.  On the other hand, the strong-pairing phase is topologically trivial, and, in the simplest case, the two phases are separated by a transition at zero chemical potential.  For example, the Pfaffian state \cite{MR} is a weak-pairing phase, while the Haldane-Rezayi state \cite{HR}, which is an $l=-2$ state, was found to lie at the weak-strong transition point.  It is useful to keep these results in mind as we consider composite boson pairing in this chapter. 

%%%%%%%%%%%%%%%%%%%%%%%%%%%%%%%%%%%%%%%%%%%%%%%%%%%%%%%%%%%%%%%%%%%%%%%%%%%%%%%
\section{Adsorption on Carbon Nanotubes}
\label{sec:intro_tubes}
%%%%%%%%%%%%%%%%%%%%%%%%%%%%%%%%%%%%%%%%%%%%%%%%%%%%%%%%%%%%%%%%%%%%%%%%%%%%%%

Monolayer adsorption of noble gases onto graphite sheets has proven to
be an interesting problem both theoretically and
experimentally \cite{Bretz,Schick,Murthy}.  Many of the observed
features can be understood within a lattice gas model, where the
underlying hexagonal substrate layer forms a triangular lattice of
preferred adsorption sites.  An equivalent formulation is in the
language of spin models on a triangular lattice, where the repulsion
between adsorbed atoms in neighboring sites translates into an
antiferromagnetic Ising coupling.  The frustration of the couplings by
the triangular lattice leads to the rich phase diagram of the
monolayer adsorption problem \cite{Schick}. Introducing hopping adds
quantum fluctuations, further enriching the phase
diagram \cite{Murthy,Moessner}.

In this Chapter we address what happens if, in addition to the
triangular lattice frustration, one has an extra geometric frustration
due to periodic boundary conditions.  In fact, such a system is
physically realized by a single walled carbon
nanotube \cite{Dresselhaus}, which may be viewed as a rolled graphite
sheet.  In this context, adsorption has been the subject of growing
experimental and theoretical interest \cite{Stan,Cole} spurred by
potential applications. Stan and Cole \cite{Stan} have considered the
limit of non-interacting adatoms at low density, finding that they are
localized radially near a nanotube's surface at a distance comparable
to that in flat graphite ($\sim 3{\mbox \AA}$). In that work, it was
sufficient to omit the hexagonal structure of the substrate.  However,
the corrugation potential selects the hexagon centers as additional
commensurate localization points \cite{Bretz}. In view of the
similarity to flat graphite, we include both the substrate lattice and
adatom interactions and consider a wider range of densities. In fact,
very recently, it has been shown \cite{ColeNew} that the adsorbate
stays within a cylindrical shell for fillings less than $\approx 0.1
/{\mbox \AA}^2$ (or $\approx 0.5$ adatom/hexagon), justifying the
densities studied here.

Our adsorption model is equivalent to a new sort of XXZ Heisenberg quantum spin tube, which is type of spin ladder with periodic boundary conditions.  A simple example with highly anisotropic couplings was considered recently in references \cite{Andrei-Cabra}.   We find density plateau structures for armchair, zig-zag and chiral nanotubes. In the language of spin systems, the density plateaus correspond to magnetization plateaus. The zig-zag tubes turn out to be special, and have extensive zero temperature entropy plateaus in the classical limit. Quantum effects lift the degeneracy, leaving gapless excitations described by a $c=1$ conformal field theory with compactification radius quantized by the tube
circumference.  This is an interesting conformal symmetry because the only other systems in nature with a quantized compactification radius, that we are aware of, are the chiral edge states in the FQHE \cite{Edge}.

\chapter{Lowest Landau Level I:  Composite Fermions}
\label{chap:Fermions}

We begin this chapter by developing the Haldane-Pasquier formalism in Section \ref{sec:Formalism_Fermions}. 

In the rest of the chapter we apply the composite fermion formalism to bosons in the lowest Landau level.  In particular, we obtain an effective theory for an incompressible quantum Hall liquid of bosons with one attached vortex at general filling.  As discussed in the introduction, theories based on the flux attachment, or Chern-Simons approach \cite{Zhang}, have not been entirely satisfactory.  More recently there have been several attempts to obtain an effective theory microscopically \cite{SM,HP,Lee}, which too have had difficulties.  In this work we avoid the Chern-Simons approach and follow an alternative that was developed for the Fermi liquid-like state of bosons at $\nu=1$ \cite{ReadHalf}.

The effective filling factor in our composite fermion model can be obtained as follows.  As we will see below, the composite feels an effective magnetic field $B=B_L+B_R$, where $B_L$ is the physical field felt by the underlying particle and $B_R$ is an arbitrary field felt by the vortex.  The total charge of the composite is $q^*=1+B_R/B_L$.  In the introduction we showed that the filling factor can be written as $\nu=\rho\Phi_0/B_L$, where $\rho$ is the density of the underlying particles and $\Phi_0$ is the magnetic flux quantum.  Since the composite particles must have the same density as the underlying particles, the effective filling is $\nu_{\rm eff}=\rho\Phi_0/B$, which can be rewritten as
\bea
\nu_{\rm eff} = \frac{\nu}{q^*}~,
\label{eq:Jain}
\eea
The original Jain construction \cite{Jain} begins with particles of charge $+1$ at $\nu=1/p$ and attaches $\widetilde\phi$ flux tubes.  On the average, the flux tubes renormalize the real magnetic field by $B_L\rightarrow q^*B_L$ with $q^*=1-\widetilde\phi\nu$.  In our framework, $q^*$ can be any real number, so there is a family of theories, i.e. anyons \cite{FradkinBook}, for {\it any} given $\nu$ parameterized by $q^*$ and an integer $\nu_{\rm eff}$.  

Strictly speaking, Jain's model does not require fermionic statistics for the underlying particles, so the bosonic case is a valid quantum Hall state.  In fact, there have been recent theoretical proposals that quantum Hall liquids of bosons are realizable in rotating Bose-Einstein condensates \cite{Gunn}.  In this case, the magnetic field is due to the angular velocity ${\bf\omega}$; roughly, the velocity is modified in the rotating frame by ${\bf v}\rightarrow{\bf v}+{\bf\omega}\times{\bf r}$, which is like minimal coupling of a magnetic field ($\nabla\times({\bf\omega}\times{\bf r})$ is a constant in the same direction as ${\bf\omega}$).  

In section \ref{sec:HF} we obtain an effective mass by extracting a kinetic term from the interaction Hamiltonian.  This problem has been central to the recent work in references \cite{SM,HP}.  The simplest starting point is a Hartree-Fock approximation.  Next, we use a conserving approximation, which restores the constraints, in order to calculate the correct density-density response function.  The self-consistent diagramatics consist of summing ring and ladder diagrams; it is essentially the same framework that we use for paired states of fermions in Chapter \ref{chap:BCS} of this thesis.

%%%%%%%%%%%%%%%%%%%%%%%%%%%%%%%%%%%%%%%%%%%%%%%%%%%%%%%%%%%%%%%%%%%%%%%%%%%%%%%
\section{Formalism}
\label{sec:Formalism_Fermions}
%%%%%%%%%%%%%%%%%%%%%%%%%%%%%%%%%%%%%%%%%%%%%%%%%%%%%%%%%%%%%%%%%%%%%%%%%%%%%%%

%%%%%%%%%%%%%%%%%%%%%%%%%%%%%%%%%%%%%%%%%%%%%%%%%%%%%%%%%%%%%%%%%%%%%%%%%%%%%%%
\subsection{Single Particle in a Magnetic Field}
\label{sec:SingleParticle}
%%%%%%%%%%%%%%%%%%%%%%%%%%%%%%%%%%%%%%%%%%%%%%%%%%%%%%%%%%%%%%%%%%%%%%%%%%%%%%%

A convenient framework for a quantum particle in two dimensions is an operator description.  We begin with the simplest case of a single charged particle in two dimensions and a perpendicular magnetic field. 

The Hamiltonian is 
\begin{equation}
  H = \frac{1}{2m}({\bf p}-q{\bf A})^2,
\label{eq:SingleParticleHamiltonian}
\end{equation}
where ${\bf p}$ is the canonical momentum, ${\bf A}$ is the vector potential for a magnetic field $B$, in the $\hat{\bf z}$ direction, and $q$ and $m$ are the charge and mass.  The units are set to $\hbar = c = 1$.

The kinetic momentum is defined by
\begin{equation}
  {\bf \pi} = {\bf p} - q{\bf A}
\label{eq:pi}
\end{equation}
with the corresponding commutator
\begin{equation}
	[\pi_\mu,\pi_\nu] = i\epsilon_{\mu\nu}qB
\label{eq:piCommutator}
\end{equation}
where $\mu$ and $\nu$ are space indices, $x$ and $y$, and $\epsilon_{\mu\nu}$ is the Levi-Civita symbol.  The dynamics of $\pi$ follow simply,
\bea
\stackrel{.}{\pi}_\mu = i[H,\pi_\mu]=\omega_c\epsilon_{\mu\nu}\pi_\nu~,
\label{eq:precession}
\eea
where $\omega_c=qB/m$ is the cyclotron frequency.  Therefore $\pi$ precesses.

As the particle executes cyclotron motion it is located by the guiding center operator
\begin{equation}
	{\bf R} = {\bf r} + \hat{\bf z}\times{\bf \pi}\frac{1}{qB}
\label{eq:R}
\end{equation}
which obeys
\begin{equation}
	[R_\mu,R_\nu] = -i\epsilon_{\mu\nu}\frac{1}{qB}~.
\label{eq:RCommutator}
\end{equation}
The operators ${\bf\pi}$ and ${\bf R}$ commute.  Projection to the lowest Landau level is accomplished by replacing the particle's coordinates by the guiding center.  Note that the coordinates no longer commute, a consequence of frozen degrees of freedom.  The appropriate generator of translations, or ``pseudomomentum'', is defined by
\begin{equation}
  {\bf K} = qB\hat{\bf z}\times{\bf R}~,
\label{K}
\end{equation}
and obeys $[K_\mu,K_\nu] = i\epsilon_{\mu\nu}\frac{1}{qB}$.  The planar coordinates of a quantum particle in a magnetic field are a well-known example of non-commutative space \cite{NonCommutativeSpace}.  

Noting the commutation relations of $\pi$ and $R$, we can define two independent harmonic oscillator operators:
\begin{eqnarray}
	a &=& \sqrt{\frac{1}{2qB}}~\overline{K}\,,\mbox{~~~~}a^\dagger=\sqrt{\frac{1}{2qB}}~K \\
	b &=& \sqrt{\frac{1}{2qB}}~\pi\,,\mbox{~~~~~}b^\dagger=\sqrt{\frac{1}{2qB}}~\overline{\pi}
\label{eq:ab}
\end{eqnarray}
where $\pi=\pi_x+i\pi_y$ and $\overline{\pi}=\pi_x-i\pi_y$, and similarly for $K$.  The Hamiltonian then becomes the familiar harmonic oscillator Hamiltonian:
\begin{eqnarray}
	H = \omega_c\left(b^\dagger b +\frac{1}{2}\right)~.
\label{eq:H_HO}
\end{eqnarray}
Its eigenstates are known as Landau levels (LL).  Each level is macroscopically degenerate since $a$ commutes with $b$ and drops out of the Hamiltonian.  The complete set of eigenstates is labeled by two integers, $m$ and $n$: 
\begin{eqnarray}
  \left|n,m\rangle\right.=
  \frac{a^{\dagger m}}{\sqrt{m!}}\frac{b^{\dagger n}}{\sqrt{n!}}\left|0,0\rangle\right.
\label{eq:nm}
\end{eqnarray}
In this convention, ${\bf\pi}$ (or $b$) is a purely inter-Landau level operator, and ${\bf K}$ (or $a$) is intra-Landau level.

To obtain the wavefunctions, we will use complex coordinates, ${\bf r} = z = x+iy$, and the symmetric gauge, ${\bf A} = -\frac{1}{2}{\bf r}\times{\bf B} = \frac{1}{2}B(-y,x)$.  The single particle operators become
\begin{eqnarray}
	\pi &=& -2i\partial_{\overline{z}}-\frac{qB}{2}iz \\
	K   &=& -2i\partial_{\overline{z}}+\frac{qB}{2}iz
\label{eqn:diff_form1}
\end{eqnarray}
If we restrict ourselves to the lowest Landau level (LLL), then the wavefunctions, $\psi_{0,m}({\bf r})$, are annihilated by $b$.  The general solution to this first order partial differential equation (up to gauge transformations) is
\be
  \psi_{0,m}({\bf r})=f_m(z)e^{-|z|^2/4\ell_B^2}~,
\ee
where $\ell_B^2=1/|qB|$ is the magnetic length.  The great simplification is that $f_m(z)$ must be an analytic function in $z$.  Further requiring that $a\psi_{0,0}=0$ yields $f_0$ and the rest of the $f_m$'s are generated by applying $a^\dagger$.  The result is the set of states spanning the LLL:
\be
u_m(z)=\frac{1}{\sqrt{2\pi2^mm!\ell_B^{m+2}}}\,z^m e^{-|z|^2/4\ell_B^2}~.
\label{eq:u_m}
\ee
It should be remarked that the intra-LL ladder operator has a very simple action on the $u_m$: $a^\dagger=z/\sqrt{2}\ell_B$ and $a=(\ell_B/\sqrt{2})\partial_z$.

%%%%%%%%%%%%%%%%%%%%%%%%%%%%%%%%%%%%%%%%%%%%%%%%%%%%%%%%%%%%%%%%%%%%%%%%%%%%%
\subsection{Two Particles in a Magnetic Field}
\label{sec:TwoParticles}
%%%%%%%%%%%%%%%%%%%%%%%%%%%%%%%%%%%%%%%%%%%%%%%%%%%%%%%%%%%%%%%%%%%%%%%%%%%%%

In this section, we introduce a bound state of two oppositely charged particles in a perpendicular magnetic field.  This formalism will be useful in interpreting the Haldane-Pasquier approach.  

Each component is characterized by its charge $q_i$ and the magnetic field that it feels, $B_i$.  For convenience, we define
\bea 
B_i=q_iB
\eea
We will assume that $B_1>0$ and $B_2<0$, guaranteeing the existence of a bound state.  In these units the charge is dimensionless.  If we fix $q_1=1$ then the total charge is 
\bea
q^*=\frac{B}{B_1}
\label{eq:q_eff}
\eea
There are two sets of guiding centers, $R_{\mu i}$, and pseudomomenta, $K_{\mu i}$, which are defined in  the same way as in section \ref{sec:SingleParticle}.  The algebra consists of two copies of the single particle, for example
\bea
  [R_{\mu i}, R_{\nu j}] = -i\epsilon_{\mu\nu}\delta_{ij}\frac{1}{B_i}~,
\label{eq:RCommutator2}
\eea
and so on. 

It turns out that this algebra can be mapped exactly into a single particle in an effective magnetic field $B$ \cite{ReadUnpub}, 
\bea
B = B_1+B_2~.
\nonumber
\label{eq:B_eff}
\eea
If we construct the effective translation and momentum operators by
\begin{eqnarray}
K &=& K_1 + K_2\\
\pi &=& \sqrt{-B_1 B_2}\left(\frac{1}{B_2}K_2 - \frac{1}{B_1}K_1\right)
\label{eq:PiK}
\end{eqnarray}
then the commutators of this algebra are exactly that of a single particle, equations (\ref{eq:piCommutator}, \ref{eq:RCommutator}, \ref{K}).  For example $[\pi_\mu,\pi_\nu]=i\epsilon_{\mu\nu}B$.  The physical picture becomes clearer once we define an effective position $\bf r$ by eqn. (\ref{eq:R}).  Solving for $\bf r$, we find
\bea
{\bf r} &=& -\hat z\times(K+\pi)\ell_B^2 \non\\
        &=& \frac{B_1{\bf R}_1+B_2{\bf R}_2}{B}-({\bf R}_1-{\bf                
            R}_2)\frac{\sqrt{-B_1B_2}}{B}~.
\label{eq:rEff}
\eea 
In the limit $B\rightarrow 0$, $\bf r$ becomes $({\bf R}_1+{\bf R}_2)/2$.  This is not surprising from a classical point of view; two opposite but equal charges travel in a straight line due to ${\bf E}\times{\bf B}$ drift, the ``guiding center'' moves off to infinity, and their position is given by a point exactly midway between them.
 
To proceed with the wavefunctions of the composite, label the real-space coordinates of each particle by $z$ and $\eta$.  The differential operators are given by
\begin{eqnarray}
K_1 = -2i\partial_{\overline{z}} + \frac{i}{2\ell_{B_1}^2}z \\
K_2 = -2i\partial_{\overline{\eta}} - \frac{i}{2\ell_{B_2}^2}\eta
\end{eqnarray}
Note the relative minus sign, which comes from assuming $B_1>0$ and $B_2<0$.  The ladder operators are given in terms of $\pi$ and $K$ by equation (\ref{eq:ab}).  The lowest eigenfuction is determined from $a\psi_{0,0} = b\psi_{0,0} = 0$:
\begin{eqnarray}
&\frac{i}{\sqrt{2B}}\left[\left(-2\partial_{z} -
\frac{1}{2\ell_{B_1}^2}\overline{z}\right)+\left(-2\partial_{\eta}
+ \frac{1}{2\ell_{B_2}^2}\overline{\eta}\right)\right]\psi_{0,0} = 0
\nonumber \\
&\frac{i}{\sqrt{2B}}\left[\frac{1}{B_1}\left(-2\partial_{\overline{z}} +
\frac{1}{2\ell_{B_1}^2}z\right)-\frac{1}{B_2}\left(-2\partial_{\overline{\eta}}
- \frac{1}{2\ell_{B_2}^2}\eta\right)\right]\psi_{0,0} = 0
\label{eq:abPsi}
\end{eqnarray}
By analogy with a single particle, we expect that the solution is an analytic function in $z$ and $\eta$ times two Gaussian factors.  Indeed, if we choose 
\begin{eqnarray}
\psi_{0,0}(z,\eta) = \phi(z,\overline\eta)e^{-|z|^2/4\ell_{B_1}^2}
e^{-|\eta|^2/4\ell_{B_2}^2} \nonumber
\end{eqnarray}
then equations (\ref{eq:abPsi}) are solved by 
\bea
\phi(z,\eta) = \frac{1}{2\pi\ell_{B_1}\ell_{B_2}}\,e^{z\overline\eta/2\ell_{B_2}^2}~.
\label{eq:phiz}
\eea 
Note the asymmetry between $B_1\leftrightarrow B_2$, stemming from the sign of $B$.  We have implicitly assumed that $B>0$, but if $B<0$ then the Gaussian factor in eqn. (\ref{eq:phiz}) would be $\ell_{B_1}^2$.  

The complete set of states is generated by $a^\dagger$ and $b^\dagger$ just as it was for one particle in eqn. (\ref{eq:nm}), 
\bea 
\psi_{\la\mu}(z,\overline{\eta})=\langle z,\ol\eta|\la\mu\rangle=
\frac{b^{\dagger\la}}{\sqrt{\la!}} 
   \frac{a^{\dagger\mu}}{\sqrt{\mu!}}\psi_{0,0}(z,\overline{\eta})~.
\label{eq:Psi_mulambda}
\eea
The $\psi_{\la\mu}$ are linear combinations of the independent particle basis
\bea
  \langle z,\ol\eta|mn\rangle=u_m(z)\overline{v_n(\eta)}~,
\label{eq:TwoParticleBasis}
\eea
where $u$ and $v$ are LLL single particle states corresponding to the two magnetic lengths $\ell_{B_1}$ and $\ell_{B_2}$, as in eqn. (\ref{eq:u_m}).  

%%%%%%%%%%%%%%%%%%%%%%%%%%%%%%%%%%%%%%%%%%%%%%%%%%%%%%%%%%%%%%%%%%%%%%%
\subsection{Fock Space, Operators, and Constraints}
\label{sec:Fock_Physical}
%%%%%%%%%%%%%%%%%%%%%%%%%%%%%%%%%%%%%%%%%%%%%%%%%%%%%%%%%%%%%%%%%%%%%%%

Having constructed the basis functions for a particle-vortex pair in the previous section, we move on to the Fock space for a many-particle system. 

We begin with canonical fermionic or bosonic operators which are matrices with two indices, $c_{mn}$, with (anti-)commutation relations
\bea
  [c^{}_{mn},c^\dagger_{n^\prime m^\prime}]_{_\pm}
  =\delta_{mm^\prime}\delta_{nn^\prime}
\label{eq:AntiCommutator}
\eea
The left index, $m$, runs from $1$ to $N_\phi$, the number of available states in the LLL.  The right index, $n$, runs $1$ through $N$, which we will interpret later as the number of vortices.  In the thermodynamic limit, the filling factor is $\nu=N/N_\phi$.  Strictly speaking, this construction must be carried out on a finite geometry (e.g. $m$ runs from $1$ to $N_\phi+1$ on the sphere), but we will ignore this subtlety here since we will only be interested in the thermodynamic limit.  

The anticommutation relations are invariant under independent transformations on the left and right indices:
\bea
c\mapsto U_L c\,U_R~,
\label{eq:Unitary}
\eea
where $U_L$ and $U_R$ are $N_\phi\times N_\phi$ and $N\times N$ unitary matrices.  These transformations are generated by the left and right ``density'' operators
\bea
  \rho^R_{nn^\prime}=\sum_{m=1}^{N_\phi} c^\dagger_{nm}c_{mn^\prime} \nonumber\\
  \rho^L_{mm^\prime}=\sum_{n=1}^N c^\dagger_{nm}c_{m^\prime n}
\label{eq:Densities}
\eea 
The left density $\rho^L$ will represent the physical density, as we will see below.
The right density $\rho^R$ specifies a set of $N^2$ constraints, which we use to define a set of physical states,
\bea \left(\rho^R_{nn^\prime}-\delta_{nn^\prime}\right)|\Psi_{\mbox{\scriptsize{phys}}}\rangle = 0~.
\label{eq:Constraints}
\eea 
It will be shown shortly that the set of $|\Psi_{\rm phys}\rangle$ do indeed give the correct Fock space.  

Because the $\rho^R$ generate the unitary group $\mbox U(N)_R$ and there is a phase factor, $\mbox U(1)$, common to both $\rho^R$ and $\rho^L$, the physical states must be singlets under $\mbox{SU}(N)_R$.  The physical states solving this constraint are linear combinations of
\bea
|\Psi_{\mbox{\scriptsize{phys}}}^{m_1\cdots m_N}\rangle = 
\sum_{n_1\cdots n_N}\epsilon^{n_1\cdots n_N}
c^\dagger_{n_1m_1}c^\dagger_{n_2m_2}\cdots c^\dagger_{n_Nm_N}
|0\rangle~,
\label{eq:PhysStates}
\eea
where $|0\rangle$ is the vacuum with no fermions.  The Levi-Civita symbol $\epsilon$ ensures that these states are singlets under $\mbox{SU}(N)_R$.  If the $c$'s are fermions their anticommutation relations ensure that $|\Psi_{\mbox{\scriptsize{phys}}}^{m_1\cdots m_N}\rangle$ is symmetric under the interchange of any pair $m_i\leftrightarrow m_j$.  Therefore, the physical space is equivalent to $N$ bosons each of which can occupy any one of $N_\phi$ states, i.e. the Fock space of bosons at filling $\nu=N/N_\phi$.  Had the $c$'s been bosonic operators rather than fermionic, the result would have been a Fock space of fermions at the same filling.

We now construct the many body operators.  By mapping into a single particle, we showed that the effective magnetic length is
\bea
\frac{1}{\ell_B^2}=\frac{1}{\ell_{B_L}^2}-\frac{1}{\ell_{B_R}^2}
\label{eq:effL}
\eea 
As a reminder of the physical picture, we have made the notation change from $1,2$ to $L,R$.  Our convention guarantees that $\ell_B$ is positive since $\ell_{B_L}<\ell_{B_R}$ (equivalently, $|B_L|>|B_R|$ and $B=B_L+B_R>0$). 

In real space the matter field is defined by
\bea
c(z,\ol\eta)=\sum_{mn}u^L_m(z)\ol{u^R_n(\eta)}~c_{mn}
\label{eq:cuv}
\eea
where $u^L_m(z)=z^m\exp\{-|z|^2/4\ell_{B_L}^2\}$ and $u^R_m(\eta)=z^m\exp\{-|\eta|^2/4\ell_{B_R}^2\}$ (apart from normalizations).  A unitary transformation connects the $|mn\rangle$ independent particle basis to the $|\mu\la\rangle$ bound state basis in section \ref{sec:TwoParticles}.  Accordingly, the fermions $c_{mn}$ transform into
\bea
c_{\mu\la}=\sum_{mn}c_{mn}\langle mn|\mu\la\rangle~,
\label{eq:cBasis}
\eea
where 
\bea
|\mu\la\rangle=\frac{a^{\dagger\mu}}{\sqrt{\mu !}}\frac{b^{\dagger\la}}{\sqrt{\la !}}~|0,0\rangle~.
\label{eq:MuLa}
\eea
The overlap $\langle mn|\mu\la\rangle$ is obtained from the definitions in eqns. (\ref{eq:Psi_mulambda}) and (\ref{eq:TwoParticleBasis}).  To write the density operators, we observe that they take a plane wave form in the operator language:
$\hat\rho^R_{\bq}=\sum_ie^{i\bq\cdot{\bf R}_{i,R}}$ and $\hat\rho^L_{\bq}=\sum_ie^{i\bq\cdot{\bf R}_{i,L}}$,
where ${\bf R}_{i,R}$ and ${\bf R}_{i,L}$ are the two guiding center coordinates of the $i$'th particle.\footnote{I was reminded by R.~Shankar that he had guessed the same density expressions based on a small $\bq$ limit of the Chern-Simons formulation \cite{Shankar99}.}  In second quantization, the left density becomes
\bea
\hat\rho^L_\bq = \sum_{\mu\la,\,\mu^\prime\la^\prime} c^\dagger_{\la\mu}c_{\mu^\prime\la^\prime} \langle\mu\la|\,e^{i\bq\cdot{\bf R}_L}|\mu^\prime\la^\prime\rangle~
\label{eq:rhoMatrixOp}
\eea
and similarly for $\hat\rho^R_\bq$.  The matrix element can be calculated by solving for ${\bf R}_{R,L}$ in terms of ${\bf K}$, $\pi$, giving
\bea
{\bf R}_L &=& \wedge ({\bf K}+\frac{\ell_{B_L}}{\ell_{B_R}}\,{\bf\pi})\ell_B^2 \non\\
{\bf R}_R &=& \wedge ({\bf K}+\frac{\ell_{B_R}}{\ell_{B_L}}\,{\bf\pi})\ell_B^2~.
\label{eq:RLRR}
\eea
We have introduced the shorthand notation, $\wedge{\bf a} = -\hat z\times{\bf a}$ (for the vector ${\bf a}$).  Because $[{\bf K},{\bf\pi}]=0$, the plane wave factors into an intra- and an inter-Landau level piece.  We write
\bea
\hat\rho^L_\bq &=& \sum_{\mu\la,\,\mu^\prime\la^\prime}\rho_\bq(\mu|\mu^\prime) \rho^L_\bq(\la|\la^\prime)\,c^\dagger_{\la\mu}c_{\mu^\prime\la^\prime} \non\\
\hat\rho^R_\bq &=& \sum_{\mu\la,\,\mu^\prime\la^\prime}\rho_\bq(\mu|\mu^\prime) \rho^R_\bq(\la|\la^\prime)\,c^\dagger_{\la\mu}c_{\mu^\prime\la^\prime}~, 
\label{eq:RLRRf}
\eea
where the $\rho$-coefficients are defined by 
\bea
\rho_\bq(\mu|\mu^\prime) &=& \langle\mu|\exp\left\{i\ell_B^2\,\bq\wedge{\bf K}\right\}|\mu^\prime\rangle \non\\
\rho^L_\bq(\la|\la^\prime) &=& \langle\la|\exp\left\{i\ell_B^2\frac{\ell_{B_L}}{\ell_{B_R}}\,\bq\wedge{\bf \pi}\right\}|\la^\prime\rangle \\
\rho^R_\bq(\la|\la^\prime) &=& \langle\la|\exp\left\{i\ell_B^2\frac{\ell_{B_R}}{\ell_{B_L}}\,\bq\wedge{\bf \pi}\right\}|\la^\prime\rangle~. \non
\label{eq:fDef}
\eea
The operation ${\bf a}\wedge{\bf b}$ stands for ${\bf a}\cdot\wedge{\bf b}$.  Note that $\hat\rho_\bq$ is identical to $\hat\tau_\bq$ for a particle in field $B$ and therefore follow the same orthonormality properties and commutation relations.  $\rho^L_\bq$ and $\rho^R_\bq$ are similar but with an additional factor of $\ell_{B_L}/\ell_{B_R}$ or $\ell_{B_R}/\ell_{B_L}$ in the phase of the commutator.  
%Perhaps it is not surprising that the $f_\bq$-coefficients have commutation properties similar to the magnetic translations $\tau_\bq$ (cf. equation (\ref{eq:TauComm})).  For example, if we view $f_\bq(\mu|\mu^\prime)$ as a matrix in $\mu$, then 
%\bea
%\hat f_\bk \hat f_{\bk^\prime} = \hat f_{\bk+\bk^\prime}~e^{\frac{1}{2}i\ell_B^2\bk\wedge\bk^\prime}
%\label{eq:fComm}
%\eea
%and similarly for $\hat f^{L,R}$, but with factors of $\ell_{B_L}/\ell_{B_R}$ in the phase.

Since $K$ and $\pi$ are nothing other than harmonic oscillator operators, the $\hat\rho$'s can be calculated explicitly.  The one that we need later is $\hat\rho^L$, so we use it as an example.  First, rewrite $i\bq\wedge\pi$ in complex coordinates as $\frac{1}{2}(\ol q\pi-q\ol\pi)$.  Secondly, recall that $b\sim\pi$ and $|\la\rangle\sim b^{\dagger\la}|0\rangle$, which reduces the calculation to a harmonic oscillator matrix element.  The rest is straightforward and we quote the final result:
\bea
\rho^L_{\frac{\sqrt{2}}{\ell_B}\frac{\ell_{B_R}}{\ell_{B_L}}\bq}(\la|\la^\prime)= 
\frac{1}{\sqrt{\la !}}\frac{1}{\sqrt{\la^\prime !}}~e^{q\ol q/2}~(-\partial_{\ol q})^\la (\partial_q)^{\la^\prime}~e^{-q\ol q}~.
\label{eq:f_explicit}
\eea
A bunch of factors were absorbed into $\bf q$ on the left-hand side to avoid repetitiously writing them on the right.

The final step in the construction is the Hamiltonian.  In coordinate space it involves only the diagonal components of $\hat\rho^L$ \cite{ReadHalf},
\bea
H=\frac{1}{2}\int d^2z_1 d^2z_2~V({\bf r}_1-{\bf r}_2)~:\rho^L(z_1,\ol z_1)\rho^L(z_2,\ol z_2):
\label{eq:H1}
\eea
Or in Fourier space (Appendix),
\bea
H=\frac{1}{2}\int\frac{d^2\bq}{(2\pi)^2}~V(\bq)\,e^{-|q|^2/2\ell_{B_L}^2}\, :\hat\rho^L_\bq~\hat\rho^L_{-\bq}:
\label{eq:HFourier}
\eea
where $V(\bq)$ is the ordinary Fourier transform of $V({\bf r})$.  As required, this Hamiltonian is both translationally and rotationally invariant.  By construction, the right density $\hat\rho^R$ is a constant of the motion because the Hamiltonian acts only on the left indices, i.e.
\bea
[H,~\hat\rho^R_\bq]=0~.
\label{eq:Hinv}
\eea 
The system of Hamiltonian plus constraints is the starting point for Read's analysis of composite bosons at $\nu=1$ \cite{ReadHalf}.  

%%%%%%%%%%%%%%%%%%%%%%%%%%%%%%%%%%%%%%%%%%%%%%%%%%%%%%%%%%%%%%%%%%%%%%%%%%%%%%%
\section{Hartree-Fock Approximation}
\label{sec:HF}
\begin{figure}[htb]
\noindent
\center
\epsfxsize=1.5 in
\epsfbox{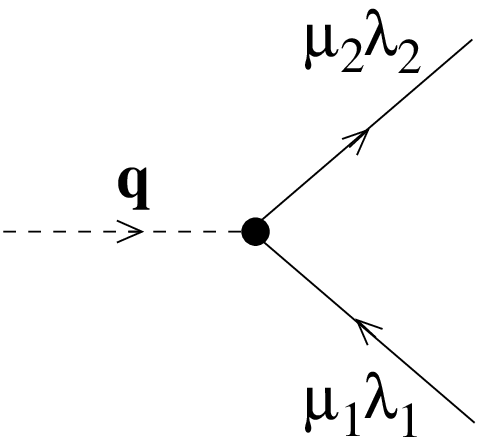}
\caption{\small The vertex $\rho_\bq(\mu_1|\,\mu_2)\rho^L_\bq(\la_1|\la_2)$.  The dotted line represents the interaction $\widetilde V(\bq)$}
\label{fig:vertex}
\end{figure}

%%%%%%%%%%%%%%%%%%%%%%%%%%%%%%%%%%%%%%%%%%%%%%%%%%%%%%%%%%%%%%%%%%%%%%%%%%%%%%%      
For convenience, we restate here the Hamiltonian that we derived in the last section.
\bea
H=
\frac{1}{2}\sum_{\mu_i,\,\la_i}\int\frac{d^2\bq}{(2\pi)^2}\,\widetilde V(\bq)\,F_\bq(\mu_1\la_1,\mu_3\la_3|\,\mu_2\la_2,\mu_4\la_4)\,
c_{\la_1\mu_1}^\dagger c_{\la_3\mu_3}^\dagger c_{\mu_4\la_4} c_{\mu_2\la_2}~,
\label{eq:H_fermions}
\eea
where the matrix element is given by
\bea
F_\bq(\mu_1\la_1,\mu_3\la_3|\,\mu_2\la_2,\mu_4\la_4)\!=\!
\rho_\bq(\mu_1|\mu_2)\rho_{-\bq}(\mu_3|\mu_4)\rho^L_\bq(\la_1|\la_2)\rho^L_{-\bq}(\la_3|\la_4) 
\label{eq:F}
\eea
and $\widetilde V(\bq)=V(\bq)e^{-|\bq|^2/2\ell^2_{B_L}}$ is the apodized potential.  The vertex is shown in Fig. \ref{fig:vertex}

The Hartree-Fock (HF) approximation consists of replacing pairs of fermion operators by their expectation value at zero temperature:
\bea
\langle c^\dagger_{\mu\la}c_{\mu^\prime\la^\prime}\rangle_{_0} = 
\delta_{\mu\mu^\prime}\delta_{\la\la^\prime}\,\Theta(\la_{max}-\la)~,
\label{eq:T0expectation}
\eea
which fills $\la_{max}+1$ Landau levels of composite bosons.  In Jain's mapping \cite{Jain}, this corresponds to an effective filling fraction $\nu_{\rm eff}=\la_{max}+1$.  At finite temperature, the $\Theta$-function is replaced by the Fermi distribution $f(\varepsilon_\la-\mu_c)$, where $\mu_c$ is the chemical potential.  Since the chemical potential is restricted to $\varepsilon_{\la_{max}}<\mu_c<\varepsilon_{\la_{max}+1}$ but is otherwise arbitrary, we will drop it in the following. 

Expanding the Hamiltonian around this ground state allows us to sum over the intra-level indices $\mu_i$.  In the following, we ignore the direct term, which only shifts the chemical potential.  In the exchange term, completeness of the $|\mu\rangle$ basis within a Landau level, $\sum_\mu |\mu\rangle\langle\mu |=1$, gives $\delta_{\mu_1\mu_4}$ or $\delta_{\mu_2\mu_3}$.  Completeness of the intra-LL basis is tantamount to translation invariance.  Rotation invariance, on the other hand, shows up in $\widetilde V(\bq)$, which is required to be a function of only $|\bq|$ for an isotropic system.  This gives terms diagonal in $\la$, $\delta_{\la_1\la_4}$ or $\delta_{\la_2\la_3}$, since the others vanish when we consider the explicit expression for $\hat f^L$ in equation (\ref{eq:f_explicit}).  The HF Hamiltonian is now
\bea
H_0=\sum_{\mu\la}\varepsilon_\la\,c^\dagger_{\la\mu}c_{\mu\la}
\label{eq:H_HF}
\eea
with the exchange energy
\bea
\varepsilon_\la=-\int\frac{d^2\bq}{(2\pi)^2}\,\widetilde V(\bq)\,\sum_{\la^\prime=0}^{\la_{max}} 
|\rho^L_\bq(\la|\la^\prime)|^2~.
\label{eq:e_HF}
\eea
For concreteness, we choose the simplest non-trivial case: we fill only the LLL, $\la_{max}=0$, and take a hard-core repulsive interaction, $V(\bq)=V(0)$.  Using the explicit form of $\hat f^L_\bq$ from equation (\ref{eq:f_explicit}) the energy becomes
\bea
\varepsilon_\la &=& -\left(2\frac{\ell_{B_L}^2}{\ell_{B_R}^2}\,\ell_B^2\right)^\la\frac{1}{\la!} \int\frac{d^2\bq}{(2\pi)^2}\,V(\bq)\,e^{-\frac{1}{2}\ell_B^2|\bq|^2}|\bq|^{2\la} \non \\
&=& -\left(\frac{\ell_{B_L}}{\ell_{B_R}}\right)^{2\la}\,V(0)\,\ol\rho~,
\label{eq:e_0}
\eea
where $\ol\rho=1/2\pi\ell_B^2$ is the density of the composite fermions.  Note how the apodized potential now has $\ell_B^2$, not $\ell_{B_L}^2$, in the Gaussian, which is a consequence of the relation $1/\ell_B^2=1/\ell_{B_L}^2\!-\!1/\ell_{B_R}^2$.  Because $\ell_{B_L}^2<\ell_{B_R}^2$, the energy vanishes asymptotically as $\la\rightarrow\infty$.  Furthermore, the exponential form implies a linear dependence on $\la$ at small $\la$, exactly the kind of behavior that one would expect for non-interacting particles in a magnetic field. The cyclotron frequency, from equation (\ref{eq:H_HO}), is given by $\omega_c=1/m\ell_B^2$ so we can identify an effective mass with the gap $\Delta$ by $1/m^* = (\varepsilon_1-\varepsilon_0)\ell_B^2$.  More generally when $\la_{max}>0$, the low energy physics is dominated by transitions between the highest occupied LL and the lowest unoccupied one, which gives
\bea
\frac{1}{m^*}&=&(\varepsilon_{\la_{max}+1}-\varepsilon_{\la_{max}})\,\ell_B^2\non\\
&=& \ell_B^2\Delta~.
\label{eq:m_eff}
\eea 
Further justification for this interpretation will emerge as we consider fluctuations around the ground state.  At any rate, our calculation provides a framework to calculate $m^*$ in the LLL.  Note that equation (\ref{eq:e_HF}) shows that $m^*$ has contributions from the lowest LL's (of composite particles) and identifies it with a particular integral over the interaction.

As a stand-alone approximation, HF does not preserve the constraints because the commutator $[H_0,\hat\rho^R_\bq]\neq 0$ for all $\bq\neq 0$.  In the next section we augment HF in a fully self-consistent manner to restore this symmetry.

\hyphenation{Identity}
%%%%%%%%%%%%%%%%%%%%%%%%%%%%%%%%%%%%%%%%%%%%%%%%%%%%%%%%
\section{Conserving Approximation and the Ward Identity}
\label{sec:ConservingApprox}
%%%%%%%%%%%%%%%%%%%%%%%%%%%%%%%%%%%%%%%%%%%%%%%%%%%%%%%%
\hyphenation{Iden-tity}

Our goal is to find a perturbative series such that all correlation function that involve the constraint $\hat\rho^R_\bq-\ol\rho\,\delta_{\bq,0}$ vanish.  In other words, the constraints would vanish to any order of approximation.  This is guaranteed by an exact (non-perturbative) Ward identity, which is derived below.  We illustrate the method by calculating the response functions, or generalized susceptibilities, of $\hat\rho^R-\hat\rho^R$, $\hat\rho^R-\hat\rho^L$, and $\hat\rho^L-\hat\rho^L$.  The latter is related to the physical quantity of interest, the compressibility.

It is well known in the theory of metals \cite{BCS,FW} that if a Hartree-Fock approximation is used for the two-particle Green's function, then a fully self-consistent approximation that conserves charge includes ladder and bubble diagrams in the response functions.  In fact this method preserves the constraints in the composite boson problem at $\nu=1$ as well \cite{ReadHalf}, and we will show that it works here, too.

In imaginary time, the response functions take the form
\bea
\chi^{AB}(\bq,i\omega_n)=\langle\hat\rho^A_\bq(i\omega_n)\hat\rho^B_{-\bq}(-i\omega_n)\rangle~,
\label{eq:Chi_def}
\eea
where $A, B$ stand for $R$ or $L$ and $\omega_n=2n\pi/\beta$ are Matsubara frequencies.  We implicitly keep only the connected part, thus dropping a term containing $\langle\hat\rho^A\rangle$'s.  The fundamental diagrams are those that are irreducible, i.e. those that cannot be separated by cutting an interaction line.  For a short range interaction, these are the qualitatively relevant pieces \cite{BCS,FW}, so we will not perform the bubble sums explicitly here.  In any case, they are easily obtained as geometric series of the irreducible parts \cite{ReadHalf,FW}.

The form of the conserving approximation in our case states that the irreducible response functions, $\chi^{AB}_{\mbox{\scriptsize{irr}}}$, are to be calculated by including the ladder series with the HF Green's function lines.  We sum the series by solving Dyson's equation; the next few equations will describe the structure of the theory.  

First, let us recall the HF Green's function \cite{FW},
\bea
{\cal G}_0(\la,i\omega_\nu)&=&\frac{1}{i\omega_\nu-(\varepsilon_\la-\mu)}~, \\
\varepsilon_\la &=& -\frac{1}{\beta}\sum_\nu\sum_{\la^\prime}\int\frac{d^2\bq}{(2\pi)^2}\, \widetilde V(\bq)\, \rho^L_\bq(\la|\la^\prime)\rho^L_{-\bq}(\la^\prime|\la)\, {\cal G}(\la^\prime,i\omega_\nu) \non\\
&=& -\sum_{\la^\prime}\int\frac{d^2\bq}{(2\pi)^2}\widetilde V(\bq)\,|\rho^L_\bq(\la|\la^\prime)|^2\,f(\varepsilon_{\la^\prime}-\mu_c)~,
\label{eq:Greens_function}
\eea
where $\omega_\nu=(2\nu+1)\pi/\beta$ is a fermionic Matsubara frequency, and $f(\varepsilon_{\la^\prime}-\mu_c)$ is the Fermi distribution with respect to the LL index ($\mu_c$ and $\nu$ should not be confused with the intra-LL index and the filling fraction).  

Second, the renormalization of the vertices, $\La^A$, by the ladder series can be written as a matrix Dyson equation:
\bea
\lefteqn{\La^A_{\mu\la,\,\mu^\prime\la^\prime}(\bq,i\omega_n) = \rho^A_\bq(\mu\la|\,\mu^\prime\la^\prime)\,-}\\
& &-\sum_{\mu_i,\,\la_i}
\int\frac{d^2\bk}{(2\pi)^2}\widetilde V(\bk)F_\bk(\mu\la,\mu_2\la_2|\,\mu_1\la_1,\mu^\prime\la^\prime)
{\cal D}_{\la_1\la_2}(i\omega_n)\La^A_{\mu_1\la_1,\,\mu_2\la_2}(\bq,i\omega_n)\non
\label{eq:Dyson_Vertex_Full}
\eea
where $F_\bk$ has been defined in equation (\ref{eq:F}) and ${\cal D}$ is the frequency sum over the internal Green's functions,
\bea
{\cal D}_{\la_1\la_2}(i\omega_n)&=&\frac{1}{\beta}\sum_\nu{\cal G}_0(\la_2,i\omega_\nu+i\omega_n){\cal G}_0(\la_1,i\omega_\nu)\\
&=&\frac{f(\varepsilon_{\la_2}-\mu_c)-f(\varepsilon_{\la_1}-\mu_c)}
{\varepsilon_{\la_2}-\varepsilon_{\la_1}-i\omega_n}~.
\label{eq:D}
\eea
Because the Green's function is independent of the intra-LL index $\mu$, it is convenient to define a purely inter-LL vertex, $\widetilde\La$ by
\bea
\La^A_{\mu\la,\mu^\prime\la^\prime}(\bq,i\omega_n)=
\rho_\bq(\mu|\mu^\prime)\,\widetilde\La^A_{\la\la^\prime}(\bq,i\omega_n) 
\label{eq:Vertex}
\eea
so that Dyson's equation becomes
\bea
\label{eq:Dyson_Vertex}
\lefteqn{\widetilde\La^A_{\la\la^\prime}(\bq,i\omega_n)=\rho^A_\bq(\la|\la^\prime)\,-}\\
& &-\int\frac{d^2\bk}{(2\pi)^2}\widetilde V(\bk)\rho^L_\bk(\la|\la_1)\rho^L_{-\bk}(\la_2|\la^\prime)
{\cal D}_{\la_1\la_2}(i\omega_n)\,
e^{i\bq\wedge\bk\ell_B^2}\,\widetilde\La^A_{\la_1\la_2}(\bq,i\omega_n).\non
\eea
The phase factor is due to the magnetic translation commutator algebra---see the discussion immediately following equation (\ref{eq:fDef}).

\begin{figure}[htb]
\noindent
\center
\epsfxsize=3 in
\epsfbox{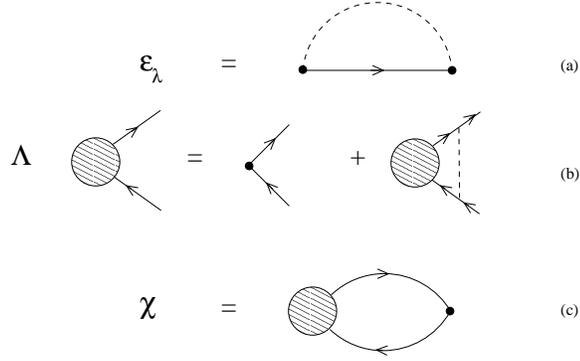}
\caption{\small Diagramatics of the conserving approximation.  (a) The exchange self energy; (b) the ladder series renormalization of the vertices, $\La$; and (c) the response functions in terms of $\La$.}
\label{fig:HF_vertices}
\end{figure}

The response functions are given in terms of the renormalized vertices by
\bea
\chi^{AB}_{\mbox{\scriptsize{irr}}}(\bq,i\omega_n)=
-\rho_0\sum_{\la_i}\widetilde\La^A_{\la_1\la_2}(\bq,i\omega_n){\cal D}_{\la_1\la_2}(i\omega_n) \rho^B_{-\bq}(\la_2|\la_1)~,
\label{eq:chiAB}
\eea
where $\rho_0=1/2\pi\ell_B^2$ is the density of particles per LL, coming from the trace over $\mu$'s.  The diagramatic structure is shown schematically in fig. \ref{fig:HF_vertices}

The key to showing that all response function containing $\La^R$ vanish, i.e  $\chi^{RR}_{\mbox{\scriptsize{irr}}}=\chi^{LR}_{\mbox{\scriptsize{irr}}}=\chi^{RL}_{\mbox{\scriptsize{irr}}}=0$, is the Ward identity for $\La^R$, which we now derive.  The first principles derivation follows standard field theoretic techniques \cite{BCS}.  Consider the exact vertex in real time:
\bea
\La^R_{\mu_1\la_1,\,\mu_2\la_2}(\bq,t,t_1,t_2) &=& 
\langle{\cal T}\left\{\hat\rho^R_\bq(t)c_{\mu_1\la_1}(t_1)c^\dagger_{\la_2\mu_2}(t_2)\right\}\rangle~,
\label{eq:expectation}
\eea
where ${\cal T}$ is the time ordering symbol.  By taking the time derivative $\partial_t$ of both sides, using $\partial_t\hat\rho^R_\bq=0$, and then Fourier transforming back to frequency space, we find the exact Ward identity
\bea
\label{eq:WI}
i\omega_n\tilde\La^R_{\la_1\la_2}(\bq,i\omega_n) =
{\cal G}^{-1}(\la_1,i\omega_n\!+\!i\omega_\nu)\rho^R_\bq(\la_1|\la_2)-
\rho^R_\bq(\la_1|\la_2){\cal G}^{-1}(\la_2,i\omega_\nu)~.
\eea
Here, ${\cal G}$ is the exact Green's function, and $i\omega_\nu$ on the right-hand side cancels identically, but is introduced for convenience.  The two terms are due to differentiating the time ordering; physically, they are due to $c^\dagger(t_1)$ and $c(t_2)$ acting as sources in equation (\ref{eq:expectation}).  For the particular HF and ladder series that we use here, we can verify the Ward identity by substituting ${\cal G}_0$ for the exact ${\cal G}$ and plugging the whole expression into the right-hand side of the Dyson equation (\ref{eq:Dyson_Vertex}).  Upon using the definition of ${\cal G}_0$ from equation (\ref{eq:Greens_function}), we find that the ladder series satisfies the Ward identity.

Therefore our diagramatic scheme preserves the constraints, and we can be sure that the physical quantities that we calculate in this approximation will be consistent. 

%%%%%%%%%%%%%%%%%%%%%%%%%%%%%%%%%%%%%%%%%%%%%%%%%%%%%%%%
\section{Response Functions}
\label{sec:Response}
%%%%%%%%%%%%%%%%%%%%%%%%%%%%%%%%%%%%%%%%%%%%%%%%%%%%%%%%

It is now straightforward to use the Ward identity in the response functions, equation (\ref{eq:chiAB}), to verify that
\bea
\chi^{RR}_{\mbox{\scriptsize{irr}}}(\bq,i\omega_n)=
\chi^{RL}_{\mbox{\scriptsize{irr}}}(\bq,i\omega_n)=
\chi^{LR}_{\mbox{\scriptsize{irr}}}(\bq,i\omega_n)= 0
\eea
and in fact for any correlator containing $\La^R$.  There is one proviso in this procedure, that is discussed in detail in reference \cite{ReadHalf}, having to do with the constraints at $i\omega_n=0$.  Our procedure is only valid at non-zero frequencies because we divided the Ward identity by $i\omega_n$ to isolate $\La^R$.  A complete proof requires more care, but we will not pursue this here.  In any case, there is no problem with taking the limit $\omega\rightarrow 0$, which requires only small but non-zero frequencies.

It remains to calculate the physical density-density response, $\chi^{LL}$.  To this end, we first rewrite the vertex in a more symmetric fashion by introducing the scattering matrix, $\Ga$, for the ladder series.  Although the internal Green's functions do not carry momentum, which comes in only through the vertices, we can nonetheless absorb some of the momentum dependence into $\Ga$ by taking advantage of translation invariance again.  We define
\bea
\!\!\!\widetilde\Ga_{\la_1\la^\prime_1;\,\la_2\la^\prime_2}(\bq,i\omega_n)\!=\!\!\!
\sum_{\mu_i,\mu^\prime_i} \rho_\bq(\mu^\prime_1|\mu_1)
\Ga_{\mu_1\la_1,\mu_1^\prime\la^\prime_1;\,\mu_2\la_2,\,\mu_2^\prime\la^\prime_2}(i\omega_n) 
\rho_{-\bq}(\mu_2|\mu^\prime_2)
\label{eq:Gamma1}
\eea
The Dyson equation for scattering, also known as the Bethe-Salpeter equation, takes the form
\bea
\lefteqn{\widetilde\Ga_{\la_1\la^\prime_1;\,\la_2\la^\prime_2}(\bq,i\omega_n)\!=\!
\int\frac{d^2\bk}{(2\pi)^2}\widetilde V(\bk)\rho^L_\bk(\la_1|\la_2)\rho^L_{-\bk}(\la_2^\prime|\la_1^\prime)\,
e^{i\bq\wedge\bk\ell_B^2} -}\\
&&\!\!\!-\!\sum_{\la\la^\prime}\int\frac{d^2\bk}{(2\pi)^2}\widetilde V(\bk)\rho^L_\bk(\la|\la_2)\rho^L_{-\bk}(\la_2^\prime|\la^\prime)\,
e^{i\bq\wedge\bk\ell_B^2}\,{\cal D}_{\la\la^\prime}(i\omega_n)
\widetilde\Ga_{\la_1\la^\prime_1;\,\la\la^\prime}(\bq,i\omega_n).\non
\label{eq:Dyson_Scattering}
\eea
It is convenient to view $\widetilde\Ga_{\la_1\la^\prime_1;\,\la\la^\prime}$ as a vector with components labeled by $\la\la^\prime$, while $\la_1\la_1^\prime$ and $(\bq,i\omega_n)$ are parameters.  Then the problem reduces to inverting a matrix in the indices $(\la_2\la_2^\prime;\,\la\la^\prime)$.

The formal structure of this matrix equation is elucidated by reducing it to the eigenvalue equation,
\bea
\sum_{\la\la^\prime}{\bf M}_{\la_2\la_2^\prime;\,\la\la^\prime}(\bq,i\omega_n)
\widetilde A_{\la_1\la^\prime_1;\,\la\la^\prime}(\bq,i\omega_n)\!=\! u_{\la_1\la_1^\prime}(\bq,i\omega_n) \widetilde A_{\la_1\la^\prime_1;\,\la_2\la_2^\prime}(\bq,i\omega_n)~,
\eea
where the kernel is
\bea
\!\!\!\!\!{\bf M}_{\la_2\la_2^\prime;\,\la\la^\prime}(\bq,i\omega_n)\!=\!
\delta_{\la\la_2}\delta_{\la^\prime\la_2^\prime}
\!\!+\!\!\int\frac{d^2\bk}{(2\pi)^2}\widetilde V(\bk)\rho^L_\bk(\la|\la_2)\rho^L_{-\bk}(\la_2^\prime|\la^\prime)\,
e^{i\bq\wedge\bk\ell_B^2}\,{\cal D}_{\la\la^\prime}(i\omega_n)
\eea  
and $u$, $\widetilde A$ are the eigenvalues, eigenvectors.  We have obtained an exact zero eigenvalue solution of this equation at $i\omega_n=0$.  Using the properties of the commutatator $[\rho^L_\bk,\rho^R_\bq]$ (c.f. equation (\ref{eq:fDef}) and the immediately following discussion) and the definitions of $\varepsilon_\la$ (\ref{eq:Greens_function}) and of ${\cal D}_{\la\la^\prime}$ (\ref{eq:D}), we find that
\bea
\widetilde A_{\la_1\la^\prime_1;\,\la\la^\prime}(\bq,0)&=&
\varepsilon_\la\,\rho^R_\bq(\la|\la^\prime)-\rho^R_\bq(\la|\la^\prime)\,\varepsilon_{\la^\prime}~,\\
u_{\la_1\la_1^\prime}(\bq,0)&=&0~.
\eea  
The similarity of this solution to the Ward identity (\ref{eq:WI}) suggests that the existence of the scattering zero mode is related to the vanishing of correlators containing $\hat\rho^R$.

\begin{figure}[htb]
\noindent
\center
\epsfxsize=3 in
\epsfbox{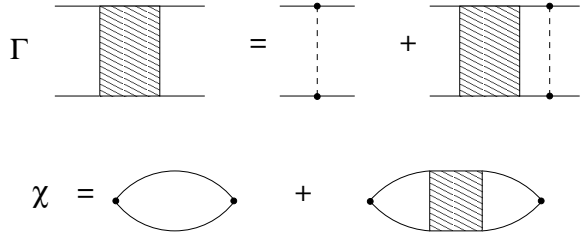}
\caption{\small (a) Ladder series for the scattering matrix $\Ga$ and (b) response functions in terms of $\Ga$.}
\label{fig:Gamma}
\end{figure}

Another advantage of the Bethe-Salpeter equation is that we can rewrite the response functions symmetrically,
\bea
\lefteqn{\chi^{AB}_{\mbox{\scriptsize{irr}}}(\bq,i\omega_n)=
\chi^{AB}_0(\bq,i\omega_n)+}\\
& &+\rho_0\sum_{\la_i\la^\prime_i}\rho^A_\bq(\la^\prime_1|\la_1)
{\cal D}_{\la_1\la^\prime_1}(i\omega_n)
\widetilde\Ga_{\la_1\la^\prime_1;\,\la_2\la^\prime_2}(\bq,i\omega_n)
{\cal D}_{\la_2\la^\prime_2}(i\omega_n)
\rho^B_{-\bq}(\la_2|\la^\prime_2)\,,\non
\label{eq:chi_Gamma}
\eea
where $\chi^{AB}_0$ is the bare bubble
\bea
\chi^{AB}_0(\bq,i\omega_n)=-\rho_0\sum_{\la\la^\prime}
\rho^A_\bq(\la|\la^\prime)
{\cal D}_{\la\la^\prime}(i\omega_n)
\rho^B_{-\bq}(\la^\prime|\la)
\label{eq:BareBubble}
\eea
Fig. \ref{fig:Gamma} illustrates the summation.

To obtain the momentum expansion of $\chi^{LL}$, consider the expansion of $\hat\rho^L$ from its definition in eqn. (\ref{eq:fDef}),
\bea
\rho^L_\bq (\la|\la^\prime) = \delta_{\la\la^\prime}+\ell^2_B\frac{\ell_{B_L}}{\ell_{B_R}}\, \langle\la|\,i\bq\wedge\pi\,|\rangle+{\cal O}(\bq^2)~.
\label{eq:f_expanded}
\eea
The first term, diagonal in $\la\la^\prime$, cannot contribute to the response because the transition amplitude ${\cal D}_{\la\la^\prime}$ is purely an inter-LL operator.  The expanded response becomes
\pagebreak
\bea
\lefteqn{\chi^{LL}_{\mbox{\scriptsize{irr}}}(\bq,i\omega_n)=
-\alpha\sum_{\la\la^\prime}\langle\la^\prime|\bq\wedge\pi|\la\rangle
{\cal D}_{\la\la^\prime}(i\omega_n)\langle\la|\bq\wedge\pi|\la^\prime\rangle\,+} \\
& &+\alpha\sum_{\la_i\la^\prime_i}\langle\la^\prime_1|\bq\wedge\pi|\la_1\rangle
{\cal D}_{\la_1\la^\prime_1}(i\omega_n)
\widetilde\Ga_{\la_1\la^\prime_1;\,\la_2\la^\prime_2}(\bq,i\omega_n)
{\cal D}_{\la_2\la^\prime_2}(i\omega_n)
\langle\la_2|\bq\wedge\pi|\la^\prime_2\rangle+\ldots~,\non
\label{eq:chi_q}
\eea
where $\alpha$ is an overall constant.  

To obtain $\chi^{LL}_{\mbox{\scriptsize{irr}}}$ through ${\cal O}(\bq^2)$, we need only $\widetilde\Ga(0,i\omega_n)$.  At $\bq=0$, the Bethe-Salpeter equation is
\bea
\widetilde\Ga_{\la_1\la^\prime_1;\,\la_2\la^\prime_2}(0,i\omega_n)=
\widetilde V_{\la_2\la^\prime_2;\,\la_1\la^\prime_1} -
\sum_{\la\la^\prime}\widetilde V_{\la_2\la^\prime_2;\,\la\la^\prime}
{\cal D}_{\la\la^\prime}(i\omega_n)
\widetilde\Ga_{\la_1\la^\prime_1;\,\la\la^\prime}(0,i\omega_n)\,,
\label{eq:Gamma_0}
\eea
where the interaction matrix element is
\bea
\widetilde V_{\la_2\la^\prime_2;\,\la_1\la^\prime_1}=\int\frac{d^2\bk}{(2\pi)^2} \widetilde V(\bk)\rho^L_\bk(\la_1|\la_2)\rho^L_{-\bk}(\la_2^\prime|\la_1^\prime)~.
\label{eq:V_matrix}
\eea
Rotation invariance at $\bq=0$ requires that the matrix elements vanish unless
\bea
\la_1+\la^\prime_2=\la^\prime_1+\la_2~.\non
\eea
Now let us make two further restrictions that afford an exact solution for $\widetilde\Ga$.  First, we choose $\la_{max}=0$ as we did above to illustrate the exchange energy.  Secondly, we work at zero temperature where the Fermi function is $f(\varepsilon_\la-\mu)=\Theta(\la_{max}-\la)$, restricting the ${\cal D}$-amplitude to
\bea
{\cal D}_{0\la}(i\omega_n)&=&-\frac{1}{\Delta_\la-i\omega_n} \non \\
{\cal D}_{\la0}(i\omega_n)&=&-\frac{1}{\Delta_\la+i\omega_n} \\
\Delta_\la &\equiv& \varepsilon_\la-\varepsilon_0 \non
\label{eq:D_01}
\eea    
Along with rotation invariance, these restrictions allow us to solve equation (\ref{eq:Gamma_0}) for the scattering matrix (at $\bq=0$) 
\bea
\widetilde\Ga_{0\la;\,0\la}&=&\frac{\widetilde V_{0\la;\,0\la}}
{1+\widetilde V_{0\la;\,0\la}{\cal D}_{0\la}} \non\\
\widetilde\Ga_{\la0;\,\la0}&=&\frac{\widetilde V_{\la0;\,\la0}}
{1+\widetilde V_{\la0;\,\la0}{\cal D}_{\la0}}~.
\label{Gamma_la}
\eea
The two channels above represent a particle in the $0$'th LL propagating on one leg of the ladder diagram and a particle in the $\la$'th LL on the other, and {\it vice versa}.  These channels do not mix in our example.  In fact, we will only need $\la=1$ for the lowest order term in $\chi^{LL}$ because of the vertices $\langle\la|\bq\wedge\pi|\la^\prime\rangle$.  At this point we find a crucial identity for $\la=1$:
\bea
\widetilde V_{01;\,01}=\widetilde V_{10;\,10} = \Delta_1~,
\label{eq:V0_id}
\eea
which is easily proven by comparing equations (\ref{eq:V_matrix}) and (\ref{eq:e_HF}).  Plugging equations (\ref{eq:D_01})-(\ref{eq:V0_id}) into equation (\ref{eq:chi_q}) for the response function, we find
\bea
\label{eq:chi_q0}
\chi^{LL}_{\rm irr}(\bq,i\omega_n)\!\!\!&=&\!\!\!
-|\langle0|\pi|1\rangle|^2\left\{{\cal D}_{01}\!+\!{\cal D}_{01}\!-\!
\frac{{\cal D}_{01}\Delta_1{\cal D}_{01}}{1+\Delta_1{\cal D}_{01}}-
\frac{{\cal D}_{10}\Delta_1{\cal D}_{10}}{1+\Delta_1{\cal D}_{10}}
\right\}|\bq|^2+{\cal O}(|\bq|^4)\non\\
&=& 0+{\cal O}(|\bq|^4)
\eea
(an overall factor has been left out).  Thus, the lowest order term in the density-density response is of order $|\bq|^4$.

This is the main physical result that we wanted to reproduce within our composite fermion framework.  Its main content is that the system is incompressible, i.e. the compressibility, $\kappa$, vanishes.  The connection of compressibility to the density-density response is contained in the definition \cite{PinesNozieres}
\bea
\kappa = \mathop{\mbox{lim}}_{\bq\rightarrow 0}\,\chi^{LL}(\bq,0)~.
\eea  
For the irreducible diagrams that we have considered so far, $\chi^{LL}=\chi^{LL}_{\rm irr}$, so that $\kappa=0$.  We expect that the order of limits is consistent with our calculation of the ladder series, which is a Taylor expansion around $\bq=0$ at finite frequency.

The full response function $\chi^{LL}$ can be obtained from the irreducible part by a bubble summation \cite{BCS,FW,PinesNozieres}
\bea
\chi^{LL} = \frac{\chi^{LL}_{\rm irr}}{1+\widetilde V(\bq)\chi^{LL}_{\rm irr}}~.
\eea
For a short range interaction, such as ours, this geometric sum has no qualitative effect, so that $\kappa$ remains at zero.  A well-known early work by Girvin, MacDonald, and Platzman \cite{GMP} contains general arguments for the momentum dependence of the density response function for incompressible liquids in the LLL, and is consistent with our result.    

%Its content is that the system is incompressible, as was argued early on by Girvin, MacDonald, and Platzman (GMP) \cite{GMP}.  For completeness, we briefly summarize their reasoning here. By analogy to the single mode approximation (SMA) theory of liquid Helium, the excitation spectrum, $\Delta(k)$, is given by $\Delta(k)=\ol f(k)/\ol s(k)$, where the overline denotes projection to the LLL.  $f(k)$ is the oscillator strength $f(\bk)=\int_0^\infty\omega S(k,\omega)$ and $s(k)$ is the static structure factor $s(k)=\int_0^\infty S(k,\omega)$, where $S(k,\omega)$ is the dynamic structure factor.  GMP then show that the projected static structure factor $\ol s(k)\sim |k|^4$ for any liquid ground state in the LLL.  Similarly, the projected oscillator strength also behaves like $\ol f(k)\sim |k|^4$, because the $|k|^2$ term is saturated by the inter-LL cyclotron mode (according to Kohn's theorem).  Thus, the excitation spectrum is gapped and the system is incompressible.  As GMP point out, the condition $s(k)\sim |k|^4$ is necessary for the existence of a gap, but it is also sufficient only within the SMA. 

%In terms of our calculation, the long distance behavior of $\ol s(k)$ and $\chi^{RR}$ is identical because $\ol s(k)$ is proportional to the frequency integral over the imaginary part of $\chi^{LL}_{\rm irr}(k,\omega)$ \cite{Mahan}.  Therefore, our result is consistent with an incompressible liquid (so long as SMA is asymptotically valid).               
Before we close this section and move on to the effective field theory, the interaction-gap identity of equation (\ref{eq:V0_id}) is worth a couple more words.  A slightly more general case is when $\la_{max}>0$,
\bea
\widetilde V_{\la,\la+1;\,\la,\la+1}=
\widetilde V_{\la+1,\la;\,\la+1,\la} = 
\varepsilon_{\la+1}-\varepsilon_{\la}~,
\label{eq:V_id}
\eea
where $\la\equiv\la_{max}$.  We expect that the compressibility will vanish again, although we have not performed this calculation explicitly.  This identity seems to have an analog in the Fermi liquid-like state of bosons at $\nu=1$ \cite{ReadHalf}.  There, the system has a divergent compressibility due to a fixed Landau parameter, $F_1=-1$, and $m^*$ is also coming from an integral of the interaction below the Fermi surface.  In both cases, the identities lead to the correct compressibility because there is no bare kinetic term due to the LLL projection.

The form of the left and right density response functions, suggest a physical interpretation of the vertices.  To lowest order in $\bq$, all response functions vanish, and we can take any linear combination $\rho^L_\bq-x\rho^R_\bq$ for the physical response without changing the result.  Suppose we choose the weighted combination
\bea
\rho^L_\bq\rightarrow \rho^L_\bq+\frac{B_R}{B_L}\rho^R_\bq~.
\label{eq:RhoL_new}
\eea
In operator language the densities are $\rho^A_\bq=e^{i\bq\cdot{{\bf R}_A}}$.  Using the operator mapping in equations (\ref{eq:PiK}) and (\ref{eq:rEff}), the momentum expansion of the new vertex leaves
\bea
\rho^L_\bq&\rightarrow& \frac{B}{B_L} + \frac{1}{B_L}i\bq\wedge{\bf K} + {\cal O}(\bq^2)\non\\
&=& \frac{B}{B_L}(1+\bq\cdot{\bf r}) - \frac{1}{B_L}i\bq\wedge\pi+{\cal O}(\bq^2)~. 
\label{eq:RhoL_new_expanded}
\eea
The first term is the first order term of a plane wave for the composite particle with charge $q^*=B/B_L$ at position ${\bf r}$, which is consistent with the two-particle mapping of section \ref{sec:TwoParticles}.  The second term is interpreted as the dipole moment.  The charge of the composite does not show up in the response functions, but presumably would come out if backflow corrections are included as in the work of Lopez and Fradkin \cite{LopezFradkin}. 

If the expansion is exponentiated, we obtain
\bea
\rho^L_\bq\rightarrow e^{i\bq\cdot{\bf r}}\left(\frac{B}{B_L}-\ell_{B_L}^2i\bq\wedge\pi\right)~.
\label{eq:Rho_L_exp}
\eea
This form agrees with the work of Shankar \cite{SM}, which starts from a different approach using the Chern-Simons theory at the outset.  As the first line of eqn. (\ref{eq:RhoL_new_expanded}) shows, this particular choice of $x$ makes the physical density a purely intra-LL operator.  As such, it is obvious that the density-density response vanishes to ${\cal O}(\bq^2)$ in the ladder apporximation, since the vertices contain only inter-LL transitions.  Further, our weighted combination agrees with Read's \cite{ReadHalf} density for $\nu=1$ where $B_R/B_L=-1$.  Nonetheless, we stress that our derivation does not specify $x$ by itself. 

%%%%%%%%%%%%%%%%%%%%%%%%%%%%%%%%%%%%%%%%%%%%%%%%%%%%%%%%
\section{Effective Theory}
\label{sec:Eff_Theory}
%%%%%%%%%%%%%%%%%%%%%%%%%%%%%%%%%%%%%%%%%%%%%%%%%%%%%%%%

In this section we will show that the ladder series can be replaced with a dynamic gauge field, yielding an effective theory much like that for bosons at $\nu=1$.  It is not a Chern-Simons field theory, but the familiar relation of density to the curl of a gauge field will appear.

In accordance with the previous section, the low energy physics of bosons at $\nu\neq 1$ is that of composite fermions filling an integral number, $\nu_{\rm eff}=\nu/q^*$, of Landau levels.  Let us take an ordinary fermion field $c({\bf x},t)$ in a static magnetic field $B=\nabla\wedge{\bf A}$ such that exactly $\nu_{\rm eff}$ levels are filled, and couple it to a nondynamic gauge field ${\bf a}$: 
\bea
H_{\mbox{\scriptsize{eff}}} = 
\int dtd^2{\bf x}\,\frac{1}{2m^*}\left|(-i\nabla-{\bf A}-{\bf a})\,c\,\right|^2-\mu c^\dagger c\,.
\label{eq:H_eff}
\eea
The fermion density is $\ol\rho=\nu_{\rm eff}/2\pi\ell_B^2$, or equivalently $\ol\rho=\nu/2\pi\ell_{B_L}^2$.  The chemical potential $\mu$ is tuned to lie between the uppermost filled LL and the lowest empty one.  $m^*$ is the only parameter in the theory and is obtained from the smallest energy gap of the original problem as defined in equation (\ref{eq:m_eff}).  The action of the theory is an action for ${\bf a}$ as well as for $c,c^\dagger$, but in contrast to Chern-Simons theory, there is no kinetic term for ${\bf a}$; it is a ``strongly coupled'' gauge field in the language of field theory.  There is also the term $a_0c^\dagger c$, however we will choose the temporal gauge in which $a_0=0$ at finite frequency, so that its fluctuations do not affect the response.

The gauge symmetry of $H_{\rm eff}$ is ordinary ${\rm U}(1)$, which can be viewed as the long distance limit of the global ${\rm U}(N)_R$ symmetry of the right coordinates.  The gauge invariant density of this model, $c^\dagger c$, is identified with the constraint $\rho^R$, which fixes 
\bea
\rho^R = c^\dagger c=\ol\rho~.
\label{eq:rhoRcc}  
\eea
This condition is the long distance limit of the full constraint that was constructed in Section \ref{sec:Fock_Physical}.  To obtain an expression for the physical density, consider the gauge invariant momentum density before ${\bf a}$ is included,
\bea
{\bf g}({\bf r}) = \frac{1}{2i}\left\{c^\dagger(\nabla-i{\bf A})c-\left[(\nabla+i{\bf A})c^\dagger\right]c\right\}~.
\eea 
The single-particle version is the $\pi$ operator of Section \ref{sec:SingleParticle}.  This allows us to rewrite the density suggestively.  The first term in equation (\ref{eq:Rho_L_exp}), $q^*e^{i\bq\cdot{\bf r}}$, is a plane wave for a charge of magnitude $q^*$; at tree level its expectation value can be replaced by $\ol\rho$, which leaves 
\bea
\rho = \ol\rho-q^*\ell^2_B\nabla\wedge{\bf g}~,
\label{eq:rho_phys}
\eea
where $q^*=B/B_L$.  Similarly, at tree level the gauge potential is related to the momentum density by ${\bf a}={\bf g}/\ol\rho$.  Therefore the physical density becomes
\bea
\rho = \ol\rho-\frac{\nu}{2\pi}\,\nabla\wedge{\bf a}~,
\label{eq:CS2}
\eea 
where we used $\nu_{\rm eff}=\nu/q^*$.  Equation (\ref{eq:CS2}) is precisely the fluctuation piece of the Chern-Simons equation (\ref{eq:CS}), despite the absence of a kinetic term for ${\bf a}$\,!  It should be borne in mind, however, that we have imposed the particular linear combination $\rho^L-x\rho^R$ with $x=B/B_L$, which is responsible for this appealing result; in principle, any coefficient of $\nabla\wedge\bf a$ is obtainable in this way.  

Let us now consider the correlation functions.  The basic conductivities are $\sigma_{ij} = \langle\g_i\g_j\rangle/m^{*2}$.  Because of the Onsager relation $\sigma_{xy}=-\sigma_{yx}$ and isotropy $\sigma_{xx}=\sigma_{yy}$, it is convenient to use complex coordinates $g=\g_x+i\g_y$ and $\ol g=\g_x-i\g_y$ so that the expectation values $\langle gg\rangle$ and $\langle\ol g\,\ol g\rangle$ vanish.  Our aim is to compare the susceptibility
\bea
\label{eq:chi_RPA_q}
\chi^{LL}(q)=\langle\delta\rho(q)\,\delta\rho(-q)\rangle
&=&\ell_{B_L}^4\langle\bq\wedge{\bf g}(q)\cdot\bq\wedge{\bf g}(-q)\rangle \\ 
&=&\ell_{B_L}^4\frac{|\bq|^2}{4}\,\langle g(q)\ol g(-q)+\ol g(q)g(-q)\rangle\non
\eea
to the ladder series in the previous section.  We will show that within a random phase approximation (RPA), the responses are identical (at least to ${\cal O}(q^2)$). 

\begin{figure}[htb]
\noindent
\center
\epsfxsize=3 in
\epsfbox{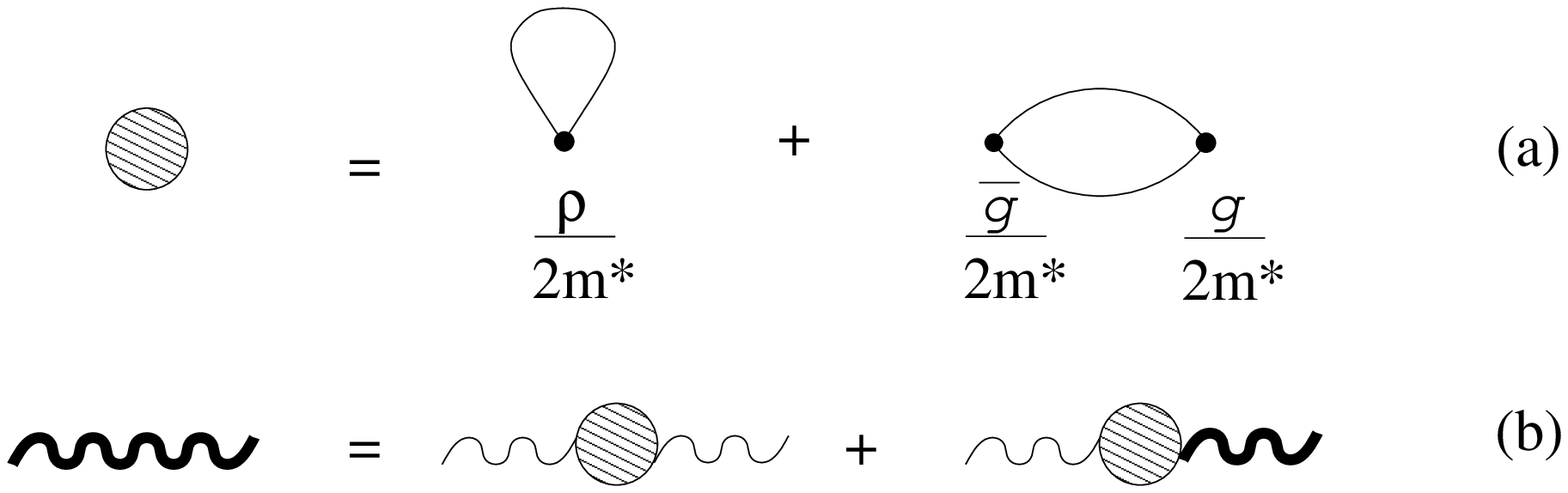}
\caption{\small Bubble summation for the gauge field propagator.  (a) The shaded circle includes the diamagnetic coupling and the $\langle g\ol g\rangle_{_0}$ bubble; (b) The thick wavy line represents $\langle a\ol a\rangle$}
\label{fig:aa}
\end{figure}

The RPA has been applied in the context of the quantum Hall effect by several authors \cite{ReadHalf,LopezFradkin}.  It is a bubble sum for the gauge field fluctuations.  The basic terms, shown in fig. \ref{fig:aa}, consist of a diamagnetic and a bubble piece.
The gauge field correlator is
\bea
\langle a(q)\ol a(-q)\rangle = \left[-\frac{\ol\rho}{m^*}+
\frac{1}{2m^{*2}}\langle g(q)\ol g(-q)\rangle_{_0}\right]^{-1}~,
\label{eq:aa}
\eea 
where $\langle g(q)\ol g(-q)\rangle_{_0}$ is the bare bubble, which we can evaluate in the single particle basis:
\bea
\langle g(q)\ol g(-q)\rangle_{_0} = 
-\sum_{\mu\la,\,\mu^\prime\la^\prime}\langle\mu\la|\pi|\,\mu^\prime\la^\prime\rangle{\cal D}_{\la\la^\prime}(i\omega_n)
\langle\mu^\prime\la^\prime|\ol\pi|\mu\la\rangle~.
\label{eq:BareBubble_RPA}
\eea
The calculation of $\langle\ol a(q)a(-q)\rangle_{_0}$ is analogous, but with $\langle\ol g(q)g(-q)\rangle_{_0}$~.  The matrix elements of $\pi$ produce a factor $2\nu_{\rm eff}\ell_B^2$ since $\pi=\sqrt{2}\ell_B\,b$ is the inter-LL ladder operator and ${\cal D}_{\la\la^\prime}$ connects only states on opposite sides of the Fermi surface, which restricts $\la, \la^\prime$ to $\la_{max}, \la_{max}+1$.  The sum over $\mu$'s gives the density per LL, $1/2\pi\ell_B^2$, with the end result
\bea
\frac{1}{2m^{*2}}\langle g(q)\ol g(-q)\rangle_{_0} &=& -\frac{\ol\rho}{m^*}{\cal D}_{_{\la_{max},\,\la_{max}+1}}(i\omega_n)\Delta\non\\
&\equiv&\frac{\ol\rho}{m^*}\frac{\Delta}{\Delta+i\omega_n}~,
\label{eq:gg}
\eea
where one factor of $m^*$ has been replaced by $1/\ell_B^2\Delta$.

\begin{figure}[htb]
\noindent
\center
\epsfxsize=3 in
\epsfbox{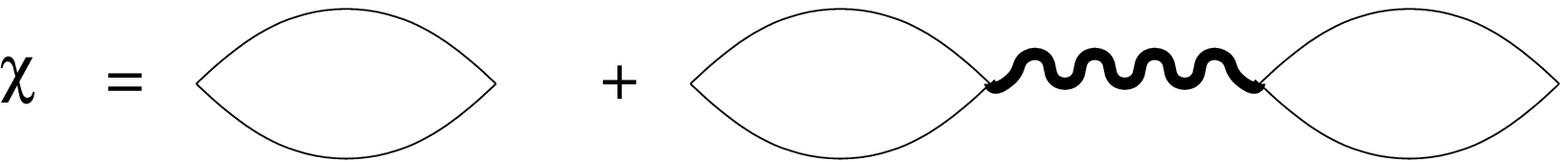}
\caption{\small $\chi^{LL}$ as a bubble sum.  The wavy line is the gauge field propagator from equation (\ref{eq:aa}).  See also fig. \ref{fig:aa}.}
\label{fig:chi}
\end{figure}

Now, the RPA $g-\ol g$ response consists of the bubble sum shown in fig. \ref{fig:chi}.
\bea
\label{eq:RPA_pi}
\langle g(q)\ol g(-q)\rangle &=&
\langle g(q)\ol g(-q)\rangle_{_0} - \\
&& \mbox{}-2\,\langle g(q)\ol g(-q)\rangle_{_0}\frac{1}{2m^*}\,\langle a(q)\ol a(-q)\rangle
\frac{1}{2m^*}\,\langle g(q)\ol g(-q)\rangle_{_0}\non
\eea
Using $\langle\ol gg\rangle = \ol{\langle g\ol g\rangle}$ and the identities (\ref{eq:aa}), (\ref{eq:gg}) in equation (\ref{eq:RPA_pi}) and plugging the result into the expression for $\chi^{LL}$, equation (\ref{eq:chi_RPA_q}), we find that
\bea
\chi^{LL}(q) = 0 + {\cal O}(q^4)~.
\eea
This is the same result as in the conserving approximation of the last section.  The structure of the RPA is such that, the gauge field propagator replaces the scattering matrix $\Ga$, and the two channels $\widetilde\Ga_{01;01}$, $\widetilde\Ga_{10;10}$ correspond to $\langle a\ol a\rangle$, $\langle \ol a a\rangle$.  Diagramatically, {\it the second bubble term in Fig. \ref{fig:chi} is exactly the ladder sum in Fig. \ref{fig:Gamma}(b)}.

Another way to test the effective theory is by integrating out the fermions.  Since there are ordinary fermions filling $\nu_{\mbox{\scriptsize{eff}}}$ Landau levels, we expect a Chern-Simons term in the effective action for ${\bf a}$.

The RPA prescription in Fig. \ref{fig:aa} implies that
\bea
{\cal L}[{\bf a}]=-\frac{1}{2m^{*2}}\sum_{i,j}a_i\langle g_ig_j\rangle_{_0}a_j + \frac{\ol\rho}{2m^*}\sum_i a_i a_i~,
\label{eq:L_eff_aa}
\eea 
The diamagnetic piece combines with the bare bubble amplitude in equation (\ref{eq:gg}) to give $\frac{\ol\rho}{m^*}\frac{\omega_n^2}{\Delta^2+\omega_n^2}a_ia_i$.  In the limit $\omega\rightarrow 0$, this diagonal term vanishes.  On the other hand, the cross term is proportional to $\omega_n$ in the same limit, leaving
\bea
{\cal L}[{\bf a}] = -\frac{\nu_{\rm eff}}{4\pi}\epsilon^{ij}a_i\omega_n a_j~,\non
\label{eq:L_eff_aa2}
\eea 
where $\epsilon^{ij}$ is the Levi-Civita symbol.  Since the original problem was gauge invariant, the complete Lagrangian must contain the scalar potential $a_0$:
\bea
{\cal L}[a] = -\frac{\nu_{\rm eff}}{4\pi}i\epsilon^{\mu\nu\la}
a_\mu\partial_\nu a_\la~,
\label{eq:L_eff_CS}
\eea
where $\mu,\nu,\la=t,x,y$ and $\partial_t\equiv i\omega_n$.  This form is correct to leading order in $\bq$, $\omega$ and shows the correct Hall conductivity of the composite fermions.

%%%%%%%%%%%%%%%%%%%%%%%%%%%%%%%%%%%%%%%%%%%%%%%%%%%%%%%%
%\section{Conclusion and Discussion}
%\label{sec:Conclusion_Fermion}
%%%%%%%%%%%%%%%%%%%%%%%%%%%%%%%%%%%%%%%%%%%%%%%%%%%%%%%%

%%%%%%%%%%%%%%%%%%%%%%%%%%%%%%%%%%%%%%%%%%%%%%%%%%%%%%%%%%%%%%%%%%%%%%%%
\chapter{Lowest Landau Level II:  Composite Bosons}
\label{chap:Bosons}
%%%%%%%%%%%%%%%%%%%%%%%%%%%%%%%%%%%%%%%%%%%%%%%%%%%%%%%%%%%%%%%%%%%%%%%%

In the previous chapter, we considered composite fermions.  In this one, we will consider composite bosons.  In the former case, we derived an effective theory from first principles.  However, this is not possible for more than one attached vortex and we follow a phenomenological approach instead.  The main feature that we find is the magneto-roton excitation, which was predicted and analyzed by several authors \cite{GMP,ParkJain}.  Our analysis provides a physical picture of this excitation.

%%%%%%%%%%%%%%%%%%%%%%%%%%%%%%%%%%%%%%%%%%%%%%%%%%%%%%%%%%%%%%%%%%%%%%%%%%%%%
\section{Formalism}
\label{sec:Formalism_Bosons}
%%%%%%%%%%%%%%%%%%%%%%%%%%%%%%%%%%%%%%%%%%%%%%%%%%%%%%%%%%%%%%%%%%%%%%%%%%%%%

In this section we generalize the Haldane-Pasquier formalism by considering $p$ objects attached to the underlying particle.  The physical picture that will emerge is similar, with the composite particle being a bound state of $p$ vortices and one particle.  We specialize to $\nu=1/p$ so that each vortex carries charge $1/p$ to maintain neutrality and the magnetic lengths are related by
\bea
\ell^2_{B_R}=p\ell^2_{B_L}~.
\eea
%%%%%%%%%%%%%%%%%%%%%%%%%%%%%%%%%%%%%%%%%%%%%%%%%%%%%%%%%%%%%%%%%%%%%%%%%%%%%
\subsection{Fock Space}
\label{sec:Fock_P}
%%%%%%%%%%%%%%%%%%%%%%%%%%%%%%%%%%%%%%%%%%%%%%%%%%%%%%%%%%%%%%%%%%%%%%%%%%%%%

For fractional fillings $\nu$ the number of vortices in the composite particle is $p$.  The matter operators are now $p+1$ rank tensors $c_{m;\,n_1\cdots n_p}$, with  
\bea
  [c_{m;\,n_1\cdots n_p},c^\dagger_{n^\prime_1\cdots n^\prime_p;\,m^\prime}]_{_\pm}
  =\delta_{mm^\prime}\delta_{n_1n^\prime_1}\cdots \delta_{n_p n^\prime_p}
\label{eq:AntiCommutatorP}
\eea
$m$ runs from $1$ to $N_\phi$ and the $n$'s from $1$ to $N$.  The left (physical) density is obtained by tracing over all the right indices
\bea
\rho^L_{mm^\prime}=\sum_{n_1\cdots n_p}c^\dagger_{n_1\cdots n_p;\,m}c_{m;\,n_1\cdots n_p}
\label{eq:RhoLeftP}
\eea
However, there are now $p$ right densities:
\bea
\rho^{R_a}_{nn^\prime}=\sum_{m;\,n_j,\,\hat n_a} c^\dagger_{n_1\cdots n_{i-1}\,n\,n_{i+1}\cdots n_p;\,m} c_{m;\,n_1\cdots n_{i-1}\,n^\prime\,n_{i+1}\cdots n_p}~,
\label{eq:RhoRightP}
\eea                
where $a=1,\ldots,p$ and $\hat n_a$ implies that there is no sum over $n_a$.  Similarly, there are $p$ copies of the constraints in eqn. (\ref{eq:Constraints}), which require $|\Psi^{m_1\cdots m_N}_{\mbox{\scriptsize{phys}}}\rangle$ to be a singlet in each of the $p$ right indices.  A straightforward extension of the $p=1$ case (eqn. (\ref{eq:PhysStates})) shows that the basis is   
\bea
|\Psi_{\rm phys}^{m_1\cdots m_N}\rangle =  
 \sum_{\alpha_i\,\beta_i\cdots\gamma_i}\epsilon^{\alpha_1\beta_1\cdots\gamma_1}\cdots \epsilon^{\alpha_p\beta_p\cdots\gamma_p}
c^\dagger_{\alpha_1\cdots \alpha_p\,;\,m_1} c^\dagger_{\beta_1\cdots\beta_p\,;\,m_2}\cdots c^\dagger_{\gamma_1\cdots\gamma_p\,;\,m_N}
|0\rangle~,
\label{eq:PhysStatesP}
\eea
where there are $N$ Greek indices of the type $\alpha\beta\cdots\gamma$.  The Levi-Civita symbols and anticommutation relations of the $c$'s ensure symmetry under the interchange of any pair of physical indices $m_i\leftrightarrow m_j$, and we are left with a bosonic Fock space of $N$ particles in $N_\phi$ orbitals.  Again, had we started with bosonic $c$'s, we would have ended up with a Fock space of composite Fermions.

%%%%%%%%%%%%%%%%%%%%%%%%%%%%%%%%%%%%%%%%%%%%%%%%%%%%%%%%%%%%%%%%%%%%%%%%
\subsection{Physical Operators and Constraints}
\label{sec:Pvortices}
%%%%%%%%%%%%%%%%%%%%%%%%%%%%%%%%%%%%%%%%%%%%%%%%%%%%%%%%%%%%%%%%%%%%%%%%

The field operator has $p$ right coordinates $\eta_i$, 
\bea
c(z,\ol\eta_1,\ldots,\ol\eta_p)=\sum_{m,\,n_i}u^L_m(z)\ol{u^R_{n_1}(\eta_1)}\cdots 
\ol{u^R_{n_p}(\eta_p)}~c_{m;\,n_1\cdots n_p}~. 
\label{eq:c_p}
\eea
It is convenient to change the vortex coordinates to the so-called ``center-of-mass'' (or Jacobi) coordinates, which have been used in few-body problems in the context of atomic physics \cite{FanoGreen}.  This reduces the $p$-complex to one center-of-mass coordinate $\xi_{cm}=\frac{1}{p}(\eta_1+\cdots\eta_p)$ and $p-1$ relative coordinates $\xi_\alpha$ ($\alpha=1,\cdots,p-1$), which are linear combinations of the $\eta_i$.  We will denote the linear transformation by
\bea
\xi_\al=R_{\al i}\eta_i~,
\label{eq:JTransf}
\eea
where $\al=0,\ldots,p-1$ and $\al=0$ stands for $cm$.  Generally, Greek indices will be used for the $\xi$'s and Latin indices for the $\eta$'s.  

One of the nice properties of Jacobi coordinates is that $\sqrt{p}R$ is orthogonal:
\bea
|\xi_{cm}|^2+|\xi_1|^2+\cdots|\xi_{p-1}|^2=\frac{1}{p}(|\eta_1|^2+\cdots +|\eta_p|^2)
\label{eq:JacobiNorm}
\eea
For the special where each vortex carries charge $1/p$ ($\nu=1/p$), we can use this property to set the magnetic length of each $\xi$ to $\ell_{B_L}^2\equiv\ell_{B_R}^2/p$.  The utility of this transformation is that {\it all coordinates now have only one magnetic length}, $\ell_{B_L}$.  

The particular way in which Jacobi coordinates are constructed is well-illustrated by two special cases, $p=2$ and $p=3$, both of which we will utilize below.  For two vortices there is only one relative coordinate, so the Jacobi system is
\bea
\xi_{cm}&=&\frac{1}{2}(\eta_1+\eta_2) \non \\
\xi_1 &=& \frac{1}{2}(\eta_1-\eta_2)
\label{eq:p2}
\eea
The normalizations are chosen so as to preserve the normalization in eqn. (\ref{eq:JacobiNorm}).  Specifically, the $1/2$ factor in $\xi_1$ is the reduced ``mass'', $m$ (with $1/m=1/m_1+1/m_2$) of the two vortices---each vortex is taken to have unit ``mass'' ($m_i=1$).  

\begin{figure}[htb]
\noindent
\center
\epsfxsize=2.5 in
\epsfbox{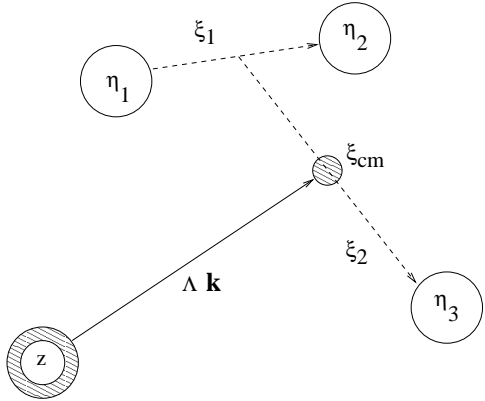}
\caption{\small The composite with three vortices attached.  The underlying particle is at $z$ with charge $+1$, and the vortices are at $\eta_i$ with charge $-1/3$.  The (unnormalized) Jacobi vectors are dotted lines.  $z$ and the center of mass of the vortices, $\xi_{cm}$, are connected by $\wedge\bk$, which we will later interpret as a dipole moment.  Since the composite is neutral, it drifts with momentum $\bk$.}
\label{fig:composite}
\end{figure}

For $p=3$, the coordinates are arranged so that $\xi_1$ is a vector from $\eta_1$ to $\eta_2$, and $\xi_2$ connects $\eta_3$ to the center of mass of $\eta_1$ and $\eta_2$.  Each $\xi$ is normalized by the square root of the reduced mass of the two objects which it connects.  Fig. \ref{fig:composite} illustrates this construction.
The particular linear combinations are
\bea
\left(\begin{array}[1]{c}\xi_{cm}\\ \xi_1\\ \xi_2 \end{array}\right)=
\frac{1}{\sqrt{3}}
\left(\begin{array}[3]{ccc}
\frac{1}{\sqrt{3}} & \frac{1}{\sqrt{3}} & \frac{1}{\sqrt{3}} \\
\frac{1}{\sqrt{2}} & -\frac{1}{\sqrt{2}} & 0 \\
\frac{1}{\sqrt{6}} & \frac{1}{\sqrt{6}} & -\sqrt{\frac{2}{3}}
\end{array} \right)
\left(\begin{array}[1]{c}\eta_{1}\\ \eta_2\\ \eta_3 \end{array}\right)~.
\label{eq:p3}
\eea
There is a certain degree of freedom inherent in assigning the $\xi$.  For instance, we could have chosen $\xi_1$ to connect $\eta_1$ to $\eta_3$, and $\xi_2$ to connect $\eta_2$ to the center of mass of $\eta_{1,2}$.  However, each choice obeys the normalization condition in eqn. (\ref{eq:JacobiNorm}).  Viewed classically, $|\xi_{cm}|^2$ is a constant in the absence of external forces, so the normalization implies that $|\xi_1|^2+\cdots+|\xi_{p-1}|^2$ is fixed and that all choices of Jacobi sets can be transformed into each other by a member of ${\rm SO}(p-1)$.  Although this is a fundamental and generally useful property of the coordinate system, we will not use it in this thesis, but point it out for completeness.  The most important property for us is that eqn. (\ref{eq:JacobiNorm}) scales the $\xi$ in such a way that all coordinates have the same magnetic length.  

For higher values of $p$, the Jacobi coordinates can be constructed recursively by grouping the vortices into pairs and connecting the centers of mass, and then repeating the process with the centers of mass.  The details of the general procedure are described elsewhere \cite{FanoGreen}; however, here we only need the $p=2,3$ cases.  

Now, the combination of the particle at $z$ and the center of mass of the vortices at $\xi_{cm}$ is just like a particle and vortex of equal but opposite charge, which can be treated by the noncommutative Fourier transform (Appendix).  The transformed field operators become  
\bea
c_\bk(\{\ol\xi\})=\int d^2z\prod_{i=1}^p d^2\eta^\prime_i~c(z,\ol\eta^\prime_1\cdots\ol\eta^\prime_p)~\tau_\bk(\xi^\prime_{cm},\ol z)\prod_{\al=1}^{p-1}\delta(\xi^\prime_\al,\ol\xi_\al)~.
\label{eq:c_tau_cm}
\eea
The $\delta$-functions implement the Jacobi coordinate transformation of eqn. (\ref{eq:JTransf}) through $\xi_\al^\prime=R_{\alpha i}\eta_i^\prime$.  Since the $\xi_\al$ themselves are coordinates with a magnetic length $\ell_{B_L}$, the field can be cast into the complementary ``spin'' basis
\bea
c_\bk(\{\ol\xi\})=\sum_{\sigma_\al}c_{\bk\,\sigma_1\cdots\sigma_p}\, \ol{u^L_{\sigma_1}(\xi_1)}\cdots \ol{u^L_{\sigma_{p-1}}(\xi_{p-1})}~.
\label{eq:c_sigma}
\eea
The quantum numbers $\sigma_\al$ are nonnegative integers.  They are angular momenta of the relative coordinates, but we will refer to them simply as ``spin''.  In this chapter we consider spinless fermions and bosons so there should be no chance for confusion.

The left density is integrated over all right coordinates and has the same momentum structure as in the $p=1$ case:
\bea
\hat\rho^L_\bq = \int\frac{d^2\bk}{(2\pi)^2}\sum_{\sigma_\al}~e^{\frac{1}{2}i\bk\wedge\bq\ell_{B_L}^2}~ c^\dagger_{\bk-\frac{1}{2}\bq,\,\sigma_1\cdots\sigma_{p-1}}\, c_{\bk+\frac{1}{2}\bq,\,\sigma_1\cdots\sigma_{p-1}}~.
\label{eq:RhoL_p}
\eea
The right densities are more complicated, 
\bea
\lefteqn{\hat\rho^{R_i}_\bq =} \\
& &\int\frac{d^2\bk}{(2\pi)^2}\int\prod_{\al=1}^{p-1}d^2\xi_\al d^2\xi^\prime_\al~ \tau_\bq(\xi^\prime_{cm}\!-\!\eta^\prime_i, \ol\xi_{cm}\!-\!\ol\eta_i) e^{\frac{1}{2}i\bk\wedge\bq\ell_{B_L}^2}~c^\dagger_{\bk-\frac{1}{2}\bq}(\{\xi\})\, c_{\bk+\frac{1}{2}\bq}(\{\ol\xi^\prime\}) \non
\label{eq:RhoR_p}
\eea
Notice the magnetic translation in $\xi_{cm}-\eta_i$, which is the vector from the center-of-mass to the $i$'th vortex.

Finally, the Hamiltonian looks just like the unequally charged case, eqn. (\ref{eq:HFourier}), in the previous chapter,
\bea
H=\frac{1}{2}\int\frac{d^2\bq}{(2\pi)^2}~V(\bq)\,e^{-|q|^2/2\ell_{B_L}^2}\,:\hat\rho^L_\bq~\hat\rho^L_{-\bq}:
\label{eq:H_unequal}
\eea 
Again, only ``left'' operators appear since $H$ is a physical quantity and $\hat\rho^{R_i}_\bq$ are constants of the motion.

%%%%%%%%%%%%%%%%%%%%%%%%%%%%%%%%%%%%%%%%%%%%%%%%%%%%%%%%%%%%%%%%%%%%%%%%%%%%%%%%%%%
\subsection{Remarks on Many-Particle Wavefunctions}
\label{sec:Wavefunctions}
%%%%%%%%%%%%%%%%%%%%%%%%%%%%%%%%%%%%%%%%%%%%%%%%%%%%%%%%%%%%%%%%%%%%%%%%%%%%%%%%%%%

In this section, we augment the physical picture of the composite fermions and bosons by outlining the many-body wavefunctions that they can describe.  In particular, we recover Laughlin's function at $\nu=1/p$ \cite{Laughlin83}.

%and some paired wavefunctions \cite{MR,Greiter,RR}, which are the focus of the FQHE portion of this thesis.

Consider first $N$ bosons at $\nu=1/p$.  There are $N$ particle coordinates $z_i$ and $pN$ vortex coordinates $\ol\eta_{s,i}$ with $s=1,\ldots,p$ and $i=1,\ldots,N$.  The constraints (Section \ref{sec:Pvortices}) require that the $\ol\eta_{s,i}$ dependence of the wavefunction be that of a full Landau level for each $s$ \cite{ReadHalf},
\bea
\Psi_{\mbox{\scriptsize{phys}}}(z_1,\ol\eta_{1,1},\ldots,\ol\eta_{p,1},\ldots,z_N,\ol\eta_{1,N},\ldots,\ol\eta_{p,N})=f(z_1,\ldots,z_N)\prod_{s;\,i<j} (\ol\eta_{s,i}-\ol\eta_{s,j})\non
\label{eq:SpatialPsi}
\eea
The last factor is the product of $p$ Laughlin-Jastrow factors (the Gaussian factors have been left off).  Another way of writing it is the product of $p$ Slater determinants of the matrices $m^s_{ij}=\{u^R_j(\ol\eta_{s,i})\}$.  Each determinant is a Vandermonde determinant and is the unique totally antisymmetric wavefuction annihilated by the corresponding constraint.  

The simplest ground state of composite bosons at zero temperature is the single-particle condensate
\bea
\langle c_{\bk\,\sigma_1\cdots\sigma_{p-1}}\rangle = \sqrt{\ol\rho}~\delta_{\bk,0} \prod_\al\delta_{\sigma_\al,0}~.
\label{eq:Condensate1}
\eea
In coordinate space this can be rewritten suggestively as  
\bea
\langle\psi_0(z,\ol\eta_1,\ldots,\ol\eta_p)\rangle = \prod_s\tilde\delta(z,\ol\eta_s)~, 
\label{eq:Condensate2}
\eea
where $\widetilde\delta$ is the delta function with the magnetic length of the vortices, $\ell_{B_R}$.  This puts the vortices on top of the particle, as one would expect from the lowest state due to electrostatic attraction.  Projection onto the physical basis gives
\bea
\!\!\!\!f(z_1,\ldots,z_N)\!=\!
\int\!\prod_{s,\,i}d^2\eta_{s,i}~e^{-|\eta_{s,i}|^2/4\ell^2_{B_R}}
\!\!\prod_{s;\,i<j}\!(\eta_{s,i}-\eta_{s,j})~
\!\!\prod_i^N\langle\psi_0(z_i,\ol\eta_{1,i},\ldots,\ol\eta_{p,i})\rangle.
\label{eq:CoordinateProjection}
\eea   
Using the condensate wavefunction (\ref{eq:Condensate2}), $f$ is the product of $p$ identical factors,
\bea
f(z_1,\ldots,z_N)=
\left[\int\prod_{i}d^2\eta_{i}~e^{-|\eta_{i}|^2/4\ell^2_{B_R}}
\prod_{i<j}\!(\eta_{i}-\eta_{j})~
\prod_i\widetilde\delta(z_i,\ol\eta_i)\right]^p~.
\label{eq:CoordinateProjectionFactor}
\eea   
Each factor is a Slater determinant of the matrix $m_{ij}=\{u^R_i(z_j)\}$, which is the same as a Vandermonde determinant of the matrix $v_{ij}=z_j^i$ times an overall factor.  The end result is exactly the Laughlin state at $\nu=1/p$,
\bea
f_p(z_1,\ldots,z_N)=\prod_{i<j}(z_i-z_j)^p
\prod_ie^{-|z_i|^2/4\ell^2_{B_L}}.
\eea
Thus, we reinterpret the Laughlin state as a composite boson condensate.  

However, there is a slight surprise if $f_p$ is the unique characteristic of a Laughlin state.  It turns out that the condensate is not required to carry the spin $\sigma_\al=0$; any linear combination of the internal states $\{u^L_{\sigma_\al}(\ol\xi_\al)\}$ will project onto the same $f_p$.  In other words, the simple condensate in eqn. (\ref{eq:Condensate1}) is generalized to 
\bea
\label{eq:DegCondensate}
\langle c_\bk(\ol\xi_1,\ldots,\ol\xi_{p-1})=\sqrt{\ol\rho}\,\delta_{\bk,0}\,
u(\ol\xi_1,\ldots,\ol\xi_{p-1})\prod_\al e^{-|\xi_\al|^2/4\ell^2_{B_L}}~,
\eea
where $u$ is a suitably normalized anti-analytic function of the $\ol\xi_\al$.  In terms of composite bosons, the Laughlin state is infinitely degenerate (for all $p\neq 1$).  We will come back to this question in Section \ref{sec:Energy_Functional} when we impose the constraints and model the fluctuations around the ground state.

Despite the infinitely degenerate ground state, the fundamental quasiholes can be represented unambiguously.  We modify one delta function in equation (\ref{eq:Condensate2}) by $\widetilde{\delta}(z,\ol\eta_s)\rightarrow\sum_m u^L_{m+1}(z)\ol{u^R_m(\eta_s)}$, moving the particle at $z$ radially from the origin by one angular momentum unit.  The effect on $f_p$ is an overall factor of $\prod_i z_i$, which is the form of the quasihole wavefunction given by Laughlin \cite{Laughlin83}.  This can be seen by writing one of the factors in eqn. (\ref{eq:CoordinateProjectionFactor}) as a Slater determinant; the shifts $u^L_m(z_i)\rightarrow u^L_{m+1}(z_i)$ are equivalent to multiplying every column $i$ by $z_i$ and rescaling every row $m$ by an overall numerical factor (due to the normalization of $u_m$), which only changes the determinant by an overall constant. Quasiholes can be moved around by applying magnetic translations $\hat\tau$, which are described in the Appendix. 

%Similarly, we can recast various paired states in the LLL as paired states of composite particles.  Following standard Bardeen-Cooper-Schrieffer (BCS) theory \cite{BCS}, the ground state is, in the grand canonical ensemble,
%\bea
%|\Psi_{\mbox{\scriptsize{BCS}}}\rangle\propto\prod_\bk e^{g_\bk c^\dagger_\bk c^\dagger_{-\bk}}|0\rangle~,
%\label{eq:BCScondensate}
%\eea
%where $g_\bk$ is the Fourier transform of the pair wavefunction.  Now, take the $c_\bk$ to be composite fermions or bosons at $\nu=1$ and fix the number of particles to be $N$.  For femions, $g$ is required to be antisymetric, and the choice $g(z)=1/z$ gives the Pfaffian \cite{MR,ReadGreen}
%\bea
%f_{\mbox{\scriptsize{Pfaff}}}(z_1,\ldots,z_N)={\cal{A}}\left\{\prod_{i=1}^N\frac{1}{(z_i-z_{i+1})}\right\}~
%\prod_{i<j}(z_i-z_j)~,
%\eea 
%where ${\cal A}$ is the antisymmetry operator.  This wavefunction is valid at long distances.

%For bosons at $\nu=1/2$, $g$ is required to be symmetric and the choice $g=1/z^2$ yields the Haffnian \cite{RR,GRR}
%\bea
%f_{\mbox{\scriptsize{Haff}}}(z_1,\ldots,z_N)={\cal{S}}\left\{\prod_{i=1}^N\frac{1}{(z_i-z_{i+1})^2}\right\}~
%\prod_{i<j}(z_i-z_j)^2~, \non\\
%\eea
%where ${\cal S}$ symmetrizes the coordinates.  If spin is included, then BCS pairing of bosons, $\langle c_{\bk\uparrow}c_{\bk\downarrow}\rangle$ gives the Permanent.  We will consider the fluctuations around the Permanent and the Haffnian ground states in Part \ref{prt:PERM}.

%%%%%%%%%%%%%%%%%%%%%%%%%%%%%%%%%%%%%%%%%%%%%%%%%%%%%%%%%%%%%%%%%%%%%%%%
\section{Energy Functional}
\label{sec:Energy_Functional}
%%%%%%%%%%%%%%%%%%%%%%%%%%%%%%%%%%%%%%%%%%%%%%%%%%%%%%%%%%%%%%%%%%%%%%%%

Section \ref{sec:Wavefunctions} explained how the Laughlin wavefunction is a condensate of composite bosons.  The salient feature was the degeneracy of the ground state.  Let us rephrase this in terms of the Hamiltonian (\ref{eq:H_unequal}), which is a function of only the left density.  $\hat\rho^L$ is   
\bea
\hat\rho^L_\bq = \int\frac{d^2\bk}{(2\pi)^2}\sum_{\sigma_\al}~e^{\frac{1}{2}i\bk\wedge\bq\ell_{B_L}^2}~ c^\dagger_{\bk-\frac{1}{2}\bq,\,\sigma_1\cdots\sigma_{p-1}}\, c_{\bk+\frac{1}{2}\bq,\,\sigma_1\cdots\sigma_{p-1}}~.
\label{eq:RhoL_p=2}
\eea
In eqn. (\ref{eq:DegCondensate}) of the previous section, we pointed out that the condensate is infinitely degenerate because the Laughlin state does not depend on the wavefunction of the internal vortex coordinates.  In other words, the expectation value of the Hamiltonian is unaffected by the internal ``spin'' state as long as the condensate is ${\langle c_{\bk,\sigma_1\cdots\sigma_{p-1}}\rangle\propto\delta_{\bk,0}}$, which is enough to give $\langle\hat\rho^L_\bk\rangle=\ol\rho$.  Unlike the composite fermion case in Chapter \ref{chap:Fermions}, there is no unique mean field ground state to expand around.  This degeneracy should disappear once the constraints are imposed.

Here we will follow an alternate route which allows us to include the constraints at the outset.  We will construct a phenomenological Lagrangian that respects all the symmetries and include the constraints as Lagrange multipliers.  The resulting Landau-Ginzburg theory can then be systematically expanded about a unique mean field solution.

The simplest rotationally invariant contribution to the Lagrangian consists of a momentum and a spin piece
\bea
\label{eq:RotInvH}
{\cal L}_M+{\cal L}_J = -\sum_{\bk,\sigma_\al}\,\left(\frac{1}{2M}|\bk|^2+\frac{J}{2}\sigma\right) c^\dagger_{\bk\sigma_1\cdots\sigma_{p-1}}c_{\bk\sigma_1\cdots\sigma_{p-1}}~,
\eea
where $M,J$ are constants and $\sigma=\mathop{\sum}_{\al=1}^{p-1}\sigma_\al$.  We may guess that the full Lagrangian should be
\bea
{\cal L} = \sum_{\bk,\sigma_\al}\,c^\dagger_{\bk\sigma_1\cdots\sigma_{p-1}}\partial_\tau c_{\bk\sigma_1\cdots\sigma_{p-1}}+{\cal L}_M + {\cal L}_J+{\cal L}_{\rm constr}~,
\eea   
where the first term is the usual time derivative \cite{FW} and ${\cal L}_{\rm constr}$ is the Lagrange multiplier term that imposes the constraints.  However, this expression is not gauge invariant.  ${\cal L}$ must be invariant under the symmetries of $c$, which preserve the commutators of eqn. (\ref{eq:AntiCommutatorP}),
\bea
c\mapsto U_L\,c\,\prod_{i=1}^p U_{R_i}~,
\label{eq:U_Symmetry}
\eea
where $U_{R_i}$ is a unitary matrix acting on the $i$'th right index and $U_L$ acts on the left index of the matrix $c_{m\,;\,n_1\cdots n_p}$.  We will find that in order to preserve these symmetries, it is necessary to introduce $p$ gauge potentials that act on the right indices.  The bare term ${\cal L}_M+{\cal L}_J$ will acquire the vector components $a^i_\mu$, $\mu=x,y$, and the Lagrange multipliers, $\la^i$, will play the role of the scalar potential $a^i_0$. Since the mass and spin terms are decoupled in ${\cal L}_M+{\cal L}_J$, we will consider their gauge invariant forms separately in the following two subsections.

%%%%%%%%%%%%%%%%%%%%%%%%%%%%%%%%%%%%%%%%%%%%%%%%%%%%%%%%%%%%%%%%%%%%%
\subsection{Mass Term}
\label{sec:M}
%%%%%%%%%%%%%%%%%%%%%%%%%%%%%%%%%%%%%%%%%%%%%%%%%%%%%%%%%%%%%%%%%%%%%

To write the mass term in coordinate space we need the transform of $i\bk$, which is the analog of $\nabla$ in ordinary space.  To this end, define the LLL coordinate operators, $Z$ and $\ol Z$
\bea
Z = z\delta(z,\ol\xi_{cm})~~~{\mbox{and}}~~~{\ol Z} = \delta(z,\ol\xi_{cm})\ol\xi_{cm}~,
\label{eq:Z_ops}
\eea
which are adjoints of each other, $Z^\dagger=\ol Z$.  In Cartesian coordinates $Z=Z_x+iZ_y$ and $Z_\mu^\dagger=Z_\mu$.  It is straightforward to verify that the $\hat\tau_\bk$ are eigenoperators of $Z,\ol Z$, that is
\bea
\left[Z\stackrel{*}{,}\hat\tau_\bk\right]=-ik\,\hat\tau_\bk~~~{\mbox{and}}~~~
\left[\ol Z\stackrel{*}{,}\hat\tau_\bk\right]=i\ol k\,\hat\tau_\bk~,
\label{eq:Z_tau}
\eea
where the $*$-commutator is defined in eqn. (\ref{eq:StarComm}).  As usual, $k=k_x+ik_y$ and $\ol k=k_x-ik_y$, so that $-\epsilon^{\mu\nu}\left[Z_\nu\stackrel{*}{,}\hat\tau_\bk\right]=k_\mu\,\hat\tau_\bk$.  Therefore, $-i\epsilon^{\mu\nu}Z_\nu$ is the analog of the ordinary derivative operator $\partial_\mu$ when acting on $\hat\tau$.  To see how $Z$ acts on the matter field, we need to consider the Fock space operators in more detail.

In general, $c_{m;\,n_1,\ldots,n_p}$ can be transformed on the left by some matrix $M_{mm^\prime}$ or on the right by some matrix $\La_{n_i^\prime n_i}$ that acts on the $i$'th right index.  In coordinate space, $M=M(z,\ol z^\prime)$ and $\La=\La_i(\eta_i^\prime,\ol\eta_i)$.  We will continue to use the $*$-operator, but its meaning must be clarified when acting on the right coordinates since there are $p$ of them.  We define
\bea
(c*\La_i)(z,\ol\eta_1,\ldots,\ol\eta_p) = 
\int d^2\eta_i^\prime\,c(z,\ol\eta_1,\ldots,\ol\eta_i^\prime,\ldots,\ol\eta_p) \La_i(\eta_i^\prime,\ol\eta_i)~.
\label{eq:StarRight}
\eea
{\it The index on the operator will always indicate the coordinate on which it acts}, rendering this notation unambiguous.  An example is $c*Z$, where $Z$ acts on the center of mass coordinate:
\bea
(c*Z)(z,\ol\eta_1,\ldots,\ol\eta_p) &=& 
\frac{1}{p}\int d^2\eta_1^\prime\,c(z,\ol\eta_1^\prime,\ldots,\ol\eta_p)
\eta_1^\prime\widetilde\delta(\eta_1^\prime,\ol\eta_1)+\cdots\non\\
&+&\frac{1}{p}\int d^2\eta_p^\prime\,c(z,\ol\eta_1,\ldots,\ol\eta_p^\prime)
\eta_p^\prime\widetilde\delta(\eta_p^\prime,\ol\eta_p)
\eea   
Let us define the coordinate operator from the right, $v^i$, by
\bea
v^i=\eta_i\widetilde\delta(\eta_i,\ol\eta_i^\prime){~~~{\rm and}~~~}
\ol v^i=\widetilde\delta(\eta_i,\ol\eta_i^\prime)\ol\eta_i^\prime~,
\label{eq:V_Star}
\eea
such that
\bea
c*Z = \frac{1}{p}\sum_i c*v^i~.
\eea  
The Cartesian components of the $v$'s are Hermitian, $v^{i\,\dagger}_\mu=v^i_\mu$.  Acting on the left, $Z*c$ is the same as before,
\bea
(Z*c)(z,\ol\eta_1,\ldots,\ol\eta_p) &=& 
\int d^2z^\prime\,z\delta(z,\ol z^\prime)c(z^\prime,\ol\eta_1,\ldots,\ol\eta_p)\non\\
&\equiv& zc(z,\ol\eta_1,\ldots,\ol\eta_p)~.
\label{eq:StarLeft}
\eea  
Now, the Fourier transform, eqn. (\ref{eq:c_tau_cm}), together with the property in eqn. (\ref{eq:Z_tau}) afford a definition of the ``derivative'' operator
\bea
\ol\partial c=-\frac{1}{2}[Z\stackrel{*}{,}c]~~~{\mbox{and}}~~~
\partial c=\frac{1}{2}[\ol Z\stackrel{*}{,}c]~.
\label{eq:Derivatives}
\eea  
We translate this definition into Cartesian coordinates by using the conventional relations $\partial=(\partial_x-i\partial_y)/2$ and $\ol\partial=(\partial_x+i\partial_y)/2$, 
\bea
\partial_\mu c=-i\epsilon^{\mu\nu}[Z_\nu\stackrel{*}{,}c]~,
\label{eq:Derivatives_Cartesian}
\eea
where $\epsilon^{\mu\nu}$ is the Levi-Civita symbol.  The non-commutative Fourier transform of $\partial_\mu c$ is $-ik_\mu c$.  Further like ordinary derivatives, $\partial_\mu$ obeys the Leibnitz property $\partial(a*b)=\partial a*b+a*\partial b$, which follows easily from the Jacobi identity $[A,[B,C]]+[C,[A,B]]+[B,[C,A]]=0$.

The unitary symmetry in eqn. (\ref{eq:U_Symmetry}) requires that the derivative operators are covariant.  The inifinitesimal version of the transformation on the right is obtained from the product of expanding each $U_{R_i}=1+i\La_i$,
\bea
c &\mapsto & c+ic*\La \non\\
c^\dagger &\mapsto& c^\dagger - i\ol\La * c^\dagger~,
\label{eq:c_Transformation}
\eea
where $\La$ is a Hermitian operator, 
\bea
\La = \sum_i \La_i(\eta_i^\prime,\ol\eta_i)
\label{eq:La}
\eea
and its conjugate is $\ol\La=\sum_i\La_i(\eta_i,\ol\eta_i^\prime)$.  A standard procedure from field theory can be used to make $\partial_\mu c$ covariant \cite{ItzyksonZuber}.  Since $Z_\mu*c$ is already covariant under right transformations, let us consider each term in $c*Z_\mu$ separately, which are of the form $c*v^i_\mu$.  First, we introduce a gauge potential $a^i_\mu(\eta_i^\prime,\ol\eta_i)$ which transforms according to
\bea
a^i_\mu\mapsto a^i_\mu-\partial_\mu\La_i+i[a^i_\mu\stackrel{*}{,}\La_i]~.
\label{eq:a_Transformation}
\eea   
The derivative of $\La_i$ is constructed like $\partial_\mu c$,
\bea
\partial_\mu\La_i=-i\epsilon^{\mu\nu}[v^i_\nu\stackrel{*}{,}\La_i]~.
\label{eq:partial_La}
\eea
Then the combination $c*(\epsilon^{\mu\nu}v^i_\nu+a^i_\mu)$ transforms according to the rules in eqns. (\ref{eq:c_Transformation}) and (\ref{eq:a_Transformation}):
\bea
\label{eq:cov}
c*(\epsilon^{\mu\nu}v^i_\nu+a^i_\mu)&\mapsto&
(c+ic*\La_i)*(\epsilon^{\mu\nu}v^i_\nu+a^i_\mu+i\epsilon^{\mu\nu}[v^i_\nu\stackrel{*}{,}\La_i]+i[a^i_\mu\stackrel{*}{,}\La_i])\non\\
&=&c*(\epsilon^{\mu\nu}v^i_\nu+a^i_\mu)+ic*(\epsilon^{\mu\nu}v^i_\nu+a^i_\mu)*\La_i+{\cal O}(\La_i^2)
\eea
showing that it is covariant.

The total contribution of the gauge potentials to $\epsilon^{\mu\nu}Z_\nu=\frac{1}{p}\sum_i\epsilon^{\mu\nu}v^i_\nu$ is just the sum of the individual gauge potentials, which we term $a^{cm}$,
\bea
a^{cm}_\mu = \frac{1}{p}\sum_i a^i_\mu~.
\eea
The covariant derivative becomes 
\bea
D_\mu\,c&=&\partial_\mu\,c-i\,c*a^{cm}_\mu \non\\
(D_\mu\,c)^\dagger&=&\partial_\mu\,c^\dagger+i\,a^{cm}_\mu*c^\dagger~.
\label{eq:CovariantD}
\eea
These derivatives obey $[D_\mu,D_\nu]=iG_{\mu\nu}$, where $G_{\mu\nu}=\partial_\mu a^{cm}_\nu\!-\!\partial_\nu a^{cm}_\mu-{i[a_\mu^{cm}\stackrel{*}{,}a_\nu^{cm}]}$ is the field strength \cite{ItzyksonZuber}.  By analogy to standard non-Abelian gauge theory, we will choose the transverse gauge for each $a^i$
\bea
\partial_\mu a^i_\mu=0~.
\label{eq:Gauge}
\eea
The derivative is again defined by eqn. (\ref{eq:partial_La}).  Using the non-commutative Fourier transform, the gauge condition is exactly $\bq\cdot a^i_\bq = 0$. 
  
The fully covariant mass term can now be written in coordinate space as
\bea
{\cal L}_M &\rightarrow& -\frac{1}{2M}{\mbox{Tr}}\left[(D_\mu c)^\dagger * D_\mu c\right] \non\\
&=& -\frac{1}{2M}{\mbox{Tr}}\left(\partial_\mu c^\dagger+ia^{cm}_\mu * c^\dagger\right) *
\left(\partial_\mu c-ic*a^{cm}_\mu\right)~.
\label{eq:h_M}
\eea 
At mean field, the saddle point of the full Lagrangian
\bea
\langle a^{cm}_\mu\rangle &=& 0\non\\
\langle c_{\bk,\sigma_1,\ldots,\sigma_{p-1}}\rangle &=& \sqrt{\ol\rho}\,\delta_{\bk,0} \prod_\al\delta_{\sigma_\al,0}~.
\label{eq:MF}
\eea
After quadratically expanding ${\cal L}_M$ about the saddle point, we find that the terms linear in $a^{cm}_\mu$ are proportional to $\partial_\mu a^{cm}_\mu$, which vanishes by the gauge choice.  The quadratic term in $a^{cm}_\mu$ decouples from the matter fluctuations, and we are left with
\bea
\delta {\cal L}_M=-\frac{1}{2M}{\mbox{Tr}}\left(\partial_\mu c^\dagger * \partial_\mu c+ \ol\rho\, a^{cm}_\mu * a^{cm}_\mu\right)~.
\label{eq:h_M_expanded}
\eea    
In the next section, we will find a similar decoupling of the matter and gauge fluctuations in the spin term.  Even if the ordinary kinetic energy $\bk^2/2M$ is generalized to an arbitrary polynomial in $\bk$, the same decoupling would hold at the Gaussian level.  Gauge fields will be felt, however, as Lagrange multipliers in the constraint terms.

%%%%%%%%%%%%%%%%%%%%%%%%%%%%%%%%%%%%%%%%%%%%%%%%%%%%%%%%%%%%%%%%%%%%%
\subsection{Spin Term}
\label{eq:J}
%%%%%%%%%%%%%%%%%%%%%%%%%%%%%%%%%%%%%%%%%%%%%%%%%%%%%%%%%%%%%%%%%%%%%

Turning now to the spin term, we will suppress the left index in this subsection.

The preceeding subsection defined the operators $v^i=\eta_i\widetilde\delta(\eta_i,\ol\eta_i^\prime)$, whose sum was related to the center of mass coordinate operator.  For the relative coordinates, the natural basis is
\bea
w^\al=-i\sum_jR_{\al,j}v^j\mbox{~~~and~~~}\ol w^\al=i\sum_jR_{\al,j}\ol v^j~,
\eea 
where $R$ is the Jacobi transformation $\xi_\al=\sum_jR_{\al,j}\eta_j$ that was constructed in Section \ref{sec:Pvortices}.  The $i$ prefactor ensures that, in Cartesian coordinates, $w^\al_\mu$ is related to $\epsilon^{\mu\nu}v^j_\nu$.  This also has the effect of making the $w^\al_\mu$ anti-Hermitian, $w^{\al\,\dagger}_\mu=-w^\al_\mu$, because $v^j_\mu$ is Hermitian.

In the single particle basis $u^L_{\sigma}(\xi_\alpha)$, $w^\alpha$ and $\ol w^\alpha$ act like $\xi_\alpha/\sqrt{2}\ell_{B_L}$ and $(\ell_{B_L}/\sqrt{2})\partial_{\xi_\alpha}$, respectively---see eqn. (\ref{eq:u_m}) and the immediatly following discussion.  Therefore, the quadratic form $\ol w^\alpha*w^\alpha$ is exactly $\sigma/2$, and the spin term can be written as
\bea
\label{eq:SpinTerm}
{\cal L}_J&=&-J\sum_\alpha {\mbox{Tr}}\left(\ol w^\alpha * c^\dagger * c * w^\alpha\right)\\
&=&-J\sum_\alpha {\mbox{Tr}}\left(w^\alpha_\mu * c^\dagger * c * w^\alpha_\mu\right)~.\non
\eea  
Cyclicity of the trace was used to separate $w*\ol w$ in the first line.  By changing to Cartesian coordinates in the second line, we ignore a constant term in the energy.  This contribution is from the commutator $[\ol w^\al\stackrel{*}{,}w^\beta]=-2\delta_{\al\beta}$ to $w^\al_\mu*w^\al_\mu={(w^\al*\ol w^\al+\ol w^\al*w^\al)/2}$.  

Just as the $\partial$'s were not covariant in the last subsection, the $w^\alpha$ are not, either.  Recall from that discussion that each $c*\epsilon^{\mu\nu}v^i_\nu$ is made covariant by $\epsilon^{\mu\nu}v^i_\nu\mapsto \epsilon^{\mu\nu}v^i_\nu+a^i_\mu$.  We thus introduce $p-1$ gauge potentials $a^\al_\mu$, 
\bea
a^\al_\mu=\sum_i R_{\al,i}a^i_\mu~
\eea
and the covariant operators $W^\alpha_\mu$,
\bea
c* W^\al_\mu &=& c*w^\alpha_\mu+c*a^\alpha_\mu \non\\
W^{\al\,\dagger}_\mu * c^\dagger &=& -w^\alpha_\mu*c^\dagger+a^\alpha_\mu *c^\dagger~.
\label{eq:Covariant_W}
\eea 
As in the previous section, we work in the Cartesian basis $\mu=x,y$ with $W_x=(W+\ol W)/2$ and $W_y = (W-\ol W)/2i$.  The correct spin term is now
\bea
\label{eq:h_J}
{\cal L}_J &\rightarrow& -J\sum_\alpha {\mbox{Tr}}\left(W^{\al\,\dagger}_\mu*c^\dagger * c * W^\alpha_\mu\right)
\non\\
&=& J\sum_\alpha {\mbox{Tr}} \left(w^\alpha_\mu*c^\dagger-a^\alpha_\mu *c^\dagger\right)*
\left(c*w^\alpha_\mu+c*a^\alpha_\mu\right)~.
\eea
The part linear in $a^\al_\mu$ is equal to 
\bea
-J\,{\mbox{Tr}}\,c^\dagger * c *[w^\alpha_\mu\stackrel{*}{,}a^\alpha_\mu]~.
\label{eq:a_linear}
\eea
The commutator is a sum of terms of the form $\epsilon^{\mu\nu}[v^i_\nu\stackrel{*}{,}a^i_\mu]\equiv i\partial_\mu a^i_\mu$ because ${[v^i_\nu\stackrel{*}{,}a^j_\mu]=0}$ unless $i=j$.  According to the transverse gauge choice, all terms in eqn. (\ref{eq:a_linear}) vanish.  In fact we can prove a more general statement for a polynomial spin dispersion.  Any term in the energy of the form $\sigma_\alpha^n$ can be written as
\bea
\sigma_\alpha^n\,c^\dagger_{\sigma_1\cdots\sigma_{p-1}} 
c_{\sigma_1\cdots\sigma_{p-1}} = 
{\mbox{Tr}}\,c^\dagger * c * (W^{\alpha\,\dagger}_\mu * W^\alpha_\mu)^n~.
\eea 
The linear piece in $a^\alpha_\mu$ always looks like (\ref{eq:a_linear}) and vanishes in the transverse gauge.  For our purposes, we will stick with the simplest non-trivial example, $J\sigma$.

The saddle point is
\bea
\langle a^\alpha_\mu\rangle &=& 0 \non\\
\langle c_{\sigma_1\cdots\sigma_{p-1}}\rangle &=&\prod_\alpha \delta_{\sigma_\alpha, 0}
\eea
where the left coordinates are suppressed.  Quadratic expansion about the saddle point gives the perturbation of ${\cal L}_J$ as
\bea
\delta {\cal L}_J=-J\sum_\alpha{\mbox{Tr}}\left(w^\alpha_\mu * c^\dagger * c * w^\alpha_\mu +
\ol\rho\,a^\alpha_\mu*a^\alpha_\mu\right)~.
\label{eq:h_J_expanded}
\eea
The second term is also proportional to $\sum_i{\mbox {Tr}}\left(a^i_\mu*a^i_\mu\right)$ because $\sqrt{p}R$ is orthogonal.  It remains to treat the constraints gauge invariantly, the subject of the next section.

%%%%%%%%%%%%%%%%%%%%%%%%%%%%%%%%%%%%%%%%%%%%%%%%%%%%%%%%%%%%%%%%%%%%%
\section{Constraints}
\label{sec:ConstraintsL}
%%%%%%%%%%%%%%%%%%%%%%%%%%%%%%%%%%%%%%%%%%%%%%%%%%%%%%%%%%%%%%%%%%%%%

We will treat the constraints by including $p$ Lagrange multipliers, $\la^i(\eta_i,\ol\eta_i^\prime)$.  The Lagrangian then includes the term
\bea
{\cal L}_{\mbox{\scriptsize{constr}}} = 
i\sum_{i=1}^p {\mbox{Tr}}\,\la^i * \rho^{R_i}~.
\label{eq:LagrangeMultiplier}
\eea
The $\rho^{R_i}$ were constructed in eqn. (\ref{eq:RhoR_p}).  Gauge invariance under the transformation in eqn. (\ref{eq:c_Transformation}) requires that 
\bea
\la^i\mapsto\la^i+i[\La_i\stackrel{*}{,}\la^i]~.
\non
\eea
The saddle point for $\la^i$ is
\bea
\langle\la^i(\eta_i,\ol\eta_i^\prime)\rangle =\widetilde\delta(\eta_i,\ol\eta_i^\prime)
\label{eq:LambdaSaddle}
\eea
or $\langle\la^i_\bq\rangle=\delta_{\bq,0}$ in momentum space.  Expansion of the fluctuations is analytically feasible if the $\rho^{R_i}_\bq$ are expanded in powers of $\bq$.  It can be done exactly for $p=2$, but we will also consider $p=3$ to finite order.

At $p=2$ there is only one relative coordinate, $\xi=(\eta_1-\eta_2)/2$.  The saddle point is 
\bea
\langle c_\bq(\ol\xi)\rangle &=& \sqrt{\ol\rho}\,\delta_{\bq,0}\,\ol{u^L_0(\xi)}
\non\\
\langle\la^i_\bq\rangle &=& \delta_{\bq,0}
\label{eq:Saddle_p=2}
\eea
The displacement of the vortices from the center of mass was $\xi_{cm}-\eta_1=-\xi$ and $\xi_{cm}-\eta_2=\xi$ according to the Jacobi mapping in eqn. (\ref{eq:p2}).  Plugging this into the right densities, eqn. (\ref{eq:RhoR_p}), and expanding in powers of $c$ about the saddle point we get
\bea
\delta\rho^{R_1}_\bq &=& \sqrt{\frac{\ol\rho}{2\pi}} \int d^2\xi\,
c_\bq(\ol\xi) e^{-\frac{1}{4}|\xi|^2-\frac{i}{2}\ol q\,\xi-\frac{1}{4}|q|^2} \non\\
&+& \sqrt{\frac{\ol\rho}{2\pi}} \int d^2\xi\,
c^\dagger_{-\bq}(\xi)e^{-\frac{1}{4}|\xi|^2-\frac{i}{2} q\,\ol\xi-\frac{1}{4}|q|^2}\,, \non\\
\delta\rho^{R_2}_\bq &=& \sqrt{\frac{\ol\rho}{2\pi}} \int d^2\xi\,
c_\bq(\ol\xi) e^{-\frac{1}{4}|\xi|^2+\frac{i}{2}\ol q\,\xi-\frac{1}{4}|q|^2} \non\\
&+& \sqrt{\frac{\ol\rho}{2\pi}} \int d^2\xi\,
c^\dagger_{-\bq}(\xi)e^{-\frac{1}{4}|\xi|^2+\frac{i}{2} q\,\ol\xi-\frac{1}{4}|q|^2}\,,
\eea
where $c$ is the fluctuation about $\langle c\rangle$ and $\delta\rho^R \equiv \rho^R-\ol\rho$.  $\ell_{B_L}^2$ is set to one above, and in what follows.  If we make use of the identity
\bea
\frac{1}{2\pi}e^{-\frac{1}{4}|\xi|^2-\frac{i}{2}\ol q\,\xi-\frac{1}{4}|q|^2} = \delta(\xi,-i\ol q)\equiv\sum_{\sigma=0}^{\infty} u_\sigma(\xi)\ol{u_\sigma(iq)}~,
\label{eq:id}
\eea
then
\bea
\delta\rho^{R_1}_\bq &=& \sqrt{p}\sum_\sigma c_{\bq\sigma}u_\sigma(-i\ol q)+
c^\dagger_{-\bq\sigma}u_\sigma(-iq)~,\non\\
\delta\rho^{R_2}_\bq &=& \sqrt{p}\sum_\sigma c_{\bq\sigma}u_\sigma(i\ol q)+
c^\dagger_{-\bq\sigma}u_\sigma(iq)~.
\label{eq:RhoR_expansion}
\eea
This is an explicit expansion in powers of $q$ since $u_\sigma(q)\sim q^\sigma$ (aside from an overall Gaussian factor).  The superscript $L$ has been dropped from $u$ as we are setting $\ell_{B_L}=1$.

Moving on to $p=3$, there are two relative coordinates, $\xi_{1,2}$ that are given by eqn. (\ref{eq:p3}).  The matter field is $c_\bq(\ol\xi_1,\ol\xi_2)$ with expectation value
\bea
\langle c_\bq(\ol\xi_1,\ol\xi_2)\rangle &=& \sqrt{\ol\rho}\,\delta_{\bq,0}\,
\ol{u_0(\xi_1)}\,\ol{u_0(\xi_2)}~.\non
\eea
As in the $p=2$ case, we obtain $\xi_{cm}-\eta_i$ in terms of the Jacobi coordinates from eqn. (\ref{eq:p3}), put the result into the definition of $\rho^{R_i}$ in eqn. (\ref{eq:RhoR_p}) and expand about the saddle point in powers of $c$.  For example, $\xi_{cm}-\eta_1=-\sqrt{\frac{3}{2}}\xi_1-\sqrt{\frac{1}{2}}\xi_2$, which gives
\bea
\lefteqn{\delta\rho^{R_1}_\bq =}\\
& &\int d^2\xi_1 d^2\xi_2\, c_\bq(\ol\xi_1,\ol\xi_2)
u_0(\xi_1)u_0(\xi_2) 
e^{-\frac{1}{4}|\sqrt{\frac{3}{2}}\xi_1+\sqrt{\frac{1}{2}}\xi_2|^2
-\frac{1}{2}\ol q(\sqrt{\frac{3}{2}}\xi_1+\sqrt{\frac{1}{2}}\xi_2)-
\frac{1}{4}|q|^2} +h.c.\non\\
& &= 2\pi\int d^2\xi_1 d^2\xi_2\, c_\bq(\ol\xi_1,\ol\xi_2)\,
\delta(\xi_1,-i\ol q\,\sqrt{3/2})\,\delta(\xi_2,-i\ol q\,\sqrt{1/2})+h.c.~,\non
\eea
where we used the delta function identity, eqn. (\ref{eq:id}).  The procedure is analogous for $\rho^{R_{2,3}}$, and we obtain in spin space
\bea
\delta\rho^{R_1}_\bq &=& \sqrt{p}\sum_{\sigma_1\sigma_2}
c_{\bq\sigma_1\sigma_2}u_{\sigma_1}(-i\ol q\,\sqrt{3/2})u_{\sigma_2}(-i\ol q\,\sqrt{1/2})+ \non\\
& & \sqrt{p}\sum_{\sigma_1\sigma_2}
c^\dagger_{-\bq\sigma_1\sigma_2}u_{\sigma_1}(-iq\,\sqrt{3/2})u_{\sigma_2}(-iq\,\sqrt{1/2})~;
\non\\
\delta\rho^{R_2}_\bq &=& \sqrt{p}\sum_{\sigma_1\sigma_2}
c_{\bq\sigma_1\sigma_2}u_{\sigma_1}(i\ol q\,\sqrt{3/2})u_{\sigma_2}(-i\ol q\,\sqrt{1/2})+ \non\\
& & \sqrt{p}\sum_{\sigma_1\sigma_2}
c^\dagger_{-\bq\sigma_1\sigma_2}u_{\sigma_1}(iq\,\sqrt{3/2})u_{\sigma_2}(-iq\,\sqrt{1/2})~;
\non\\
\delta\rho^{R_3}_\bq &=& \sqrt{2\pi p}\sum_{\sigma_2}
c_{\bq 0 \sigma_2}u_{\sigma_2}(i\ol q\,\sqrt{2})+\sqrt{2\pi p}\sum_{\sigma_2}
c^\dagger_{-\bq 0 \sigma_2}u_{\sigma_2}(iq\,\sqrt{2})~.
\non
\label{eq:Constraints_P=3}
\eea

%%%%%%%%%%%%%%%%%%%%%%%%%%%%%%%%%%%%%%%%%%%%%%%%%%%%%%%%%%%%%%%%%%%%%
\section{Density Response and the Magnetoroton}
\label{eq:Spectrum}
%%%%%%%%%%%%%%%%%%%%%%%%%%%%%%%%%%%%%%%%%%%%%%%%%%%%%%%%%%%%%%%%%%%%%
 
The results of the last three sections add into the fluctuations of the Lagrangian density,
\bea
\delta{\cal L} = {\mbox{Tr}}~(c^\dagger * \partial_\tau c) - \delta {\cal L}_M - \delta {\cal L}_J - 
i\sum_{i=1}^p\,\delta\la^i_{-\bq}\,\delta\,\rho^{R_i}_\bq~,
\label{eq:delta_L}
\eea
where the first term is the usual Berry phase, $\tau$ being complex time, and the other pieces are given by eqns. (\ref{eq:h_M_expanded}), (\ref{eq:h_J_expanded}), and ({\ref{eq:LagrangeMultiplier}).  

We can drop the fluctuations in the vector potentials, $a^{cm}$ and $a^i$ since they are decoupled and will not affect the physical quantities.  Then in momentum space, $\delta {\cal L}_J+\delta {\cal L}_M$ become
\bea
\delta {\cal L}_M+\delta {\cal L}_J = \sum_{\sigma_\alpha}E_{\bk,\sigma}\, c^\dagger_{\bk\sigma_1\cdots\sigma_{p-1}}\,c_{\bk\sigma_1\cdots\sigma_{p-1}}~,
\label{eq:delta_h}
\eea
where
\bea
E_{\bk,\sigma} &=&\varepsilon_\bk+J_\sigma \non\\
\varepsilon_\bk &=& \frac{k^2}{2M}\non\\
J_\sigma &=& J\sigma~.
\eea
As observed previously, both $\varepsilon_\bk$ and $J_\sigma$ can be arbitrary polynomials in $\bk$ and $\sigma$, respectively, but the present form suffices for our purposes.

Let us rewrite the Lagrangian as
\bea
\delta{\cal L} &=& \delta\phi^\dagger_k\,{\bf G}^{-1}_k\,\delta\phi_k\\
\delta\phi_k &=& \left(\begin{array}[1]{c}
\delta\la^i_k\\
c^\dagger_{-k\sigma}\\
c_{k\sigma}
\end{array}\right)
\label{eq:L_propagator}
\eea 
The propagator matrix ${\bf G}$ carries the physical information, including the correlation functions and the spectrum of excitations via the zeros of its determinant.  We consider the $p=2,3$ cases in the following sections.

%%%%%%%%%%%%%%%%%%%%%%%%%%%%%%%%%%%%%%%%%%%%%%%%%%%%%%%%%%%%%%%%%%%%
\subsection{The case $p=2$}
\label{sec:p=2}
%%%%%%%%%%%%%%%%%%%%%%%%%%%%%%%%%%%%%%%%%%%%%%%%%%%%%%%%%%%%%%%%%%%%

It is a lengthy, but straightforward, calculation to obtain the determinant by Gauss-Jordan elimination.  It proves convenient to use the basis $\la^\pm=\la^1\pm\la^2$, and we find, 
\bea
{\mbox{Det}}\,{\bf G}^{-1} = P_{\mbox{\scriptsize{even}}}P_{\mbox{\scriptsize{odd}}}~,
\eea
where
\bea
P_{\mbox{\scriptsize{even}}} \!\!&=&\!\! \left\{\prod_{\sigma=0}^\infty\left[E^2_{\bk,2\sigma}-\omega_n^2\right]\right\} 
\left\{\sum_{\sigma=0}^\infty\frac{\left|u_{2\sigma}\right|^2
E_{\bk,2\sigma}}{E_{\bk,2\sigma}^2-i\omega_n^2}   \right\}\\
P_{\mbox{\scriptsize{odd}}} \!\!&=&\!\! \left\{\prod_{\sigma=0}^\infty\left[E^2_{\bk,2\sigma+1}-\omega_n^2\right]\right\} 
\left\{\sum_{\sigma=0}^\infty\frac{\left|u_{2\sigma+1}\right|^2
E_{\bk,2\sigma+1}}{E_{\bk,2\sigma+1}^2-i\omega_n^2}   \right\}\non 
\eea
and $u_\sigma\equiv u_\sigma(k)$.

The general features of the spectrum emerge already if we keep only the terms up to $\sigma=3$.  We find that the lowest mode is
%i\omega_n&=&\left[\frac{|u_0|^2\varepsilon_\bk\left(\varepsilon_\bk+2J\right)^2+ |u_2|^2\left(\varepsilon_\bk+2J\right)}{|u_0|^2\varepsilon_\bk+ |u_2|^2\left(\varepsilon_\bk+2J\right)} \right]^{1/2}~\non\\
\bea
i\omega_n=\left[E_{\bk,0}E_{\bk,2}\left(\frac{\bk^4E_{\bk,0}+8E_{\bk,2}} {\bk^4E_{\bk,2}+8E_{\bk,0}}\right)\right]^{1/2}~,
\label{eq:approx3}
\eea
which is plotted in fig. \ref{fig:spectrump=2}.  In keeping with an incompressible fluid, there is a finite gap at $\bk=0$.  In general, the modes have energy gaps equal to $\sigma J$ with $\sigma\geq 2$.  The lowest mode is a mixture of only $\sigma$=even channels, the next one up is a mixture of $\sigma$=odd, and so forth, alternating in even/odd mixtures.  Because the gap would vanish if the constraints were not included, or if $J\rightarrow 0$, the low energy physics is dominated by the internal vortex excitations of $\sigma$.  
\begin{figure}[htb]
\noindent
\center
\epsfxsize=3 in
\epsfbox{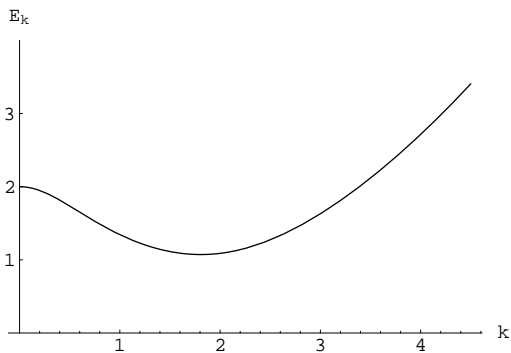}
\caption{\small Lowest mode at $\nu=1/2$ obtained from the action expanded to ${\cal O}(k^3)$.  $\bk$ is in units of inverse magnetic length.  The parameters are $m=3$ and $J=1$.}
\label{fig:spectrump=2}
\end{figure}
The main feature in fig. \ref{fig:spectrump=2} is the dip.  According to the approximation in eqn. (\ref{eq:approx3}) it appears whenever $J>1/M$.  We have plotted the spectrum by including terms up to $\sigma=4$, i.e. to ${\cal O}(q^8)$, and the shape is virtually identical to fig. \ref{fig:spectrump=2}.  In proper units, the dip is always located at $|\bk^*|\sim 1/\ell_{B_L}$.  It would seem to correspond to the magnetoroton that has been proposed by several authors \cite{GMP,ParkJain}.  However, ours is the first analytical observation of this phenomena; earlier predictions relied on numerics.  Based on the composite fermion picture, $\bk$ is proportional to the dipole moment, or separation between the particle and the center of mass of the vortices.  It is tempting to conclude that the dip is an electrostatic-like feature, occurring when the relative orbit radius of the vortices is equal to the distance from the particle to their center of mass, i.e. $\xi/2\sim\xi_{cm}$.  

The density-density response function, $\chi^{LL}(q)$ can be calculated within the above framework as well.  The connected part of this correlation function is
\bea
\chi^{LL}(q)=\langle\delta\rho^L_q\,\delta\rho^L_{-q}\rangle~,
\eea
where $q=(\bq,i\omega_n)$ and $\delta\rho^L$ is the fluctuating part of $\rho^L$ (eqn. (\ref{eq:RhoL_p=2})).  Expanding about the condensate in eqn. (\ref{eq:Saddle_p=2}) gives,
\bea
\delta\rho^L_q=\sqrt{p}\,(c_{\bq,0}u_0+c^\dagger_{-\bq,0}u_0)~.
\eea
One way to get the lowest order term in $\chi$ is to substitute the expansions for $\rho^{R_{1,2}}$.  A simple linear combination from eqn. (\ref{eq:RhoR_expansion}) yields
\bea
\frac{1}{2}\left(\delta\rho^{R_1}_\bq+\delta\rho^{R_2}_\bq\right) &=& 
\sqrt{p}\sum_{\sigma=0,2,4,\ldots}\left(c_{\bq\sigma}u_\sigma(-i\ol q) + c^\dagger_{-\bq\sigma}u_\sigma(-iq)\right)~\\
&=& \delta\rho^L_\bq + \sqrt{p}\sum_{\sigma=2,4,\ldots}\left(c_{\bq\sigma}u_\sigma(-i\ol q) + c^\dagger_{-\bq\sigma}u_\sigma(-iq)\right)~. \non
\eea
On the average, the left-hand side is zero because the constraints vanish at mean field, $\langle\delta\rho^R_q\rangle=0$, so that $\delta\rho^L_q\sim{\cal O}(|\bq|^2)$ and
\bea
\chi^{LL}(q)={\cal O}(|\bq|^4)
\eea
just as we had before in the composite fermion case.

The same result can be obtained by an explicit calculation that extracts the field correlators from eqn. (\ref{eq:L_propagator}).  An expansion in minors (Cramer's rule) of the propagator matrix ${\bf G}$ leads to
\bea
\langle c^\dagger_{\bk,0} c_{\bk,0}\rangle &=&
-\frac{\left(i\omega_n+\varepsilon_\bk\right)Q+|u_0|^2} {\left(\omega_n^2+\varepsilon_\bk^2\right)Q+2|u_0|^2\varepsilon_\bk}
\non\\
\langle c^\dagger_{\bk,0} c^\dagger_{-\bk,0}\rangle &=&
\frac{|u_0|^2} {\left(\omega_n^2+\varepsilon_\bk^2\right)Q+2|u_0|^2\varepsilon_\bk}
\eea 
where
\bea
Q=\sum_{\sigma=2,4,6,\ldots}\frac{2|u_\sigma|^2E_{\bk,\sigma}} {\omega_n^2+E_{\bk,\sigma}}~.
\eea
Plugging these correlators into $\chi^{LL}(k)=\ol\rho\,\langle(c_{\bk,0}+c^\dagger_{-\bk,0})^2\rangle$, we find 
\bea
\lim_{\bk\rightarrow 0}\,\chi^{LL}(\bk,i\omega_n=0) = -\frac{|\bk|^4}{16J}~.
\eea
This result is consistent with an incompressible quantum Hall fluid \cite{GMP}.  Note that the limit is independent of $M$ and is valid so long as $J\neq 0$, which is consistent with the vortex excitations dominating the long distance physics.  

%%%%%%%%%%%%%%%%%%%%%%%%%%%%%%%%%%%%%%%%%%%%%%%%%%%%%%%%%%%%%%%%%%%%
\subsection{The case $p=3$}
\label{sec:p=3}
%%%%%%%%%%%%%%%%%%%%%%%%%%%%%%%%%%%%%%%%%%%%%%%%%%%%%%%%%%%%%%%%%%%%

The expansions of the constraints, eqn. (\ref{eq:Constraints_P=3}), is prohibitive in general.  To get an idea of the spectrum, we cut the expansion off after $\sigma_{1,2}=2$.  The spectrum, as obtained from the zeros of ${\bf G}$ is
\be
i\omega_n=\left\{
\begin{array}{c} 
\left[E_{\bk,0}E_{\bk,2}\left(\frac{\bk^4E_{\bk,0}+4E_{\bk,2}} {\bk^4E_{\bk,2}+4E_{\bk,0}}\right)\right]^{1/2}\\
\\
\left[E_{\bk,1}E_{\bk,2}\left(\frac{\bk^4E_{\bk,1}+4E_{\bk,2}} {\bk^4E_{\bk,2}+4E_{\bk,1}}\right)\right]^{1/2}
\end{array}
\right.~.
\label{eq:spectrum_p=3}
\ee
The second branch in eqn. (\ref{eq:spectrum_p=3}) is doubly degenerate.  Presumably this degeneracy would be lifted in a better approximation.  The two modes are shown in fig. \ref{fig:spectrump=3}.  With the help of {\sc Mathematica}, we have analyzed the spectrum in more detail and find that the gaps are $J\sigma$, $\sigma\geq 2$.  A non-zero gap, of course, is consistent with an incompressible fluid.
\begin{figure}[htb]
\noindent
\center
\epsfxsize=3in
\epsfbox{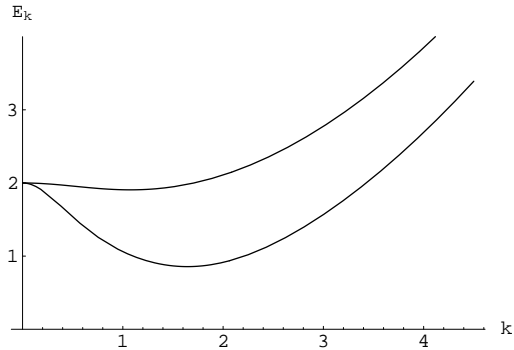}
\caption{\small Lowest two modes at $\nu=1/3$ obtained by expanding the action to ${\cal O}(k^2)$.  The higher mode is doubly degenerate.  $\bk$ is in units of inverse magnetic length.  The parameters are $m=3$ and $J=1$.}
\label{fig:spectrump=3}
\end{figure}
Once again, the main feature is the magnetoroton gap.  It seems to be a general feature for all $p$, provided that $J$ is large enough relative to $1/M$.

\chapter{Paired States of Bosons in Two Dimensions}
\label{chap:PERM}
%%%%%%%%%%%%%%%%%%%%%%%%%%%%%%%%%%%%%%%%%%%%%%%%%%%%%%%%%%%%%%%%%%%%%%%%%%%%%%%

In this chapter we analyze p- and d-wave pairing of bosons from two points of view.  First as an exact gound state of a particular Hamiltonian in the FQHE, where the bosons are really composite bosons.  And second, in the framework of BCS theory of paired states of ordinary bosons.

We find that the permanent is an anti-Skyrmion texture sitting on the transition between ferromagnetic and helical ordering, both of which are single-particle condensates.  The transition is of second order so that we can write a continuum quantum ferromagnet action \cite{ReadSachdev} for the local magnetization in its vicinity.  An analogous description on a lattice is straightforward and will be discussed briefly.  The results are compared to numerics with excellent agreement.  

In Section \ref{Haffnian} we move on to the Haffnian.  As we did for the permanent, we first construct a three-body Hamiltonian for which the trial wavefunction is exact, and from which numerical diagonalization techniques can extract the spectrum.   Unfortunately in this case we cannot calculate the spectrum analytically and we resort to an effective BCS-type hamiltonian at the outset.  A mean field analysis suggests three possible phases in the presence of an attractive channel at $l=-2$: (i) a single particle condensate, or Laughlin state, (ii) a pure pair state and (iii) a charge-density-wave phase.  We conjecture that the Haffnian lies on the special point separating (i) from (ii).  In support, our numerical evidence suggests that the Haffnian is compressible and that it may contain incipient long lange correlations in the density.
%%%%%%%%%%%%%%%%%%%%%%%%%%%%%%%%%%%%%%%%%%%%%%%%%%%%%%%%%%%%%%%%%%%%%%%%%%%%%%%%
\section{The Permanent: p-wave Pairing}
\label{Permanent}
%%%%%%%%%%%%%%%%%%%%%%%%%%%%%%%%%%%%%%%%%%%%%%%%%%%%%%%%%%%%%%%%%%%%%%%%%%%%%%%% 
%%%%%%%%%%%%%%%%%%%%%%%%%%%%%%%%%%%%%%%%%%%%%%%%%%%%%%%%%%%%%%%%%%%%%%%%%%%%%%%%
\subsection{Analytic Structure of the Ground State}
\label{PermanentStructure}
%%%%%%%%%%%%%%%%%%%%%%%%%%%%%%%%%%%%%%%%%%%%%%%%%%%%%%%%%%%%%%%%%%%%%%%%%%%%%%%%
Before discussing the actual calculations, we briefly summarize the relevant properties of the permanent state.  Detailed analysis and related states may be found elsewhere \cite{MR,RR,Yoshioka}.  For the moment we begin by choosing to put the system on a sphere with a magnetic monopole in the center \cite{Haldane83}. 

For $\nu=1/q$, the permanent state is a spin-singlet ground state of spin-1/2 fermions for $q$=odd and of spin-1/2 bosons for $q$=even.  The former case is the relevant one here, the simplest being $q=1$.  At this filling factor, the microscopic three-body Hamiltonian is the projection onto the manifold of states spanned by the closest approach of three fermions.  It takes an especially compact form on the sphere, where each particle in the lowest Landau level (LLL) has orbital angular momentum $N_{\phi}/2$:
\begin{equation}
H=V\sum_{i<j<k}P_{ijk}(\frac{3}{2}N_{\phi}-1,\frac{1}{2})~.
\label{HPermSph}
\end{equation} 
The arguments $(3N_{\phi}/2-1,1/2)$ are the total orbital angular momentum and $z$-component of spin, respectively, of three fermions at closest approach.  $P_{ijk}$ is the projection operator onto these states for each triplet of particles.  Note that the angular momentum is allowed to be half integral because the relevant single particles states are monopole harmonics \cite{WuYang}. Due to symmetry, three fermions can never have total orbital angular momentum $3N_{\phi}/2$, so $H$ is properly regarded as a projection onto triplet states of momentum {\em greater than or equal to} $3N_{\phi}/2-1$.  To specify a unique ground state, the total flux, $N_{\phi}$, through the sphere must be fixed.  At $N_{\phi}=(N-1)-1$, the permanent is the densest zero-energy eigenstate of $H$.  The densest state being the one with the smallest $N_\phi$.  For our purposes, it is most convenient to project stereographically onto the plane, where the {\it i\/}'th particle has the complex coordinate $z_{i}=x_{i}+iy_{i}$. Denoting the spin component of the $i$'th particle by $\sigma_i=\uparrow_i,\downarrow_i$, casts the permanent into the form:
\begin{equation}
\Psi_{\rm perm}(z_{1}\sigma_1,\ldots, z_{N/2}\sigma_{N/2})=\sum_{P}\prod^{N/2}_{i=1}
 \frac{\uparrow_{P(2i-1)}\downarrow_{P(2i)}-\downarrow_{P(2i-1)}\uparrow_{P(2i)}}{z_{P(2i-1)}-z_{P(2i)}}\prod_{i<j}(z_{i}-z_{j}).
\label{PsiPerm}
\end{equation}
The prefactor is the permanent of an $N\times N$ matrix, which is a determinant with the sign of the permutation omitted, and $P$ stands for all permutations of $N$ objects.  The second factor is the usual Laughlin-Jastrow ansatz with the Gaussian factors omitted.  The BCS pairing structure of the prefactor is manifest in the above form; the spatial part of the pair wavefunction is in an orbital angular momentum $l=-1$ eigenstate and the spin part is a singlet.  Therefore the prefactor is totally symmetric, representing a paired wavefunction of bosonic coordinates, which are non other than the composite bosons.  

The factors in the denominator ensure that the relative angular momentum of each disjoint pair $(z_i-z_j)$ is reduced such that each projection by $P_{ijk}$ gives zero.  In other words, $H$ penalizes those states which do not appear in the wavefunction.  Addition of one flux quantum preserves this property and there is a space of zero energy states including the Laughlin state.  Equivalently, reducing the flux by one quantum through the Laughlin state creates an anti-~Skyrmion, which is a uniform spin configuration on the sphere (a ``hedgehog'') costing zero energy.  Stated in yet another way, extra flux in the permanent creates quasiholes (or edge states), which belong to a degenerate manifold of zero-energy states \cite{RR,MilovanovicRead}.  The advantage of the $q~=~1$ Laughlin state is that it is a Slater determinant of single particle states, which in this case is a Vandermonde determinant:  
\begin{equation}
\Psi_{L}(z_{1},\cdots,z_{N})=\prod_{i}\left(\begin{array}{c}1\\0\end{array}\right)_{i}\;\;\prod_{i<j}(z_{i}-z_{j}),
\label{PsiLaughlin}
\end{equation}
where $\left(1\;0\right)_{i}$ is the spin state of the {\em i\/}'th particle, so that the total $z$~-~component of spin is $S_{z}=N/2$.  The spin-wave states correspond to superpositions of the degenerate states defined by $S_{z}=N/2-1$.  Notice that the prefactor of the Laughlin-Jastrow factor is trivially constant and symmetric under particle interchange.  By analogy to the permanent (\ref{PsiPerm}), we interpret the prefactor as a wavefunction of composite bosons.  

In the following subsection we switch to the plane and construct the spin wave excitations in the LLL using magnetic translation operators, with which we obtain the exact spin wave spectrum at all wavevectors.  We do not utilize the composite boson picture yet and all results in the next section are completely exact.
%%%%%%%%%%%%%%%%%%%%%%%%%%%%%%%%%%%%%%%%%%%%%%%%%%%%%%%%%%%%%%%%%  
\subsection{Magnetic Translations on the Plane}
\label{MagneticTranslations}
%%%%%%%%%%%%%%%%%%%%%%%%%%%%%%%%%%%%%%%%%%%%%%%%%%%%%%%%%%%%%%%%%
A single magnon mode with momentum ${\bf k}$ is represented in terms of spin density operators by $\hat{S}_{{\bf k}}^{-}=\sum_{j=1}^{N}e^{-i{\bf k}\cdot{\bf r}_{j}}\hat{\sigma}_{j}^{-}\;$, which periodically flips one spin of a completely polarized state.  However, all operators must not have any components in the higher Landau levels. To impose this constraint, it proves necessary to project all operators into the LLL using the Bargmann-Fock representation \cite{Moon,GirvinJach}.  As a consequence, a magnon is composed of flipped spins along the direction {\em perpendicular\/} to $\bk$.  

The fundamental operator that we will need is the projected one-body density operator $\overline{\rho}_{\bf k}({\bf r}_{i})=\overline{e^{-i{\bf k}\cdot{\bf r}_{i}}}$.  Here and throughout this chapter the overbar denotes projection into the LLL. This operator is discussed in detail in the Appendix.  Because translations no longer commute in the LLL due to broken time translation symmetry in the presence of a magnetic field, a charged particle picks up an Aharonov-Bohm phase as it traverses a closed circuit.  However, there is a one-to-one correspondence between the ordinary translations in zero field and the magnetic translations.  In zero field, $\rho_{{\bf k}}({\bf r}_{i})$ translates the {\em i\/}'th particle by a distance ${\bf r}$, and the corresponding magnetic translation, $\tau_{{\bf k}}(i)$, is a translation by $l_{B}^{2}\hat{\bf z}\times{\bf k}$, where $l_{B}$ is the magnetic length and $\hat{\bf z}$ is the direction of the magnetic field.  The cross-product appears as a consequence of the Lorentz force.  The explicit relation is
\begin{equation}
\tau_\bk(i)=e^{\frac{k^{2}}{4}}\overline{\rho}_\bk({\bf r}_{i})\;, 
\label{Tau}
\end{equation} 
where the magnetic length has been set to unity---as it will be throughout the chapter.  The ubiquitous Gaussian factor can be thought of as the momentum space version of the most localized wavepacket in the LLL, or the projected delta function.  The total spin and charge density operators are easily obtained from (\ref{Tau}):
%\begin{mathletters}
\begin{equation}
\overline{\rho_{{\bf k}}}=\sum_{i=1}^{N}e^{-\frac{k^{2}}{4}}\tau_{{\bf k}}(i)
\label{TotChargeDens}
\end{equation}
\begin{equation}
\overline{S}_{\bf k}^{a}=\sum_{i=1}^{N}e^{-\frac{k^{2}}{4}}\tau_{{\bf k}}(i)\sigma_{i}^{a}\;,
\label{TotSpinDens}
\end{equation}
%\end{mathletters}
where $\sigma^{a}$ is the {\em a\/}'th Pauli matrix.  The projected spin density flips one spin and translates the resulting state by $\hat{\bf z}\times{\bf k}$, which is a quasihole-quasiparticle pair.  The limit of large ${\bf k}$ corresponds to a large pair so we expect that the spin wave spectrum approaches an asymptotic value in this limit.  

The Laughlin state (\ref{PsiLaughlin}) is a Slater determinant, which allows the density-density correlation to be determined analytically \cite{GirvinPrange} at $\nu =1$.  In terms of the magnetic translations in (\ref{TotChargeDens}), the pair correlation function (projected onto the LLL) is:
\begin{eqnarray}
\langle\Psi_{L}|:\overline{\rho_{\bf q}\rho_{-\bf k}}:|\Psi_{L}\rangle =
\sum_{i\neq j}e^{-\frac{q^{2}}{4}}e^{-\frac{k^{2}}{4}}\langle\tau_{\bf q}(i)\tau_{-\bf k}(j)\rangle =
-\langle\rho\rangle^{2}\delta_{{\bf k}{\bf q}}e^{-\frac{q^{2}}{2}},\nonumber
\end{eqnarray}
or
\begin{equation}
\sum_{i\neq j}\langle\tau_{\bf q}(i)\tau_{-\bf k}(j)\rangle= 
-\langle\rho\rangle^{2}\delta_{{\bf k}{\bf q}}~,
\label{TauCorr}
\end{equation}
where the colons remove self-correlations by normal ordering.  Equation (\ref{TauCorr}) is the building block for much of the subsequent analytical results.

The non-commutativity of magnetic translations is irrelevant in (\ref{TauCorr}) because the products involve different particles.  However, the translations do not commute for the same particle.  Rather, they form a representation of the magnetic translation group with the algebra:
%\begin{mathletters}
\begin{equation}
	\left[\tau_{\bf q}(i),\tau_{\bf k}(j)\right]=
	2\mbox{i}\delta_{ij}\tau_{{\bf q}+{\bf k}}(i)\sin\frac{{\bf q}\wedge{\bf k}}{2}
\label{TauComm}
\end{equation}
\begin{equation}
	\tau_{\bf q}(i)\tau_{\bf k}(i)=
	\tau_{{\bf q}+{\bf k}}(i)e^{\frac{i}{2}{\bf q}\wedge{\bf k}},
\label{TauProd}
\end{equation}
%\end{mathletters}
where the wedge product stands for $({\bf q}\times{\bf k})\cdot\hat{\bf z}$.

Using the projected spin density operator we can easily construct a single magnon by applying the spin density operator (\ref{TotSpinDens}) to the Laughlin state: $\Psi_{\bf k}=\overline{S}_{\bf k}^{-}\Psi_{L}$.
%%%%%%%%%%%%%%%%%%%%%%%%%%%%%%%%%%%%%%%%%%%%%%%%%%%%%%%%%%%%%%%%%%%%%%%%
\subsection{Spin-Waves and Instability; Exact Results}
\label{ExactSpectrum}
%%%%%%%%%%%%%%%%%%%%%%%%%%%%%%%%%%%%%%%%%%%%%%%%%%%%%%%%%%%%%%%%%%%%%%%%
The translationally invariant version of the three-body Hamiltonian (\ref{HPermSph}) on the plane can be written as:
\begin{equation}
H=V\sum_{i\neq j\neq k}\nabla_{i}^{2}\delta ({\bf r}_{i}-{\bf r}_{j})\delta 
({\bf r}_{i}-{\bf r}_{j})
\label{HPermPlane}
\end{equation}
Of course, all observables are to be calculated after projection into the LLL. Since the Hamiltonian conserves the total spin, as well as being translationally invariant, the state with one magnon is an exact eigenstate of $\overline{H}$ with energy given by the usual expression \cite{Auerbach}:
\begin{equation}
\omega_{\bf k}= \frac{\langle\Psi_{L}|\overline{S}_{\bf 
k}^{+}\left[\overline{H},\;\overline{S}_{\bf 
k}^{-}\right]|\Psi_{L}\rangle}{\langle\Psi_{L}|\overline{S}_{\bf 
k}^{+}~\overline{S}_{\bf k}^{-}|\Psi_{L}\rangle}~.
\label{SMA}
\end{equation} 
Within the expectation value of $\overline{H}$, the two gradients reduce the orbital angular momentum of a triplet in $\Psi_{\bf k}$ and in $\Psi_{\bf k}^{*}$ by one, hence the identity with the spherical Hamiltonian (\ref{HPermSph}).

$\overline{H}$ can be rewritten in terms of the real-space density operator $\rho({\bf r})=\sum_{i}\delta ({\bf r}-{\bf r}_{i})$ in a similar manner:
\begin{eqnarray}
\overline{H}&=&V~\int d^{2}{\bf x}d^{2}{\bf x}^{\prime}d^{2}{\bf x}^{\prime\prime}:\nabla_{\bf 
x}^{2}~\overline{\rho}({\bf x})~\overline{\rho}({\bf x}^{\prime})~\overline{\rho}({\bf x}^{\prime\prime}): \delta({\bf x}-{\bf x}^{\prime)}\delta({\bf x}^{\prime}-{\bf 
x}^{\prime\prime})\nonumber\\
 &=&-V\sum_{{\bf q}{\bf p}}q^{2}e^{-\frac{q^{2}}{4}} e^{-\frac{({\bf q}-{\bf p})^{2}}{4}}e^{-\frac{p^{2}}{4}}~\sum_{i\neq j\neq k}~\tau_{-{\bf q}}(i)\tau_{{\bf q}-{\bf p}}(j)\tau_{\bf p}(k)~.
\label{11}
\end{eqnarray}
The last line is obtained by a Fourier transform and by the relation between the density and magnetic operators (\ref{TotChargeDens}).  Substitution of the last line into the energy expression (\ref{SMA}) and use of the magnetic operator algebra (\ref{TauComm}) and (\ref{TauProd}) reduces the dispersion to:
\begin{eqnarray}
\omega_{\bf k}&=
-V&\sum_{{\bf q},{\bf p}}q^{2}e^{-\frac{q^{2}}{2}} 
 e^{-\frac{({\bf q}-{\bf p})^{2}}{2}}e^{-\frac{p^{2}}{2}}\times\\
 &&\sum_{i\neq j\neq k}~
 \langle\tau_{-{\bf q}}(i)\tau_{{\bf q}-{\bf p}}(j)\tau_{\bf p}(k)\rangle_{_L}      
 \left(~e^{i{\bf p}\wedge{\bf k}}+e^{i({\bf q}-{\bf p})\wedge{\bf k}}+
 e^{-i{\bf q}\wedge{\bf k}}-3\right)~.\non
\label{12}
\end{eqnarray}
Rather than calculating the triple-density correlator directly, observe that everything inside the expectation value is already in the LLL.  Thus, the magnetic translations can be replaced by density operators according to the rule in (\ref{TotChargeDens}).  After Fourier transforming back into real space, we are left with terms like $\langle:\!\!\nabla_{\bf x}^{2}\rho({\bf x})\rho({\bf x}^{\prime})\rho({\bf x}^{\prime\prime})\!\!:\rangle\delta({\bf x}-{\bf x}^{\prime})\delta({\bf x}^{\prime}-{\bf x}^{\prime\prime}+{\bf k}\times\hat{\bf z})$, which can be evaluated by writing the density operator in second quantization, $\rho({\bf x})=\langle\hat{\psi}^{\dag}({\bf x})\hat{\psi}({\bf x})\rangle$, and using Wick's theorem.  For the spin-polarized $\nu$=1 state, the Green's function is known exactly \cite{GirvinPrange}, being, in the symmetric gauge,
\begin{equation}
\langle\hat{\psi}^{\dag}({\bf x})\hat{\psi}({\bf x}^{\prime})\rangle_{_L} 
=\langle\rho\rangle~
e^{-\frac{1}{4}|{\bf x}-{\bf x}^{\prime}|^{2}}e^{\frac{i}{2}{\bf x}\wedge{\bf 
x}^{\prime}}~.
\label{GreenFn}
\end{equation}
The end-result is the {\it exact\/} spin wave spectrum:
\begin{equation}
\omega_{\bf k}=C\left[1-e^{-k^{2}/{2}}\left(\frac{k^{2}}{2}+1\right)\right]~,
\label{Omega}
\end{equation}
where $\Omega$ is the area of the system, $\langle\rho\rangle=1/2\pi\ell_{B}^{2}$, and the asymptotic value is $C=16V\langle\rho\rangle\Omega$.  Equation (\ref{Omega}) is the central result of this section.  It should be borne in mind that, although it was obtained for the Laughlin state (\ref{PsiLaughlin}), it is also valid for the permanent (\ref{PsiPerm}), since the two are degenerate in the presence of our three-body interaction (\ref{HPermSph}).   The comparison to exact numerics is shown in Fig. \ref{fig:exact_spectrum}.
\begin{figure}[htb]
\noindent
\center
\epsfxsize=3 in
\epsfbox{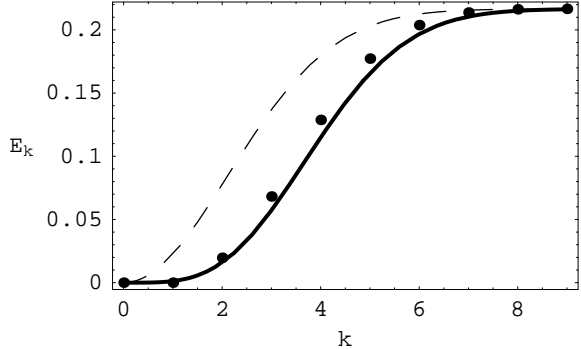}
\caption{\small Spin-Wave Dispersion: Points are numerical data for $N=10$,$N_{\phi}=9$ (computed by E. Rezayi).  Solid line is the spectrum in (\ref{Omega}), with $C$ determined by matching to the data at the highest $k$.  For comparison, the dotted line is the spectrum for a two body interaction (\ref{H2body}) with $V_{2}=0$.}
\label{fig:exact_spectrum}
\end{figure}

At large $\bk$, $\omega_{\bf k}$ approaches an asymptotic value as expected.  This is the energy of a widely separated quasihole-quasiparticle pair, as in earlier work on two-body interactions \cite{Kallin}.  However, the novel feature of (\ref{Omega}) emerges at small $\bk$, where $\omega_{\bf k}\sim~k^{4}$~, in contrast to the usual quadratic dependence.  In other words, the spin stiffness is exactly zero when the interaction is the three-body Hamiltonian (\ref{HPermPlane}).  This is precisely what was conjectured earlier based on numerical evidence \cite{RR}.  Here, however, we have an exact calculation verifying that claim, and the comparison with the original data is quite good, as shown in Fig. \ref{fig:exact_spectrum}.  Therefore, the permament is poised on the brink of an instability because $\omega_{\bf k}$ becomes negative as soon as the spin stiffness dips below zero.

The degeneracy of the {\em polarized} Laughlin state and the {\em unpolarized} permanent can also be understood in light of the long wavelength behavior of the spectrum.  At zero spin stiffness, it should cost no energy to create a slowly varying spin texture of unbounded size.  In fact numerical calculations of the spin-spin correlator $\langle S_{z}({\bf x})S_{z}({\bf y})\rangle$ on the sphere confirm that the spins are approximately anti-aligned at antipodal points, indicating an anti-Skyrmion texture with long range order (Fig. \ref{fig:S_Correlator}).  In the next subsection we will see that pairing of composite bosons predicts precisely this ordering.  Skyrmions are not new in the FQHE \cite{Sondhi,Moon} and are typically associated with excitations when extra flux is added.  Here, however, they appear at one {\it fewer} flux quantum.
\begin{figure}[htb]
\noindent
\center
\vspace{45pt}
\epsfxsize=3.5 in
\epsfbox{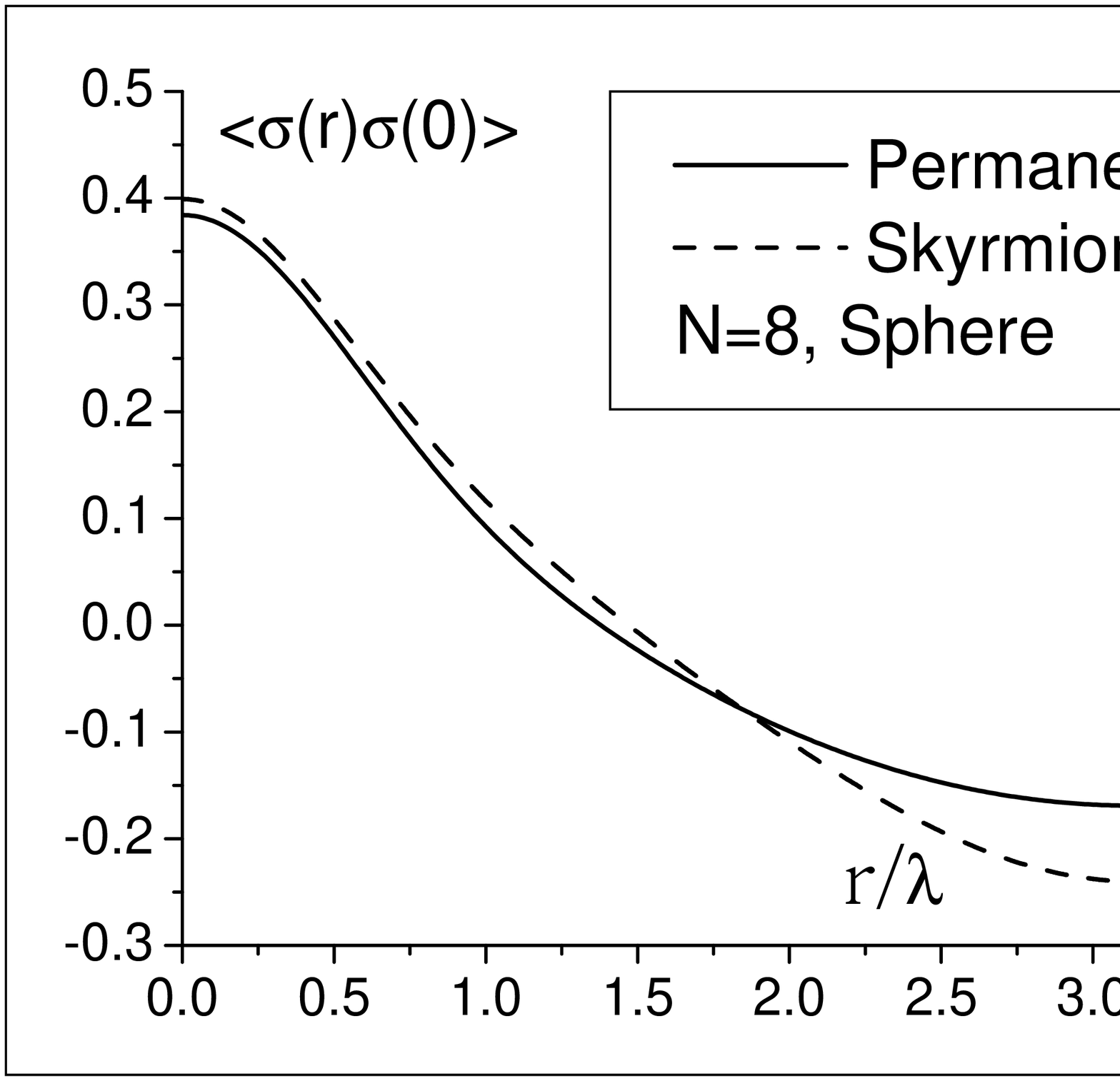}
\vspace{-45pt}
\caption{\small $\langle S_{z} S_{z}\rangle$ of the Permanent and, for comparison, of the Laughlin state with one fewer flux.  $\lambda$ is the magnetic length. (Computed by E. Rezayi)}
\label{fig:S_Correlator}
\end{figure}

The three-body spectrum obtained above cannot exhibit an instability because it is always stable---i.e. projection operators generally have no negative eigenvalues.  However, a negative quadratic term can be restored by including a short-range, two-body interaction.  For concreteness, we model such a potential by the simplest, non-trivial expansion, 
\begin{equation}
H_{2}=\sum_{i\neq j}V_{0}\delta({\bf x}_{i}-{\bf x}_{j}) + 
\frac{1}{2}V_{2}\nabla_{i}^{2}\delta({\bf x}_{i}-{\bf x}_{j})~,
\label{H2body}
\end{equation}
in place of $H$.  There is no restriction on the sign of $V_{2}$, but $V_{0}$ should be positive.  Except for the special case $V_{2}=0$, for which $H_2$ is thepseudopotential \cite{GirvinPrange} at $\nu =1$, the Laughlin state is no longer an eigenstate.  However, due to translational invariance, the state with one magnon is an eigenstate, so that its spectrum may be calculated exactly once again.  Projecting $H_{2}$ onto the LLL and following the same procedure as before yields:
\begin{equation}
\omega_{\bf k}=-\frac{1}{2}\sum_{\bf q}V_{q}e^{-q^{2}/2}\left(e^{i{\bf 
q}\wedge{\bf k}} 
+ e^{-i{\bf q}\wedge{\bf k}}-2\right)~,
\label{Omega2}
\end{equation}
where $V_{q}=V_{0}+\frac{1}{2}V_{2}q^{2}~$.  Eqn. (\ref{Omega2}) agrees with earlier calculations using many-body techniques \cite{Kallin}.  Unlike the three-body case, the coefficient of the quadratic term in the dispersion does not vanish in general.  However, by tuning the interaction parameters, the stiffness can be forced to zero, and the two- and three-body interactions will behave similarly at large distance.  For example, the ratio $V_{2}/V_{0}=-2/3$ mimics the three-body spectrum (\ref{Omega}) at small {\bf k}.  

When $V_{2}/V_{0}<-2/3$, $\omega_{\bf k}$ looks like the familiar ``Mexican Hat'' potential, with a minimum at the wavevector $Q^{2}=|V_{0}/V_{2}|$.  At zero temperature, all the magnons (or quasiparticle-quasihole pairs) condense into the momentum ${\bf Q}$. Nonetheless, the total velocity of the superfluid is zero because the up-spins condense into ${\bf Q}/2$ and the down-spins into $-{\bf Q}/2$.  We illustrate this explicitly by a Hartree-Fock function in the LLL with the variational parameter ${\bf Q}$.  At this point this is more of an ansatz than anything else, but we will justify this in the next subsection.  For the moment, consider the single particle orbitals
\begin{equation}
\phi_{m}({\bf r})=\frac{1}{\sqrt{2}}\left(\begin{array}{c}
			e^{\frac{i}{2}{\bf Q}\cdot{\bf r}}\\
			e^{-\frac{i}{2}{\bf Q}\cdot{\bf r}}\end{array}
		  \right)z^{m}e^{|z|^{2}/2}\equiv
		  \left(\begin{array}{c}\rho_{{\bf Q}/2}({\bf r})\\
		                        \rho_{-{\bf Q}/2}({\bf r})\end{array}
		                        \right)z^{m}e^{|z|^{2}/2}~,
\nonumber
\end{equation}
where $\rho_{{\bf Q}/2}({\bf r})$ is the one-body density operator that is to be projected into the LLL according to the rule in (\ref{Tau}).  The Slater determinant, $\tilde{\Psi}_{\bf Q}$, of these orbitals exhibits helical spin order and reduces to $\Psi_{L}$ at ${\bf Q}=0$.  In terms of the magnetic operators, $\widetilde{\Psi}_{\bf Q}$ is:
\begin{equation}
\widetilde{\Psi}_{\bf Q}(z_{1},\ldots,z_{N})=\prod_{i}\frac{1}{\sqrt{2}}e^{-Q^{2}/4} \left(\begin{array}{c}\tau_{{\bf Q}/2}(i)\\\tau_{-{\bf Q}/2}(i)\end{array}     \right)_{i}~\prod_{i<j}(z_{i}-z_{j})e^{-\sum_{i}|z_{i}|^{2}/4}~.
\label{17}
\end{equation}
Let us rewrite $\widetilde{\Psi}_{\bf Q}$ as the spinor operator times the purely spatial part, $\widetilde{\Psi}_{\bf Q}=T_{\bf Q}\Psi_{0}$.  The energy $E_{\bf Q}$ of $\widetilde{\Psi}_{\bf Q}$ can be calculated in exactly the same way as the spin-wave energy (by using (\ref{SMA})).  We find that $E_{\bf Q}$ is identical to $\omega_{\bf Q}$, i.e.
\begin{eqnarray}
\omega_{\bf Q}&=&\frac{\langle\overline{S}_{\bf 
Q}^{+}\left[\overline{H}_{2},\;\overline{S}_{\bf Q}^{-}\right]\rangle_{_L}} 
{\langle\overline{S}_{\bf Q}^{+}~\overline{S}_{\bf 
Q}^{-}\rangle_{_L}}=\nonumber\\
E_{\bf Q}&=&\frac{\langle\overline{T}_{\bf 
Q}^{\dagger}\left[\overline{H}_{2},\;\overline{T}_{\bf Q}\right]\rangle_{_0}} 
{\langle\overline{T}_{\bf Q}^{\dagger}~\overline{T}_{\bf Q}\rangle_{_0}}~.
\label{HFState}
\end{eqnarray}
In fact, this identity holds generally for {\em any} translationally invariant interaction that can be expressed as a product of charge density operators.  In particular, it is true for both of the three- and two-body interactions, illustrating that our picture of a magnon condensate is internally consistent.      

The wavevector {\bf Q} plays the role of an order parameter which increases continuously from zero as one crosses over from the ferromagnetic into the helical phase.  On the other hand, the magnetization undergoes a first order transition.  The boundary between the two phases is defined by $\rho_{s}=0$.  In the following subsection, we prove that helical order is incipient in the ground states of the permanent by adopting the composite boson point of view.  A similar procedure on the sphere will show that the analog of the helical ordering is exactly the anti-Skyrmion, as was suggested by the numerical data in Fig. \ref{fig:S_Correlator}. 
%%%%%%%%%%%%%%%%%%%%%%%%%%%%%%%%%%%%%%%%%%%%%%%%%%%%%%%%%%%%%%%%%%%%%%
\subsection{Spin Order of the Permanent}
\label{PwavePairing}
%%%%%%%%%%%%%%%%%%%%%%%%%%%%%%%%%%%%%%%%%%%%%%%%%%%%%%%%%%%%%%%%%%%%%%
Let us begin by recalling the general theory of paired bosons in zero magnetic field \cite{Nozieres}.  An effective Hamiltonian which captures this physics is of the BCS type:
\begin{equation}
K_{eff}=\sum_{{\bf k}\sigma}\left[ \xi_{\bf k}c^{\dagger}_{{\bf k}\sigma}c_{{\bf k}\sigma}+\frac{1}{2}\left(\Delta^{\ast}_{\bf k}c_{-{\bf k}\downarrow}c_{{\bf k}\uparrow}+\Delta_{\bf k}c^{\dagger}_{{\bf k}\uparrow}c^{\dagger}_{-{\bf k}\downarrow}\right)\right]
\label{KeffPerm}
\end{equation}
where $\xi_{\bf k}=\varepsilon_{\bf k}-\mu$ and $\varepsilon_{\bf k}$ is the single-particle kinetic energy and $\Delta_{\bf k}$ is the gap function.  In the fractional quantum Hall effect, the quasiparticles entering $K_{eff}$ are the composite bosons, $c_{{\bf k}\sigma}$, which see no magnetic field.    We assume that $\varepsilon_{\bf k}\simeq k^2/2m^{\ast}$ at small ${\bf k}$, where $m^{\ast}$ is an effective mass of the composite (see for example Chapter \ref{chap:Fermions}).  For p-wave pairing, we take $\Delta_{\bf k}$ to be an eigenfunction of rotations in ${\bf k}$ of eigenvalue $l=-1$.  At small ${\bf k}$ the generic form of the gap function is thus
\begin{equation}
\Delta_\bk\simeq\hat{\Delta}(k_x-ik_y)
\label{DeltaPerm}
\end{equation}
where $\hat{\Delta}$ is a constant.  Although there is no explicit single particle condensate in $K_{eff}$, it will be emerge naturally below as being equivalent to pure p-wave order.
   
More rigorously, one must solve the self-consistent gap equation when the interaction contains an attractive $l=-1$ channel.  Consider a non-singular interaction which has a power series expansion at short distance: $V({\bf k})=a_0+a_2k^2+\ldots$ .
The gap equation is
\begin{equation}
\Delta_{\bf k}=-\frac{1}{2}\sum_{{\bf k}^\prime}V({\bf k}-{{\bf k}^\prime})\frac{\Delta_{{\bf k}^\prime}}{E_{{\bf k}^\prime}}.
\label{GapEqn}
\end{equation}
where the quasiparticle energy is
\begin{equation}
E_{\bf k}=\sqrt{\xi_{\bf k}^2-|\Delta_{\bf k}|^2}
\label{Ek}
\end{equation}
Note the minus sign in contrast with the familiar fermion case.  This is a complicated non-local integral equation, but it separates if we assume that the gap function contains only one angular momentum channel: $\Delta_{\bf k}=|\Delta_{\bf k}|e^{il\phi}$, where $\phi$ is the polar angle of ${\bf k}$.  The interaction expands similarly into angular momenta:
\begin{equation}
V(\bk-\bk^\prime)=\sum_{m=-\infty}^{\infty}V_m(k,k^\prime)e^{-im(\phi-\phi^\prime)}
\label{Vexp}
\end{equation}
where the coefficients $V_l(k,k^\prime)$ depend only on the magnitudes of ${\bf k}$ and  ${\bf k}^\prime$.  It is straightforward to show that the leading order behavior of $V_l$ in $\bk$ is
\be
V_l(k,k^\prime)\simeq k^l
\label{Vl}
\ee
Substituting the expansion (\ref{Vexp}) into the gap equation (\ref{GapEqn}) yields precisely the p-wave gap (\ref{DeltaPerm}) at leading order in ${\bf k}$ whenever $V_{-1}(k,k^\prime)$ is negative and $V_l=0$ for $l\neq -1$.  For general {\it l}-wave pairing the gap is proportional to $(k_x-ik_y)^l$ at long distance.

The quasiparticle energy $E_{\bf k}$ contains important physical information.  When $\mu<0$ and $\hat{\Delta}$ is small enough, the gap is $E_0=\mu$ and $2E_0$ is the energy needed to break a condensed pair.  On the other hand, as the gap closes the paired state ceases to exist and a single particle condensate appears \cite{Nozieres}.  Precisely at $\mu=0$ the spectrum is unstable for any finite $\hat{\Delta}$.  However, as we will show shortly, the pure pair state is really a single particle condensate and the solution of $K_{eff}$ is fully consistent when we expand about this new minimum.  

In the absence of any single particle condensates, a pure pair state of spin-$1/2$ bosons is
\begin{equation}
|\Omega\rangle=\frac{1}{\cal N}\exp\left\{\frac{1}{2}\sum_{\bf k}g_{\bf k}c^{\dagger}_{{\bf k}\uparrow}c^{\dagger}_{-{\bf k}\downarrow}\right\} |0\rangle
\label{PsiPairK}
\end{equation} 
where $\cal N$ is the normalization given by
\begin{equation}
{\cal N}=\prod_{\bf k}\frac{1}{1-|g_{\bf k}|^2}.
\label{Norm}
\end{equation}
To ensure that $|\Omega\rangle$ is normalizable it is necessary that $|g_{\bf k}|^2<1$ for all ${\bf k}$ and p-wave order requires that $g_{\bf k}$ is antisymmetric in momentum space: $g_{\bf k}=-g_{-{\bf k}}$.  Therefore, due to bosonic statistics, the lowest allowed spin state of a pair is the spin singlet $\uparrow_i\downarrow_j - \downarrow_i\uparrow_j$ so  $|\Omega\rangle$ has total spin $S=0$.  

In real space, the (unnormalized) component of the pair wavefunction with $N$ particles ($N$ even) is 
\begin{equation}
\Psi({\bf r}_1\sigma_1,\ldots,{\bf r}_N\sigma_N)=\sum_P\prod^{N/2}_{i=1}g({\bf r}_{P(2i-1)}-{\bf r}_{P(2i)})\left(\uparrow_{P(2i-1)}\downarrow_{P(2i)}-\downarrow_{P(2i-1)}\uparrow_{P(2i)}\right)
\label{PsiPairR}
\end{equation}
where $g({\bf r})$ is the inverse Fourier transform of $g_{\bf k}$ and $P$ runs over all permuations of $N$ objects.  This is the form of a permanent of an $N\times N$ matrix, which looks like a determinant, but with the sign of $P$ omitted.  In fact, it is the analog of the Pfaffian (which is the determinant of an antisymmetric matrix) for paired fermions.

If we are dealing with the FQHE, then the bosonic operators $c_{{\bf k}\sigma}$ really originated as composite bosons.  During projection to the LLL, one typically picks up a cutoff factor on $g_{\bf k}$ of $\mbox{exp}(-l^2_B|\bk|^2/2)$, where $\ell_B$ is the magnetic length.  We will neglect this factor in all that follows with the understanding that $K_{eff}$ and $g_{\bf k}$ are valid at long distance.  With this caveat, it is easy to see that if 
\begin{equation}
g_\bk=\frac{\lambda}{(k_x+ik_y)}
\label{gk}
\end{equation}
with $\lambda$ a constant, then the asymptotic behavior at long distance of the inverse Fourier transform is 
\begin{equation}
g({\bf r})\simeq\frac{1}{z}\;.
\label{gr}
\end{equation}
Comparing this with (\ref{PsiPairR}), we find that at long distance, the pair wavefunction is precisely the permanent prefactor of the LLL state defined by $\Psi_{\rm perm}$ (\ref{PsiPerm}).

While the Laughlin-Jastrow factor is taken care of by the transformation to composite bosons, it should be borne in mind that the price of this projection is an extra constraint or a fluctuating (Chern-Simons) gauge field.  We are neglecting these effects at the mean field level.  Analogous questions for pairing of composite fermions have been raised recently \cite{Bonesteel99}.

Now let us consider the occupation number at wavevector ${\bf k}$: $\langle n_{\bf k}\rangle=\langle c^\dagger_{{\bf k}\uparrow}c_{{\bf k}\uparrow}\rangle + \langle c^\dagger_{{\bf k}\downarrow}c_{{\bf k}\downarrow}\rangle.$  From the form of $|\Omega\rangle$ (\ref{Omega}) this can be written as a function of $g_{\bf k}$ only,
\begin{equation}
\langle n_{\bf k}\rangle=\frac{|g_{\bf k}|^2}{1-|g_{\bf k}|^2} .
\label{nk}
\end{equation}
Substituting the asymptotic behavior (\ref{gk}), we can write the total number of particles as
\begin{equation}
N=\sum_\bk\langle n_\bk\rangle=\sum_\bk\frac{|\lambda|^2}{|\bk|^2-|\lambda|^2}
\label{N}
\end{equation}
The occupation numbers fall off algebraically at small ${\bf k}$ and even more quickly (exponentially) on the scale of $k>1/\ell_B$.  The condition of normalisability of the ground state guarantees that $|\lambda/k|^2<1$ for all ${\bf k}$---in other words, $|\lambda|$ must be less than the minimum wavevector, $|{\bf k}_{min}|$.  To make sense of this expression we now impose boundary conditions that compactify the plane into a torus.  For simplicity, consider an $L\times L$ torus in the $xy$-plane; generalization to $L_x\times L_y$ and modular parameter $\tau$ is straightfoward. 

On the torus there are four degenerate ground states for the permanent Hamiltonian (\ref{PsiPerm}) corresponding to periodic $(++)$ or antiperiodic, $(+-)$, $(-+)$, $(--)$, boundary conditions in each of the two directions.  When both directions are periodic $(++)$ the minimum reciprocal lattice vector allowed is ${\bf k}_{min}=0$ and $N$ becomes sharply peaked (in fact it is a delta function) at $n_0$.  One of the spin directions (say, $\uparrow$) is singled out by the correlations and we are left with nothing other than a Bose condensate with
\begin{equation}
\langle c_{{\bf k}\sigma}\rangle=\sqrt{N}\delta_{{\bf k},0}\delta_{\sigma,\uparrow} 
\label{BEC}
\end{equation}
which is the spin-polarized Laughlin state.  On the other hand, in the antiperiodic sector,$(+-)$ or $(-+)$, $|{\bf k}_{min}|=\pi/L$.  Inverting (\ref{N}), we find that $|\lambda| =|{\bf k}_{min}|-{\cal O}(1/N)$ and in the thermodynamic limit one of these two sectors is macroscopically occupied.  For example, in the $(+-)$ direction, ${\bk}_{min}=(\pi/L,0)$.  Since the total momentum of the condensate must be zero, we occupy $(+\pi/L,0)$ and $(-\pi/L,0)$ with equal probability:
\begin{equation}
\langle c_{{\bf k}\uparrow}\rangle=\sqrt{\frac{N}{2}}\delta_{{\bf k},{\bf k}_{min}}\mbox{  and  }\langle c_{{\bf k}\downarrow}\rangle=\sqrt{\frac{N}{2}}\delta_{{\bf k},-{\bf k}_{min}}
\label{BEChelical}
\end{equation}
In real space this is precisely the helical winding with ${\bf Q}=2{\bf k}_{min}$ which corresponds to the spins winding exactly once over the length of the system $L$ in the $\hat{x}$ direction.  For the remaining antiperiodic sector, $(--)$, $|{\bf k}_{min}|=\sqrt{2}\pi/L$ and the winding is along the diagonal.  

In any case, we have shown that three of the ground states of the permanent Hamiltonian are really single particle condensates with helical spin order and the remaining one is the Laughlin spin polarized state.  As we tune through the transition, one or the other long range order takes over.  This justifies our Hartree-Fock ansatz (\ref{HFState}).  In the following subsection \ref{CQFM} we shall expand about the single particle minimum by using an effective Landau-Ginzburg theory instead of $K_{eff}$. 

Finally, we repeat the analog of the above on a sphere, which provides a nice intuitive picture of the anti-Skyrmion texture.  Recall that the permanent state on the sphere has one fewer flux quantum than the Laughlin state.  Thus, composite bosons live on the surface of a sphere with a magnetic monopole of strength one.  The appropriate single particle states are monopole harmonics, $Y_{L,M}(\theta,\phi)$, with angular momentum $L\geq 1/2$ and $M$ is the magnetic quantum number in the range $-L\leq M\leq L$ \cite{WuYang}.  A pair wavefunction must be rotationally invariant just as it is translationally invariant on the torus or the plane.  The unique bilinear scalar is 
\begin{equation}
\sum_{L=-1/2}^{\infty}\sum_{M=-L}^{L}Y_{L,M}(\theta,\phi)Y_{L,-M}(\theta^\prime,\phi^\prime)f_L\langle 0,0|L,M;L,-M\rangle ,
\label{Clebsch}
\end{equation}
where $\langle 0,0|L,M;L,-M\rangle$ is the Clebsch-Gordan coefficient for coupling $|L,M\rangle$ and $|L,-M\rangle$ into the orbital singlet $|0,0\rangle$ and $f_L$ is a function of $L$ only.  This particular Clebsch-Gordan coefficient is equal to a function of $L$ times $(-1)^{L-M}$, i.e. the only $M$-dependence is in the phase factor $(-1)^M$ \cite{Edmonds}.  If we combine all of the $L$ dependence of $f_L$ and the angular momentum coupling into a single pair amplitude, $g_{_L}(-1)^M$, then we can write the many-body paired state as
\begin{equation}
|\Omega\rangle=\frac{1}{\cal N}\exp\left\{\sum_{L,M}g_{_L}(-1)^M c^{\dagger}_{{L,M}\uparrow}c^{\dagger}_{{L,-M}\downarrow}\right\}|0\rangle .
\label{OmegaSph}
\end{equation}
Note that $M$ is half-integral so the pairing amplitude $g_L(-1)^M$ is antisymmetric under $M\rightarrow -M$ which is consistent with bosonic statistics and with the planar symmetry $\bk\rightarrow -\bk$.  The relationship of angular momentum to linear momentum is $L=|{\bf k}|R$, where $R$ is the radius of the sphere (Haldane in ref. \cite{GirvinPrange}).  Therefore, by analogy to (\ref{gk}), we expect that $g\simeq 1/L$ at small $L$, although we have no explicit proof of this statement.  Fortunately, the exact form of $g_{_L}$ is not important for the following. 

The lowest Landau level of the composite bosons has only two available states: $|1/2,\pm 1/2\rangle$.  By analogy to the torus, the spins ought to condense into these two lowest states with equal probability.  In real space this says that
\begin{equation}\begin{array}{l}
\langle c_{\downarrow}(\theta,\phi)\rangle =Y_{1/2,-1/2}(\theta,\phi)=-e^{i\phi}\sqrt{1-\cos{\theta}}\\
\langle c_{\uparrow}(\theta,\phi)\rangle =Y_{1/2,1/2}(\theta,\phi)=\sqrt{1+\cos{\theta}}
\end{array}
\label{BECondSph}
\end{equation}
The expression for the first few monopole harmonics may be found in ref. \cite{WuYang}, and a factor of $\sqrt{N/2}$ has been omitted from the right hand sides of (\ref{BECondSph}).  With the above condensate it is easy to calculate the expectation of the spin density $\langle S_i\rangle=\langle c^\dagger_\alpha \sigma^{\alpha\beta}_i c_\beta\rangle$, where $\sigma_i^{\alpha\beta}$ is the $i$'th Pauli matrix with spin indices $\alpha$ and $\beta$:
\begin{equation}\begin{array}{l}
\langle S_z\rangle=-\cos\theta \\\langle S_x\rangle =-\sin\theta\cos\phi\\
\langle S_y\rangle=-\sin\theta\sin\phi
\end{array}
\label{S}
\end{equation}
This is precisely an anti-Skyrmion spin ordering as the numerical data in Fig. \ref{fig:S_Correlator} shows.

In summary, the mean field theory of composite bosons is completely consistent with the analytical and numerical results of the previous subsection.  Furthermore, it seems that pure p-wave pairing of bosons in two dimensions should be viewed as a single particle condensate.
%%%%%%%%%%%%%%%%%%%%%%%%%%%%%%%%%%%%%%%%%%%%%%%%%%%%%%%%%%%%%%%
\subsection{Effective Field Theory Near the Transition} 
\label{CQFM} 
%%%%%%%%%%%%%%%%%%%%%%%%%%%%%%%%%%%%%%%%%%%%%%%%%%%%%%%%%%%%%%%
An effective continuum quantum ferromagnetic action (CQFM) for the polarized FQHE has been proposed recently by Read and Sachdev \cite{ReadSachdev}.  The idea is to write a sigma model for the local magnetization \cite{Auerbach}, $\hat{\bf n}$.  Their model includes terms up to momentum squared, the higher order terms being irrelevant in the renormalization group sense.  We modify this CQFM by stabilizing it in the helical region ($\rho_{s}<0$) by adding terms that are quartic in momentum.  At the end of this subsection, we will briefly discuss this theory on a lattice.

The CQFM Lagrangian density is
\begin{equation}
{\cal L}_{0}[\hat{\bf n}]=i{\bf{\cal A}(\hat{\bf n})}\cdot\partial_{\tau}\hat{\bf 
n}+\frac{\rho_{s}}{2}(\nabla\hat{\bf n})^{2}~,
\label{L0}
\end{equation}
where $\tau$ is complex time and the time derivative term is the Berry phase.  ${\bf{\cal A}}$ is the monopole vector potential such that $\nabla_{\hat{\bf n}}\times{\bf{\cal A}}=\hat{\bf n}$.  Although the model defined by ${\cal L}_{0}$ is unstable when the stiffness is negative, higher order derivative terms can stabilize this region.  The symmetry broken helical phase contains an $SO(2)\times SU(2)$ residual symmetry; $SO(2)$ for rotations of ${\bf Q}$ in the plane and $SU(2)$ for the spins.  There are three terms at leading non-trivial order that obey this requirement:
\begin{eqnarray}
\partial_an_{i}\partial_a n_{i}\partial_b n_{j}\partial_b n_{j}~,~~\partial_an_{i}\partial_b n_{i}\partial_a n_{j}\partial_b n_{j}~,~~\partial_a^2n_{i}\partial_b^2n_{i}~,\non
\end{eqnarray} 
where $a,\;b=1,\;2$ and $i,\;j=1,\;2,\;3$ and repeated indices are summed over.  A 
renormalization group analysis shows that mode elimination in ${\cal L}_{0}$ generates a combination of only the first two terms: ${\cal L}_{J}=J(2\partial_a n_{i}\partial_b n_{i}\partial_a n_{j}\partial_b n_{j}-\partial_a n_{i}\partial_a n_{i}\partial_b n_{j}\partial_b n_{j})$, which is associated with spin wave scattering \cite{ReadSachdev}.  The third term, ${\cal L}_{K}=K\left(\nabla^{2}\hat{\bf n}\right)^{2}$, is associated with a second-nearest neighbor interaction on a lattice.  If only these two terms are retained, then the total CQFHM Lagrangian is given by 
\begin{eqnarray}
{\cal L}[\hat{\bf n}]={\cal L}_{0}[\hat{\bf n}]+{\cal L}_{J}[\hat{\bf n}]+{\cal L}_{K}[\hat{\bf n}]~.
\nonumber
\end{eqnarray}
Although ${\cal L}_{J}$ cannot introduce any $k^{4}$ terms into the dispersion since it arises from mode elimination, it will emerge that ${\cal L}_{K}$ is sufficient to reproduce all of the long wavelength features found in the previous subsection.  We choose to keep ${\cal L}_{J}$ at this point for added generality.

Using spherical angles, $\hat{\bf n}$ can be parameterized by $\phi$, its direction in the plane and $\theta$, the fluctuation out of the plane.  Small deviations $(\vartheta,\varphi)$ from helical ordering with wavevector ${\bf Q}$ are given by $\theta=\vartheta$, $\phi={\bf Q}\cdot{\bf x}+\varphi$.  The fluctuations in the spherical angles should obey $|\nabla\varphi|,\;|\nabla\vartheta|\ll~|{\bf Q}|$. In these coordinates, \begin{math}\hat{{\bf n}}=(\cos\vartheta\cos({\bf Q}\cdot{\bf x}+\varphi), \cos\vartheta\sin({\bf Q}\cdot{\bf x}+\varphi),-\sin\vartheta)\end{math}, and the ferromagnetic phase is recovered when ${\bf Q}=0$.  The Berry phase reduces to the simple expression, $i\vartheta\partial_{\tau}\varphi$.

The mean field energy density of the helical state is given by
\begin{equation}
E_{0}(Q^{2})=\frac{1}{2}\left[\rho_{s}Q^{2}+(J+K)Q^{4}\right]~,
\label{E0}
\end{equation}
which has the desired shape when the stiffness is negative.  The spectrum of ${\cal L}$ can be found by including fluctuations to second order in $\varphi$ and $\vartheta$.  The Green's functions $\langle\varphi(-{\bf k},-\omega)\varphi({\bf k},\omega)\rangle$ and $\langle\vartheta(-{\bf k},-\omega)\vartheta({\bf k},\omega)\rangle$ both have poles at  
\begin{eqnarray}
\lefteqn{i\omega_{\bf k}=}\\
&&\left\{\begin{array}{ll}
                       \frac{1}{2}\left\{\left[4J({\bf Q}\cdot{\bf k})^{2}+K(k^{2}-Q^{2})^{2}\right]\left[4(J+K)({\bf Q}\cdot{\bf k})^{2}+Kk^{4}\right]\right\}^{1/2}~
                       & \rho_{s}\le 0\\ 
\left(\frac{1}{2}\rho_{s}k^{2}+\frac{1}{2}Kk^{4}\right)
                       & \rho_{s}\ge 0 ,
\end{array}\right.
\nonumber
\end{eqnarray}
where {\bf Q} is the momentum which minimizes $E_{0}$.  For certain ranges of the parameters $(J,K)$, the spectrum is always positive, and ${\cal L}$ describes a stable system.  For simplicity, consider the helical state with $J=0$ and $K$ positive. Then the spin wave energy of the CQFHM simplifies to
\begin{equation}
i\omega_{\bf k}=\frac{K}{2}|k^{2}-Q^{2}|\sqrt{4\left({\bf Q}\cdot{\bf 
k}\right)^{2}+k^{4}}
\label{omegak}      
\end{equation}
As required, $\omega_{\bf k}\sim k^{4}$ when $\rho_{s}$ zero.  The three-body spectrum obtained in the previous subsection can be reproduced by identifying $K/2$ with the coefficient of the $k^{4}$ term in (\ref{omegak}).  On the other hand, when $\rho_{s}$ is negative, we obtain the reasonable behavior, at small $k$:
\begin{equation}
i\omega_{\bf k}\sim\left\{ \begin{array}{ll}
			k^{2} & \mbox{if~~}{\bf k}\perp{\bf Q}\\
			k     & \mbox{if~~}{\bf k}\parallel{\bf Q}
		       \end{array}
		 \right.
\label{omegakLim}
\end{equation}
This may be expected because the spins are alligned ferromagnetically perpendicular to $\hat{\bf Q}$ but anti-ferromagnetically parallel to $\hat{\bf Q}$.
%%%%%%%%%%%%%%%%%%%%%%%%%%%%%%%%%%%%%%%%%%%%%%%%%%%%%%%%%%%%%%%%%
\subsection{Lattice Model}
\label{Lattice}
%%%%%%%%%%%%%%%%%%%%%%%%%%%%%%%%%%%%%%%%%%%%%%%%%%%%%%%%%%%%%%%%%
Before leaving the spin waves and moving on to the charged excitations, we briefly summarize a lattice model of spin-1/2 bosons that exhibits the helical transition.  We represent the spin sector by two types of hard-core bosons hopping on a lattice with a short-range, spin-dependent interaction.  

Defining $b_{i\sigma}$ to be the bosonic destruction operator on site $i$, consider the following Hamiltonian:
\begin{equation}
H_{lat}=-t\sum_{\langle ij\rangle,\sigma}b_{i\sigma}^{\dag}b_{j\sigma}-\mu\sum_{i}n_{i} + U\sum_{i}n_{i}(n_{i}-1)+J\sum_{\langle ij\rangle}\left({\bf S}_{i}\cdot{\bf S}_{j}-\frac{1}{4}n_{i}n_{j}\right)~,
\label{Hlat}
\end{equation}
where $n_{i}$ is the total number of bosons per site, and $\langle ij\rangle$ denotes the sum over nearest neighbors.  The $U$-term is required since we are dealing with hard-core bosons; the singular gauge transformation which mapped the fermions to bosons maintains the repulsion at the same site.  The final, antiferromagnetic, term is very much like the fermion $t-J$ interaction \cite{Auerbach}.  In the continuum limit, it reduces to the rotationally invariant interaction, $|\epsilon^{\sigma\tau}\hat{\psi}_{\sigma}\nabla\hat{\psi}_{\tau}|^{2}$.  We will not discuss $H_{lat}$ further, save to point out its mean-field features.  

This two-component model is similar to the {\em one}-component lattice boson model considered earlier by M. Fisher et al. \cite{FWGF}.  In the superfluid regime, which is characterized by large $t/U$ or special values of $\mu$ where $H_{lat}$ has particle-hole symmetry, one is free to consider bose condensation of the particle fields.  If one also makes the self-consistent restriction $t/U<Jn/2U$, $n$ being the average number of bosons per site, then the free energy of the helical phase is lower than that of the ferromagnetic phase.  As in the continuum, this winding is described by the condensation:
\begin{eqnarray}
\langle b_{i\uparrow}^{\dag}\rangle&=&\sqrt{\frac{N_{\uparrow}}{N_{L}}}e^{i{\bf Q}\cdot{\bf x}_{i}}\\ 
\langle b_{i\downarrow}^{\dag}\rangle&=&\sqrt{\frac{N_{\downarrow}}{N_{L}}}e^{-i{\bf Q}\cdot{\bf x}_{i}}~,
\nonumber
\end{eqnarray}  
where $N_{\uparrow}=N_{\downarrow}=N/2$ and $N_{L}$ is the number of lattice sites.  The optimal condensate wavevector lies along the diagonal of the lattice: ${\bf Q}=Q\hat{\bf x}+Q\hat{\bf y}$ with $Q$ determined by the solution of $\cos Q=2t/Jn$ (in units of the inverse lattice constant).  

As the hopping decreases, provided that one is in a given region of $\mu/U$, the ground state crosses over into a Mott insulator and it is no longer valid to argue based on Bose condensation.  However, at points of particle-hole symmetry, the superfluid persists down to infinitesimal hopping.  For the Hall liquid, the insulating phase is most relevant since the charge excitations must be gapped.  It would be interesting to map out in detail the phase diagram of this magnetic superfluid to insulator transition.  
%%%%%%%%%%%%%%%%%%%%%%%%%%%%%%%%%%%%%%%%%%%%%%%%%%%%%%%%%%%%%%%%%
\section{The Haffnian: d-wave Pairing}
\label{Haffnian}
%%%%%%%%%%%%%%%%%%%%%%%%%%%%%%%%%%%%%%%%%%%%%%%%%%%%%%%%%%%%%%%%%
The previous section has analyzed the spin sector of the permanent in some detail.  Its spin-wave dispersion was shown to be soft, allowing a magnetic transition.  We now want to ask the question whether there is a ground state wavefunction whose density sector exhibits an analogous behavior.  To this end, we introduce a d-wave paired wavefunction, or ``Haffnian'', describing hard-core, spinless {\it bosons\/} at filling factor $\nu=1/2$.  In the FQHE, it is a d-wave paired state of composite bosons.  We will then argue that the Haffnian is compressible and sits on the phase boundary between incompressibility and non-uniform charge density order.  At present we have no effective field theory that captures this behavior. 

Although d-wave paired states have been proposed in a different contex by Wen and Wu \cite{Hf}, their Haffnian wavefunctions describe {\it in}compressible states of {\it fermions}.  As our proposal is somewhat different, in the following subsection we present the relevant constructions in some detail, mainly along the lines used to investigate non-abelian statistics \cite{RR}.      
%%%%%%%%%%%%%%%%%%%%%%%%%%%%%%%%%%%%%%%%%%%%%%%%%%%%%%%%%%%%%%%%%
\subsection{Analytic Structure of the Haffnian}
\label{HaffnianAnalytic}
%%%%%%%%%%%%%%%%%%%%%%%%%%%%%%%%%%%%%%%%%%%%%%%%%%%%%%%%%%%%%%%%%
The Haffnian builds in pairing into the Laughlin state, much like the Pfaffian of Moore and Read \cite{MR}:
\begin{equation}
\Psi_{Hf}=\sum_P\frac{1}{(z_{P(1)}-z_{P(2)})^{2}\cdots (z_{P(N-1)}-z_{P(N)})^{2}}\prod_{i<j}(z_{i}-z_{j})^{2}~.
\label{PsiHf}
\end{equation}
This describes an even number, $N$, of spinless bosons at half-filling and flux ${N_{\phi}=2(N-1)-2}$ (on the sphere).  The prefactor is known as a Haffnian in linear algebra and is also the permanent of the ${N\times N}$ ($N>2$) matrix $M_{ij}=1/(z_{i}-z_{j})^{2}~(i\neq j,~M_{ii}=0)$.  The Haffnian, Pfaffian and determinant are related by several identities, which may be found in, for instance, Greiter et al. \cite{Greiter}    

To construct the parent Hamiltonian for $\Psi_{Hf}$, it is convenient to work on the sphere, where each particle has orbital angular momentum $N_{\phi}/2$.  Using the same notation as in the permanent (\ref{HPermSph}), $H_{Hf}$ is a sum of three-body projection operators:
\begin{equation}
H_{Hf}=\sum_{i\neq j\neq k}V_{0}P_{ijk}(3N_{\phi}/2)+V_{2}P_{ijk}(3N_{\phi}/2-2)+V_{3}P_{ijk}(3N_{\phi}/2-3
)~.
\label{Ham}
\end{equation}
$\Psi_{Hf}$ is the unique zero-energy eigenstate of $H_{Hf}$ at $N_\phi$ flux.  The proof of this statement proceeds by showing that the maximum angular momentum of any triplet in $\Psi_{Hf}$ is $3N_\phi/2-4$; the details are in Section \ref{sec:HaffH}.  Note that projection onto angular momentum $3N_{\phi}/2-1$ is absent.  This is a consequence of the symmetries of Clebsch-Gordan coupling; three spinless bosons of angular momentum $L$ cannot be in a total angular momentum state of $3L-1$. 

Alternatively, $\Psi_{Hf}$ can be rewritten explicitly as a paired state of composite bosons.  By analogy with the permanent, the order parameter (on the plane or torus) is an eigenstate of angular momentum with eigenvalue $l=-2$, i.e. $\Delta\simeq\hat{\Delta}(k_x-ik_y)^2$ to leading order in ${\bf k}$.  Likewise the many-body state is a BCS wavefunction of d-wave bosons in two dimensions and the asymptotic behavior of the pair state reproduces the $1/z^2$ nature of the Haffnian prefactor (\ref{PsiHf}).

%%%%%%%%%%%%%%%%%%%%%%%%%%%%%%%%%%%%%%%%%%%%%%%%%%%%%%%%%%%%%%%%%
\subsection{The Haffnian Hamiltonian on the Sphere}
\label{sec:HaffH}
%%%%%%%%%%%%%%%%%%%%%%%%%%%%%%%%%%%%%%%%%%%%%%%%%%%%%%%%%%%%%%%%%

To see that $\Psi_{Hf}$ does not contain any triplets of total angular momentum greater that $3N_{\phi}/2-4$, we generalize Haldane's original argument for two-body interactions \cite{Haldane83}.  This argument is easily applicable to $n$-body interactions.  The factors $(z_{i}-z_{j})$ on the plane correspond to $(u_{i}v_{j}-u_{j}v_{i})$ on the sphere, with $(u_{i},v_{i})$ being the spinor coordinates of $z_{i}$.  The total angular momentum of a triplet $(ijk)$ on the sphere is {\em one-half\/} of the maximum of the coherent state operator ${\bf S}_{ijk}=\hat{\Omega}\cdot ({\bf L}_{i}+{\bf L}_{j}+{\bf L}_{k})$, where $\hat{\Omega}(\phi ,\theta)$ is any direction on the sphere and ${\bf L}_{i}$ is the angular momentum of the $i$'th particle.  $({\bf L}_{i}+{\bf L}_{j}+{\bf L}_{k})$ commutes with any factor involving only particles $i,j,k$; i.e. all factors $(u_{a}v_{b}-u_{b}v_{a})$ with $a,b\in\{i,j,k\}$.  Translating this to the plane, we can find the maximum of ${\bf S}_{ijk}$ acting on $\Psi_{Hf}$ almost by inspection.  When all factors involving {\it exactly one of\/} $z_{i},z_{j}$ or $z_{k}$ are multiplied out, the result will be a polynomial with terms of the form $z_{i}^{A}z_{j}^{B}z_{k}^{C}$.  The maximum value of $1/2 (A+B+C)$ is exactly the maximum value of the total angular momentum of the triplet $(ijk)$.  In particular, this maximum is $3N_{\phi}/2-4$ for $\Psi_{Hf}$, so it is certainly a zero energy eigenstate of $H_{Hf}$.

It must still be shown that $\Psi_{Hf}$ is the unique ground state.  To this end, we will use the method in Appendix A of Milovanovi\'{c} and Read \cite{MilovanovicRead}.  Consider the behavior of the Haffnian prefactor in (\ref{PsiHf}) as three particles $(ijk)$ approach each other, the other particles remaining far away from the three.  The Laurent series must contain terms of the form $(z_{i}-z_{j})^{q_{ij}}(z_{j}-z_{k})^{q_{jk}}(z_{k}-z_{i})^{q_{ki}}$ with $q_{ab}$ positive or negative integers.  By continuity, the total function must contain this Laurent factor for any position of the particles.  In order for $\Psi_{Hf}$ to be analytic, as it must be in the LLL, each $q$ must not be smaller than $-2$, or $Q\equiv q_{ij}+q_{jk}+q_{ki}\ge -6$.  In particular, $\Psi_{Hf}$ is annihilated by $P_{ijk}(3N_{\phi}/2)$ if the inequality is strict, $Q>-6$.  Another way to see this is on the plane: $P_{ijk}(3N_{\phi}/2)$ is the projection onto the closest approach of a triplet---all three particles are clumped at the north pole, taking the maximum $L_z$ value---which takes the form $\delta^{(2)} (z_i-z_j)\delta^{(2)} (z_j-z_k)$ on the plane.  Thus, if $Q=-6$, the delta function interaction does not annihilate $\Psi_{Hf}$.  Reducing the total angular momentum (on the sphere) by one corresponds to restricting further the particles' closest approach, or increasing the lower limit of $Q$ by one.  In this way, we obtain the requirement $Q>-3$ in order for $P_{ijk}(L)$ to annihilate the ground state whenever $L>3N_{\phi}/2-4$.  The extreme case $Q=-2$ corresponds to the densest eigenstate.  There are four possible such factors: 
\begin{eqnarray}
\frac{1}{(z_{i}-z_{j})^{2}},~ \frac{1}{(z_{i}-z_{j})(z_{k}-z_{j})},~ 
\frac{(z_{i}-z_{j})^{2}}{(z_{j}-z_{k})^{2}(z_{k}-z_{i})^{2}},~ 
\frac{(z_{i}-z_{j})}{(z_{j}-z_{k})^{2}(z_{k}-z_{i})}~.
\nonumber
\end{eqnarray}
Symmetrizing these factors with respect to $(ijk)$ leaves only terms like the first one.  Therefore $H_{Hf}$ automatically requires a pairing structure of its ground state that is given by the Haffnian.

Other zero-energy eigenstates are obtained by multplying in factors symmetric in the particle coordinates, being formally allowed since they can only increase $Q$.  These states describe quasiholes and edge excitations and are less dense than the ground state since they contain added flux.  They are enumerated explicity in the Section \ref{HaffnianQuasiholes}; here we verify that they are indeed zero-energy eigenstates of $H_{Hf}$.  Without loss of generality, pick a definite triplet, $(ijk)=(123)$, for convenience.  There are several cases to check, corresponding to the possible terms appearing in some quasihole state (\ref{DegStates}): (i) the pair $(z_1,z_2)$ is broken, (ii) the pair involving $z_3$ and another particle, say $z_4$, is broken, (iii) both pairs $(z_1,z_2)$ and $(z_3,z_4)$ are broken, and (iv) neither pair is broken.  For illustration, we check case (ii), the others being done similarly.  Applying the generalization of Haldane's argument, the maximum degree of the triplet $(123)$ is $2(N-3)+2(N-3)+2(N-3)+2n+(n-2)$.  The first three contributions come from the terms $(z_i-z_a)$, where $i=1,2,3$ and $a\neq 1,2,3$, the fourth term is due to the quasihole operator $\Phi (z_1,z_2,z_5,\ldots)$, and the last takes into account the maximum orbital quantum number of the unpaired boson, $z_{3}^{n-2}$.  Using the flux condition $N_{\phi}=2(N-1)-2+n$, leads to one-half the maximum degree (or maximum total orbital angular momentum of a triplet) being \mbox{$3N_{\phi}/2-4$}, which is consistent with requirement that it be less than \mbox{$3N_{\phi}/2-3$}.  Note that it has tacitly been assumed that $n\geq 2$; if $n=1$ then the unpaired boson is in the zeroth orbital and half of the maximum degree is $3N_{\phi}/2-3-n/2$.  The other three cases can be checked straightforwardly by such counting, proving that the quasihole states (\ref{DegStates}) are annihilated by $H_{Hf}$.

%%%%%%%%%%%%%%%%%%%%%%%%%%%%%%%%%%%%%%%%%%%%%%%%%%%%%%%%%%%%%
\subsection{Zero Energy Eigenstates}
\label{HaffnianQuasiholes}
%%%%%%%%%%%%%%%%%%%%%%%%%%%%%%%%%%%%%%%%%%%%%%%%%%%%%%%%%%%%%
In addition to the Haffnian, there are other zero energy eigenstates of $H_{Hf}$ generated by adding $n=1,2,\ldots$\,flux quanta.  The structure of these excitations is more complex than that of the familiar Laughlin quasiholes, having an infinite degeneracy in the thermodynamic limit.  In previous work \cite{RR}, this degeneracy has been suggested to provide the necessary manifold of states for nonabelian statistics, which we discuss below.  Proper construction of the zero energy states is useful for understanding the numerical spectrum of $H_{Hf}$, so we will go through them in some detail.  

The quasihole in the paired state, like in the Laughlin state, is generated by one flux, but it is built of {\it two\/} vortices at $w_1$ and $w_2$ each carrying {\it one-half\/} flux quantum.  Let $n$ be the number of flux added to the Haffnian, i.e. $N_{\phi}=2(N-1)-2+n$, with $w_1,\ldots,w_{2n}$ being the positions of the vortices.  Denoting $B/2$ as the number of broken pairs in the Haffnian and $\{m_{1},m_{2},\ldots,m_{B}\}$ as the quantum numbers of the orbitals into which the upaired bosons are placed, the explicit form for the manifold of zero-energy eigenstates is 
\begin{eqnarray}
\lefteqn{\Psi_{m_1,m_2,\ldots,m_B}(z_1,z_2,\ldots,z_N;w_1,w_2,\ldots,w_{2n}) =} 
\\
&&\sum_{\sigma\in S_N}\prod_{k=1}^{B}z_{\sigma (k)}^{m_k}
\prod_{l=1}^{(N-B)/2}\frac{\Phi (z_{\sigma (B+2l-1)},z_{\sigma 
(B+2l)};w_1,\ldots,w_{2n})}{(z_{\sigma (B+2l-1)}-z_{\sigma (B+2l)})^2}
\prod_{i<j}(z_i -z_j)^2~.\non
\label{DegStates}
\end{eqnarray}
$\Phi$ is the quasihole operator given by:
\begin{equation}
\Phi (z_1,z_2;w_1,\ldots,w_{2n})=\sum_{\tau\in S_{2n}}\prod_{r=1}^{n}
 (z_1 -w_{\tau (2r-1)})(z_2 -w_{\tau (2r)})~.
\label{qhole}
\end{equation}
It can be verified that the  $\Psi_{m_1,m_2,\ldots,m_B}(z_1,z_2,\ldots,z_N;w_1,w_2,\ldots,w_{2n})$ are in fact zero-energy eigenstates of $H_{Hf}$ (Section \ref{HaffnianAnalytic}). If no pairs are broken ($B=0$), $\Phi$ builds in two vortices---each within half of the bosons---for each of the $n$ flux.  Hence the interpretation that the vortices carry charge $1/4$, or $1/2q$ for more general filling factors.  The unpaired bosons are labeled by the orbital quantum numbers $\{m_1,\ldots,m_B\}$, which must satisfy the condition $0\le m_1\le\ldots\le m_B\le n-2$.  The upper limit follows from the constraint on $N_{\phi}$ and the ordering is simply to avoid overcounting upon symmetrization.  This is equivalent to putting $B$ bosons into $n-1$ orbitals, for which the multiplicity is \begin{equation}
\left(\begin{array}{c} B+n-2\\ B\end{array}\right)~.
\label{BosonDeg}
\end{equation}
Note that for $n=2$ all of the bosons coming from broken pairs are in the lowest orbital, $z^0$, and there are no unbroken pairs at $n=1$.  

There is an additional degeneracy coming from the positions of the quasiholes themselves, which is calculated by expanding $\Phi$ in the $w$'s using the elementary symmetric polynomials $e_{m}(w)=\sum_{i_1<\ldots <i_m}w_{i_1}\ldots w_{i_m}$. The $e_m$ have the property that, for $m=0,\ldots ,j$, they form a basis for the algebra of all symmetric polynomials in $j$ variables.  Thus, zero-energy states may also be obtained as linear combinations of the $e_m$.  When $\Phi$ is expanded in this way we obtain all the symmetric polynomials in $w_1,\ldots ,w_{2n}$ in which the degree of any $w$ is no greater than $(N-B)/2$.  The total number of linearly independent states, for a fixed $B$ and a fixed set of $m_i$'s, is at most the total number of linearly independent symmetric functions of $w$ in the expansion of $\Phi$, which establishes the upper bound on the positional degeneracy of the quasiholes.  That this is also the correct degeneracy, without overcounting, is proven elsewhere \cite{RR}.  We can now write down this number by regarding the vortices as some kind of bosonic particle, interpreting the $e_{m}(w)$ as the states for $2n$ bosons occupying the $(N-B)/2+1$ orbitals $\{1,\ldots,w^{(N-B)/2}\}$ (recall that $\Phi$ appears $(N-B)/2$ times due to the product $\prod_{l=1}^{(N-B)/2}$ in (\ref{DegStates})):
\begin{equation}
\left(\begin{array}{c} (N-B)/2+2n\\ 2n\end{array}\right)~.
\label{VortDeg}
\end{equation}
Throughout this construction it has been tacitly assumed that $N$ is even, so $B$ is necessarily even as well.  The construction for $N$ odd proceeds with only slight modification, but the counting in (\ref{BosonDeg}) and in (\ref{VortDeg}) does not change, since $B$ has the same parity as $N$.  In either case, the total number of linearly independent quasihole states is obtained by multiplying the two combinatorial factors and summing over the allowed values of $B$:
\begin{equation}
\sum_{B,(-1)^{B}=(-1)^{N}} \left(\begin{array}{c} B+n-2\\ B\end{array}\right) 
\left(\begin{array}{c} (N-B)/2+2n\\ 2n\end{array}\right)~.
\label{TotDeg}
\end{equation}
To complete this description, one should check that all states in (\ref{DegStates}) exhaust all zero-energy eigenstates at fixed $N$ and $n$ and that they are linearly independent.  Since this is somewhat involved and is discussed at length elsewhere \cite{RR}, we will omit it here.

As a special case, notice that for two added flux ($n=2$) the Laughlin state, for bosons at $\nu=1/2$, is recovered when all pairs are broken ($B=N$).  Therefore, up to two flux quanta, the Haffnian and the Laughlin states are degenerate eigenstates of $H_{Hf}$.  It is convenient to adopt the composite boson interpretation of the quasihole positional degeneracy (\ref{VortDeg}).  Rewriting this combinatorial factor as 
\begin{eqnarray}
\left(\begin{array}{c} (N-B)/2+2n\\ (N-B)/2\end{array}\right)~,
\label{L2Deg}
\end{eqnarray}
affords an interpretation as $(N-B)/2$ composite bosons in $2n+1$ orbitals.  On the sphere at $n=2$ this is the correct degeneracy for an $L=2$ angular momentum multiplet, independently of both $N$ and $B$.  The unpaired bosons are all forced into the lowest orbital $m=0$, which is manifested by the degeneracy factor (\ref{BosonDeg}) reducing to unity.  Thus, the manifold of degenerate states is composed of $N/2$ states (one for each of $B=0,2,\ldots ,N-2$) carrying angular momentum $L=2$ and one Laughlin state (for $B=N$) carrying $L=0$, which is a rotationally invariant.  The main point of the expression (\ref{L2Deg}) is that the zero energy eigenstates of $H_{Hf}$ contain {\it both} d-wave pairs {\it and} unpaired composite bosons in the $m=0$ orbital   

The degeneracy at $L=2$ is a feature that is reminescent of the permanent.  In that case, an analogous argument at $n=1$ leads to a Laughlin state at $\nu=1$ and a set of $N/2$ states carrying $L=1$.  The $(N-B)/2$ paired bosons were interpreted as spin waves \cite{RR}, which we have established to condense into helical order (\ref{BEChelical}).  Spin wave excitations are, of course, gapless, but one may expect that the $L=2$ modes in the Haffnian are massive, being density excitations.  However, we argue that this is not the case and $H_{Hf}$ is compressible, indicating that it sits right on the transition from an incompressible state to one with non-uniform charge density order.
%%%%%%%%%%%%%%%%%%%%%%%%%%%%%%%%%%%%%%%%%%%%%%%%%%%%%%%%%%%%%%%%%%%
\subsection{d-wave Pairing of Spinless Bosons}
\label{HaffnianPairing}
%%%%%%%%%%%%%%%%%%%%%%%%%%%%%%%%%%%%%%%%%%%%%%%%%%%%%%%%%%%%%%%%%%%
In the previous subsection we showed that the lowest single particle orbital may be macroscopically occupied by breaking pairs (\ref{L2Deg}).  A proper description of this system must therefore include a single particle condensate at the outset.  As soon as there is such a condensate there is also isotropic scattering out of the superfluid, which gives rise to an s-wave pair amplitude.  

On the torus or plane, the full Hamiltonian is given by
\be
K=\sum_\bk\left(\frac{k^2}{2m^\ast}-\mu\right)c^\dagger_\bk c_\bk+\frac{1}{2}\sum_{\bk,\bk^\prime,\bq}V(\bq) c^\dagger_{\bk+\frac{1}{2}\bq} c^\dagger_{\bk^\prime-\frac{1}{2}\bq} c_{\bk-\frac{1}{2}\bq} c_{\bk^\prime+\frac{1}{2}\bq}
\label{FullH}
\ee
We assume that the lowest state $\bk=0$ is macroscopically occupied with an amplitude $\Phi=\langle c_0\rangle$ and expand around this condensate: $c_\bk\rightarrow \Phi+\widetilde{c}_\bk$.  In addition there is the possibility of s- and d-wave pairing of the anomalous correlators $\langle\widetilde{c}_\bk\widetilde{c}_{-\bk}\rangle$.  The resulting effective Hamiltonian is
\begin{equation}
K_{eff}={\sum_\bk}'\left[\xi_\bk c^\dagger_\bk c_\bk +\frac{1}{2}\left(\Delta_\bk c^\dagger_\bk c^\dagger_{-\bk}+\Delta^\ast_\bk c_{-\bk}c_\bk\right)\right]~,
\label{KeffHf}
\end{equation}
where the prime on the summation indicates that the $\bk=0$ term is to be omitted when $\Phi>0$; otherwise the summation is over all wavevectors. The chemical potential is shifted by the $\langle c^\dagger_0 c_0\rangle$ contribution from expanding the interaction so that the single particle energy is now $\xi_{\bf k}\simeq{\bf k}^2/2m^\ast-\mu+2V(0)|\Phi|^2$.  For consistency, it is necessary that $V(0)$ is positive, which is a standard assumption in interacting Bose systems.   It is also common to assume that the exchange contributions do not modify $m^\ast$ significantly.  

The pair order parameter must satisfy the self-consistent gap equation in the presence of a condensate:
\be
\Delta_\bk=\Phi^2 V(\bk)-\frac{1}{2}{\sum_{\bk^\prime}}'V(\bk-\bk^\prime)\frac{\Delta_{\bk^\prime}}{E_{\bk^\prime}}
\label{GapWithPhi}
\ee
To make sense of this complicated integral equation, we shall assume that $\Delta_\bk$ is dominated by the (s+d)-wave symmetry and, furthermore, that its leading order behavior is
\begin{equation}
\Delta_\bk=\Delta^s_\bk+\Delta^d_\bk\mbox{~~~with~~~}\left\{\begin{array}{l}
\Delta^s_\bk\simeq\hat{\Delta}^s\\
\Delta^d_\bk\simeq\hat{\Delta}^d(k_x-ik_y)^2\end{array}\right.
\label{DeltaHf}
\end{equation}
where $\hat{\Delta}^{s,d}$ are constants.  Recall that the interaction can be expanded into angular momentum eigenstates (\ref{Vl}) and that each channel has the asymptotic behavior $V_l(k,k^\prime)\simeq k^l$.  In other words, it is implicit in the above form (\ref{DeltaHf}) that the long distance property of each order parameter is dictated by its respective angular momentum channel.  

The standard procedure is to diagonalize $K_{eff}$.  This is the familiar Bogoliubov transformation that replaces the original particles and holes by quasiparticles $\alpha_{\bf k}$:
\begin{equation}
\alpha_\bk=u_{\bf k}^\ast c_{\bf k}+v_{\bf k} c_{-{\bf k}}^\dagger
\label{Bog}
\end{equation}
Commutation relations are preserved if $|u_{\bf k}|^2-|v_{\bf k}|^2=1$ and symmetry requires that $u_{\bk}=u_{-\bk}$ and $v_{\bk}=v_{-\bk}$.  The following solutions for $u_\bk$ and $v_\bk$ put $K_{eff}$ into the form $K_{eff}=\sum_\bk E_\bk\alpha_\bk^\dagger\alpha_\bk$~:
\begin{eqnarray}
v_\bk^2 &=& \frac{1}{2}\left(\frac{\xi_\bk}{E_\bk}-1\right)\frac{\Delta_\bk}{|\Delta_\bk|} \non \\
u_\bk^2 &=& \frac{1}{2}\left(\frac{\xi_\bk}{E_\bk}+1\right) \label{uv}\\
E_\bk^2&=&\xi^2-|\Delta_\bk|^2~.
\label{uvEk}
\end{eqnarray}
There is a kind of gauge freedom in that both $u_\bk$ and $v_\bk$ can be multiplied by a $\bk-$dependent phase factor, without changing the physics.  We adopt the convention that $u_\bk$ is real and positive, while $v_\bk$ carries the possible d-wave phase factor.

If both s- and d-wave amplitudes are nonvanishing, an interesting feature emerges immediately.  The quasiparticle spectrum contains an anisotropic piece coming from the cross terms in $|\Delta_{\bf k}|^2$,
\begin{equation}
E_{\bf k}^2=\xi^2_{\bf k}-|\hat{\Delta}^s+\hat{\Delta}^d k^2e^{-2i\phi}|^2~,
\label{Ek2}
\end{equation}
where $k$ is the magnitude of and $\phi$ is the polar angle of ${\bf k}$.  Choosing the constants $\hat{\Delta}^{s,d}$ to be real for the moment, the anisotropy is 
\be
-2\hat{\Delta}^s\hat{\Delta}^d k^2 \cos2\phi
\label{inhomog}
\ee
which has two minima at $\phi=0,~\pi$.  There is a relative $U(1)$ phase degree of freedom between the pairing amplitudes, which is equivalent to an $SO(2)$ rotation of the minima.  In other words, the transformation $\hat{\Delta}^s\rightarrow e^{2i\alpha}\hat{\Delta}^s$ is the same as $\phi\rightarrow\phi +\alpha$.  It is interesting that this anisotropy arises not from an attraction in $V(\bk)$, which is isotropic, but rather from the coexistence of two distinct pairing symmetries in momentum space. 

A unique ground state wavefunction exists for each solution of the gap equations.  Its generic form is like that of the permanent (\ref{Omega}) but with an allowance for the single particle condensate
\begin{equation}
|\Omega\rangle=\frac{1}{\cal N}\exp\left\{\Phi c^\dagger_0+{\sum_\bk}'g_\bk c^\dagger_\bk c^\dagger_{-\bk}\right\}|0\rangle~. 
\label{OmegaHf}
\end{equation}
The normalization is the same as in Eq. (\ref{Norm}) but with an extra factor of $\exp(|\Phi|^2).$  The reciprocal space pair wavefunction, $g_\bk$, can be expressed in terms of the Bogoliubov parameters $u_\bk$ and $v_\bk$ as
\be
g_\bk=\frac{v_\bk}{u_\bk}
\label{guv}
\ee
The generic form of the $N$-body paired wavefunction has already been discussed in Eq. (\ref{PsiPairR}) and the Haffnian exists only when $\Phi=0$ since it is a pure paired state.  Furthermore, the asymptotic behavior of the real space pair wavefunction $g({\bf r})\simeq 1/z^2$ is obtained when
\begin{equation}
g_\bk=\lambda\frac{k_x^2+k_y^2}{(k_x+ik_y)^2}~.
\label{gkHf}
\end{equation}
The $N$-body state of composite bosons constructed in this way is exactly the Haffnian prefactor of Eq. (\ref{PsiHf}).  And, the same remarks following Eq. (\ref{gr}), concerning the validity of this mean field approximation, apply here. 

Unfortunately, unlike the permanent case, the occupation number $\langle n_{\bf k}\rangle=\langle c^\dagger_{\bf k}c_{\bf k}\rangle$ does not contain any new information since it is constant:
\begin{eqnarray}
\langle n_\bk \rangle=\frac{|\lambda|^2}{1-|\lambda|^2}~.
\nonumber
\end{eqnarray}
However, the required form of $g_{\bf k}$ does constrain the parameters at which the Haffnian can exist.  At small $\bk$, the asymptotic behavior of $g_\bk$ (\ref{gkHf}) is consistent with $\hat{\Delta}^s=0$ only if $u_\bk$ and $v_\bk$ have the same asymptotic behavior.

To understand where this point lies we must understand the phase diagram as derived from the self-consistency condition (\ref{GapWithPhi}).  It is simplified by assuming the asymptotic expansion for $\Delta_\bk$ in Eq. (\ref{DeltaHf}) and by expanding $V({\bf k}-{\bf k}^\prime)$ into angular momentum eigenstates as before (\ref{Vexp}).  This decomposition separates the BCS gap equation (\ref{GapWithPhi}) into two coupled integral equations,
\begin{eqnarray}
\hat{\Delta}^s&=&\Phi^2 V(\bk)-\frac{1}{2}{\sum_{\bk^\prime}}'V_0(k,k^\prime)\frac{\hat{\Delta}^s+\hat{\Delta}^d k^{\prime~2}\cos2\phi^\prime}{E_{{\bk}^\prime}}
\label{Gap2a}\\
\hat{\Delta}^d k^2&=&-\frac{1}{2}{\sum_{\bk^\prime}}'V_{-2}(k,k^\prime)\frac{\hat{\Delta}^s\cos2\phi^\prime+\hat{\Delta}^d k^{\prime~2}}{E_{{\bk}^\prime}}~.
\label{Gap2b}
\end{eqnarray}
The assumption that the s-wave amplitude is constant then implies that $\bk=0$ in the first equation to leading order.  The coefficient $V_0(k,k^\prime)$  reduces to $V(\bk^\prime)$ at $\bk=0$, and the leading order term of $V_{-2}(k,k^\prime)$ is proportional to $k^2$ so the second equation is consistent (c.f. Eq. (\ref{Vl})).  Further simplification is possible by choosing a particular gauge for the phases of the condensates.  A convenient choice is to pick a real $\hat{\Delta}^d$ and to absorb the phase of $\hat{\Delta}^s$ into $\phi^\prime$; this only rotates the inhomogeneity of the spectrum (\ref{inhomog}).  Since the phase of $\Phi$ is locked to that of $\hat{\Delta}^s$, {\it all order parameters can be chosen to be real and positive.}  Finally, since the inhomogeneity is even under parity, $\phi^\prime\rightarrow -\phi^\prime$, the odd terms $\sin2\phi^\prime$ in the numerator cancel and only $\cos2\phi^\prime$ appears.

The crudest criterion for the solutions of the gap equations is whether or not $\Phi$ vanishes.  Consider first the situation when $\Phi=0$ and the fluid consists of pure pairs.  In principle both s- and d-wave pairing can coexist since the gap equations allow this generally.  However, if we make the reasonable assumption that $V(\bk)$ is positive everywhere then only $\hat{\Delta}^d$ is allowed.  This may be seen from a closer inspection of the $\phi^\prime$-dependent part of the integrand in (\ref{Gap2a}).  Let us split it into two disjoint pieces characterized by $\cos2\phi^\prime$ negative or positive.  The corresponding $E_{\bk^\prime}$ is always larger in the former case because the inhomogeneity (\ref{inhomog}) has a negative sign.  Therefore, the total contribution of the $\hat{\Delta}^d$ term to the right hand side is always negative.  That is to say
\bea
\int_0^{2\pi}d\phi^\prime~V(\bk^\prime)\frac{\cos2\phi^\prime}{E_{\bk^\prime}}\geq0~.
\label{positivity}
\eea
Since $\hat{\Delta}^s$ is also positive in our gauge there is no consistent solution of (\ref{Gap2a}) when $\Phi=0$.  The only possibility is a pure d-wave gap that satisfies (\ref{Gap2b}) with $\hat{\Delta}^s=0$.  In this phase $\mu<0$ and $2E_0=2|\mu|$ is the gap to breaking a pair (see also \cite{Nozieres}).  Thus, when there is no single particle condensate the system is specified by the following gap equation and quasiparticle spectrum:
\bea
&\hat{\Delta}^d k^2&=-\frac{1}{2}\sum_{\bk^\prime}V_{-2}(k,k^\prime)\frac{\hat{\Delta}^d k^{\prime~2}}{E_{\bk^\prime}}\non\\
&E_\bk^2&=\mu^2+\frac{|\mu|}{m^\ast}k^2+\left[\left(\frac{1}{2m^\ast}\right)^2-\left(\hat{\Delta}^d\right)^2\right]k^4
\label{Phi=0}
\eea
$\hat{\Delta}^d$ is real in this gauge.

On the other hand, when $\Phi>0$ there is a new set of considerations, which require $V_0(k,k^\prime)=0$ for consistency.  Firstly, in order for $K_{eff}$ to be a stable approximation to the full Hamiltonian, $\langle c_0\rangle$ must be a minimum of the free energy.  The resultant constraint, sometimes known as the Gross-Pitaevsky equation, pins the chemical potential to the single particle condensate by $\mu=V(0)\Phi^2$ (in the gauge where $\Phi$ is real).  Furthermore, the spectrum is dominated by the gapless Goldstone phonon mode at low $\bk$, requiring $E_0=\xi_0^2-|\Delta_0|^2$, or 
\be
-\mu+2V(0)\Phi^2=\hat{\Delta}^s~.
\label{mu}
\ee
In conjunction with the pinning condition $\mu=V(0)\Phi^2$, this implies that $\hat{\Delta}^s=V(0)\Phi^2$.  But, using the positivity condition in Eq. (\ref{positivity}), this is consistent with the gap equation (\ref{Gap2a}) only if we reinstate the $\bk$-dependence of $\hat{\Delta}^s$ or if $V_0(k,k^\prime)$ vanishes.  We shall assume the latter as it is much more tractable.  Ultimately, this condition is traced back to the asymptotic behavior of the gap function (\ref{DeltaHf}), which is somewhat restrictive.  However, we now have a tractable and fully consistent system.  Summarizing these results for $\Phi>0$, 
\bea
\hat{\Delta}^s&=&V(0)\Phi^2=\mu \non\\
\hat{\Delta}^d k^2&=&-\frac{1}{2}{\sum_{\bk^\prime}}'V_{-2}(k,k^\prime)\frac{\hat{\Delta}^s\cos2\phi^\prime+\hat{\Delta}^d k^{\prime~2}}{E_{{\bk}^\prime}}\label{Phi>0}\\
E_\bk^2&=&2\mu\left(\frac{1}{2m^\ast}-\hat{\Delta}^d\cos2(\phi+\alpha)\right)k^2+\left[\left(\frac{1}{2m^\ast}\right)^2-\left(\hat{\Delta}^d\right)^2\right]k^4~.\non
\eea
We have used the gauge in which all order parameters are real, which shifts $\phi$ by $\alpha$, the phase of $\hat{\Delta}^s$.  The single particle condensate constraint $\hat{\Delta}^s=\mu$ has also been used in the last expression.

We can now map out the phase diagram in the space of $\mu$ and $\hat{\Delta}^d$ by looking at the spectra in Eqs.(\ref{Phi=0}) and (\ref{Phi>0}).  Fig. \ref{fig:d-wave} illustrates the possible phases.  The dotted line indicates the instability of both spectra when $\hat{\Delta}^d>1/2m^\ast$.  For clarity, in each of the four regions there is an inset of the typical spectrum $E_\bk^2$.  The spectrum in region (IV) is presumably stabilized at higher momenta (which is represented by the dashed curve) so that the instability is really a kind of ``d-wave roton''.  The spectrum in region (III) is clearly unstable, indicating a new phase.  As $\mu$ increases from (IV) into (III), the pair gap and probably the roton gap, too, go to zero.   Although we have no explicit calculations, it is probable that the phase in (III) is a charge-density-wave at the roton wavevector.  On the other hand, there are no instabilities in regions (I) and (II).  (I) is a pure, isotropic, d-wave condensate, in which the gap is half of the energy required to break a pair.  As the gap closes, and we move into (II) at fixed $\hat{\Delta}^d$, a single particle condensate develops so that all three condensates, $\Phi$, $\hat{\Delta}^s$, and $\hat{\Delta}^d$ exist; accordingly, there is a linear phonon mode and an anisotropy due to the coexistence of both pairings.  As $\hat{\Delta}^d$ increases from (II) to (III), the anisotropy is strong enough that the gap closes at finite momentum, in addition to $\bk=0$.  This is consistent with the transition from (IV) to (III).  

The Haffnian itself can only exist at $\mu=0$, which can be seen from the asymptotics of $u_\bk$ and $v_\bk$.  Since the pair amplitude $g_\bk$ is a pure $l=-2$ eigenstate at $\bk\rightarrow 0$ when the Haffnian is the asymptotic form of the many-body wavefunction, $\hat{\Delta}^s$ must vanish.  Furthermore, the requirement that $|g_\bk|=|v_\bk/u_\bk|\rightarrow 1$ as $\bk\rightarrow 0$ restricts the relative behavior of $u_\bk$ and $v_\bk$, and the explicit solution in Eq. (\ref{uvEk}) implies that $\mu=0$.  The thick line in the figure marks the transition region, $\mu=0$ with a stable spectrum ($\hat{\Delta}^d<1/2m^\ast$), where the Haffnian represents the long range behavior. 
\begin{figure}[htb]
\noindent
\center
\epsfxsize=4in
\epsfbox{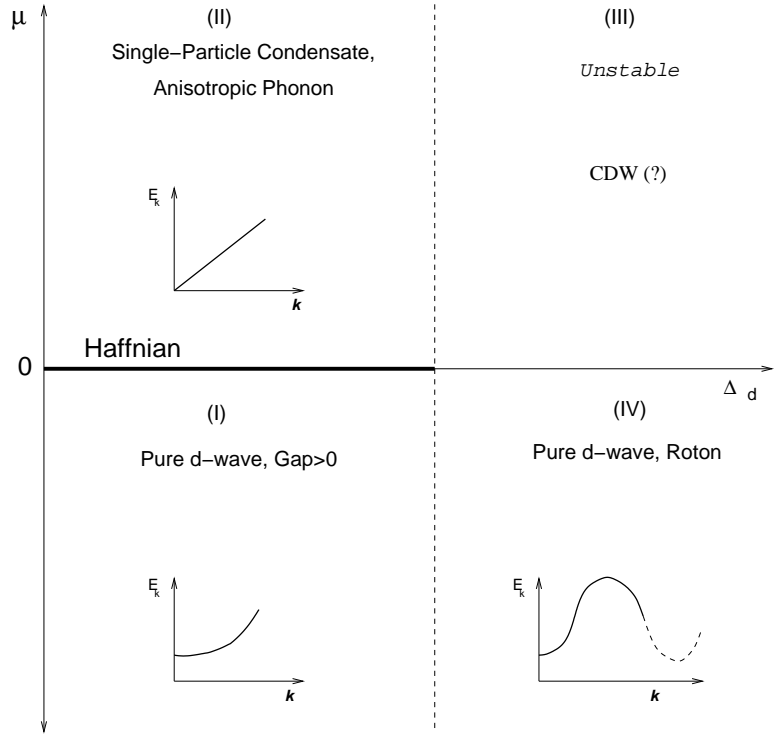}
\caption{\small Phases of bosons in two dimensions.}
\label{fig:d-wave}
\end{figure}
E.~Rezayi \cite{ER_unpub} has analyzed numerically the Haffnian Hamiltonian in eqn. \ref{PsiHf}.  He found that for $V_0$ negative the ground state is a paired state, while for $V_0$ positive it is a Laughlin state, which is a Bose condensate.  This is consistent with the Haffnian being on a phase boundary.
\chapter{Fermion Pairing: Quantum Hall Effect for Spin}
\label{chap:BCS}
%%%%%%%%%%%%%%%%%%%%%%%%%%%%%%%%%%%%%%%%%%%%%%%%%%%%%%%%%%%%%%%%%%%%%%%%%%%%%%%%%%%%%%

In this Chapter, we provide a detailed derivation of the Hall conductivity for
spin transport in the d- and p-wave pairing of fermions. In the FQHE, the fermions are really composite fermions, and we ignore gauge field fluctuations.  This is equivalent
to showing that the induced action for the system in an external gauge field
that couples to the spin is a Chern-Simons (CS) term. In the d-wave case, the system is spin-rotation invariant, so we obtain an SU(2) CS term, while in the p-wave
case, there is only a U(1) symmetry, so we find a U(1) CS term. In both cases,
the Hall spin conductivity is given by a topological invariant. Within the BCS mean field approach, using a conserving
approximation, this topological invariant is the winding number of the order parameter in momentum space and is an integer, which is the statement of quantization. We argue that the
quantization in terms of a topological invariant is more general than the
approximation used. 

%%%%%%%%%%%%%%%%%%%%%%%%%%%%%%%%%%%%%%%%%%%%%%%%%%%%%%%%%%%%%%%%%%%%%%%%%%%%%%%%%%%%
\section{BCS Hamiltonian}
\label{chap:BCS_H}
%%%%%%%%%%%%%%%%%%%%%%%%%%%%%%%%%%%%%%%%%%%%%%%%%%%%%%%%%%%%%%%%%%%%%%%%%%%%%%%%%%%%%%

Considering first the spin-singlet paired states, we use the Nambu basis where
the symmetries are transparent. Define
\be
\Psi=\frac{1}{\sqrt{2}} \left(\begin{array}{c}
       c  \\
       i\sigma_y c^\dagger  \end{array}\right),
\ee
with
\be
c=\left(\begin{array}{c}
           c_\up\\
           c_\down \end{array}\right),
\ee
so that $\Psi$ transforms as a tensor product of particle-hole and
spin-space
spinors. We consider an interacting system and approximate it as in BCS
theory, then with a minimal coupling to the gauge field, we use a conserving
approximation to obtain the spin response. In Fourier space, we should note
that
\be
\Psi_\bk=\frac{1}{\sqrt{2}} \left(\begin{array}{c}
       c_\bk  \\
       i\sigma_y c_{-\bk}^\dagger  \end{array}\right).
\ee
In the Nambu basis, the kinetic energy becomes (with $K=H-\mu N$)
\bea
K_0&=&\sum_\bk\xi_\bk^0(c_{\bk\up}^\dagger c_{\bk\up}+
c_{\bk\down}^\dagger c_{\bk\down})\non\\
&=&\sum_\bk\xi_\bk^0\Psi_\bk^\dagger(\sigma_z\otimes I)\Psi_\bk,
\eea
where $\xi_\bk^0=|\bk|^2/(2m)-\mu$ is the kinetic energy, containing the bare
mass $m$, and the products in the spinor space are understood. Products like
$\sigma_z\otimes I$ act on the Nambu spinors, with the first factor acting in
the particle-hole factor, the second in the spin-space factor.
The interaction term, for a spin-independent interaction $V$, is
\be
K_{\rm int}=\frac{1}{2}\sum_{\bk\bk'{\bf q}}V_{\bf q}:\Psi_{\bk+{\bf
q}}^\dagger(\sigma_z\otimes I)\Psi_\bk\Psi_{\bk'-{\bf q}}^\dagger
(\sigma_z\otimes I)\Psi_{\bk'}:.
\ee
Here the colons $:\ldots:$ denote normal ordering, that is all the $c^\dagger$s
are brought to the left. In the BCS-extended Hartree-Fock approximation, the
effective quasiparticle Hamiltonian (for later reference) is
\bea
K_{\rm eff}&=&\sum_\bk \Psi_\bk^\dagger\left[\xi_\bk(\sigma_z\otimes I)
   +{\rm Re}\,\Delta_\bk(\sigma_x\otimes I)\right.\non\\
&&\left.\mbox{}-{\rm Im}\,\Delta_\bk(\sigma_y\otimes I)\right]\Psi_\bk.
\eea
This is for singlet pairing, where $\Delta_{-\bk}=\Delta_\bk$, and not
just for d-wave. Here $\xi_\bk$
is $\xi_\bk^0$ plus the Hartree-Fock corrections. If we define a vector
\be
\bE_\bk=({\rm Re}\,\Delta_\bk,-{\rm Im}\,\Delta_\bk,\xi_\bk)
\ee
then the quasiparticle energy $E_\bk=|\bE_\bk|$, and
\be
K_{\rm eff}=\sum_\bk\Psi_\bk^\dagger(\bE_\bk\cdot\hbox{$\sigma$}
\otimes I)\Psi_\bk.
\ee

%%%%%%%%%%%%%%%%%%%%%%%%%%%%%%%%%%%%%%%%%%%%%%%%%%%%%%%%%%%%%%%%%%%%%%%%%%%
\section{Spin Response in a Conserving Approximation}
\label{chap:SpinResponse}
%%%%%%%%%%%%%%%%%%%%%%%%%%%%%%%%%%%%%%%%%%%%%%%%%%%%%%%%%%%%%%%%%%%%%%%%%%

In the Nambu notation, it is clear that $K=K_0+K_{\rm int}$, and $K_{\rm eff}$,
are invariant under global SU(2) rotations that act on the spin-space, that is
the second factor in the tensor products. The spin density, the integral of
which over all space is the total spin and generates such global
transformations, and the spin current densities are given by
\bea
J_0^a({\bf q})&=&\frac{1}{2}\sum_\bk\Psi_{\bk-{\bf q}/2}^\dagger(I\otimes
\sigma_a)\Psi_{\bk+{\bf q}/2}\\
J_i^a({\bf q})&=&\frac{1}{2}\sum_\bk\frac{k_i}{m}\Psi_{\bk-{\bf q}/2}^\dagger
(\sigma_z\otimes \sigma_a)\Psi_{\bk+{\bf q}/2},
\eea
where $i=x$, $y$ is the spatial index, and $a=x$, $y$, $z$ is the spin-space
index. Spin conservation implies the continuity equation, as an
operator equation,
\be
\partial J_\mu^a/\partial x_\mu =0,
\ee
where $\mu=0$, $x$, $y$, and the summation convention is in force.

So far we have not introduced a gauge field for spin. Since the spin is
conserved locally, we can turn the symmetry into a local gauge symmetry by
introducing an SU(2) vector potential, and making all derivatives covariant.
The effect on $K$ is to add the integral of
\be
A_\mu^a J_\mu^a +\frac{1}{8m}A_i^a A_i^a \Psi^\dagger(\sigma_z\otimes I)\Psi.
\ee
The gauge field is to be used solely as an external source, with which to probe
the spin response of the system, and then set to zero.

If we now consider integrating out the fermions, then we can obtain an action
in the external gauge fields, which can be expanded in powers of $A_\mu^a$. The
zeroth-order term is the free energy density, times the volume of spacetime,
and the first-order term vanishes by spin-rotation invariance. The second-order
term corresponds to linear response: the second functional derivative with
respect to $A_\mu^a$, at $A_\mu^a=0$, yields the (matrix of) linear response
functions. In particular, the spatial components yield the conductivity tensor
in the usual way. Therefore we consider the imaginary-time time-ordered
function,
\be
\Pi_{\mu\nu}^{ab}=-i\langle J_\mu^a(q)J_\nu^b(-q)
\rangle,
\ee
where time-ordering is understood, and from here on we use a convention that
$p$, $q$, etc. stand for three-vectors $p=(p_0,{\bf p})$, and further
$p_0=i\omega$ is imaginary for imaginary time.
For $\mu=\nu=i=x$ or $y$, an additional ``diamagnetic'' term
$\bar{n}\delta^{ab}/4m$ is present in $\Pi_{\mu\nu}^{ab}$, which we do not show
explicitly.
As consequences of the continuity equation and the related gauge invariance,
$\Pi_{\mu\nu}$ must be divergenceless on both variables, $q_\mu\Pi_{\mu\nu}
=q_\nu\Pi_{\mu\nu}=0$. To maintain these when using the BCS-Hartree-Fock
approximation for the equilibrium properties, one must use a conserving
approximation for the response function, which in this case means summing
ladder diagrams (compare the charge case in Ref.\ \cite{BCS},
pp.\ 224--237).  We used the identical method in treating composite bosons at $\nu\neq 1$ in Chapter \ref{chap:Fermions}, Chapter \ref{chap:Fermions}.

One begins with the BCS-Hartree-Fock approximation, which can be written in
terms of Green's functions as (we consider only zero temperature, and $\int
dp_0$ is along the imaginary $p_0$ axis throughout)
%\pagebreak
\bea
G^{-1}(p)&=&p_0-\xi_{\bf p}^0\sigma_z\otimes I -\Sigma(p),\\
\Sigma(p)&=&i\int\frac{d^3k}{(2\pi)^3}(\sigma_z\otimes I)G(k)(\sigma_z\otimes I)
V(k-q).
\eea
Note that $G(p)$ and $\Sigma(p)$ are matrices acting on the tensor product
space. The equations are solved by
\be
G^{-1}(p)=p_0-\bE_{\bf p}\cdot\hbox{$\sigma$}\otimes I,
\ee
(we write $1$ for $I\otimes I$) as one can also see {}from the effective
quasiparticle Hamiltonian $K_{\rm eff}$, and $\Delta_{\bf p}$ obeys the
standard gap equation.

In the response function, the ladder series can be summed and included by
dressing {\em one} vertex, to obtain (again not showing the diamagnetic term)
\bea
\Pi_{\mu\nu}^{ab}(q)&=&-i\int\frac{d^3p}{(2\pi)^3}
{\rm tr}\left[\gamma_\mu^a(p,p+q)G(p+q)\right.\non\\
&&\left.\times\Gamma_\nu^b(p+q,p)G(p)\right],
\eea
where $\gamma_\mu^a$ is the bare vertex,
\bea
\gamma_0^a(p,p+q)&=& \frac{1}{2}I\otimes\sigma_a,\\
\gamma_i^a(p,p+q)&=& -\frac{(p+\frac{1}{2}q)_i}{2m}\sigma_z\otimes\sigma_a,
\eea
and $\Gamma_\mu^a$ is the dressed vertex satisfying
\bea
\Gamma_\nu^b(p+q,p)&=&\gamma_\nu^b(p+q,p)+i\int\frac{d^3k}{(2\pi)^3}
\sigma_z\otimes I G(k+q)\non\\
&&\mbox{}\times\Gamma_\nu^b(k+q,k)G(k)\sigma_z\otimes I V(p-k).
\eea
At small $q$, we can obtain useful information about this function {}from the
Ward identity that results {}from the continuity equation. The particular
Ward identity we use here is an exact relation of the vertex function to the
self-energy, and the conserving approximation (the ladder series) was
constructed to ensure that it holds also for the approximated vertex and self
energy functions.

Following Schrieffer's treatment \cite{BCS}, we consider the vertex
function with external legs included:
\be
\Lambda_\mu^a(r_1,r_2,r_3)=\langle J_\mu^a(r_3)\Psi(r_1)\Psi^\dagger(r_2)
\rangle,
\ee
for spacetime coordinates $r_1$, $r_2$, $r_3$. Applying $\partial/\partial
r_{3\mu}$ to both sides and using the operator continuity equation, we obtain
the exact identity in Fourier space
\be
q_\mu\Gamma_\mu^a(p+q,p)=\frac{1}{2}I\otimes\sigma_a G^{-1}(p)
    -\frac{1}{2}G^{-1}(p+q)I\otimes\sigma_a.
\ee
Since $G^{-1}$ is trivial in the spin-space indices, it commutes with
$I\otimes\sigma_a$. Hence at $q\to0$, the right-hand side vanishes, so
$\Gamma(p+q,p)$ has no singularities as $q\to0$. This differs {}from the charge
case, for example, where this calculation (using the ladder series
approximation) leads to the discovery of the collective mode \cite{anderson58}.
Since the spin symmetry is unbroken, no collective mode is necessary to restore
this conservation law, and so there is no singularity in the vertex function
for spin.

One can verify that the Ward identity is satisfied using the BCS-Hartree-Fock
$G^{-1}$ and the ladder series for $\Gamma$. At $q=0$, this yields the
important results
\be
\Gamma_\mu^a(p,p)=-\frac{1}{2}\partial_\mu G^{-1}(p)I\otimes\sigma_a,
\label{qzerovert}
\ee
or explicitly,
\bea
\Gamma_0^a(p,p)&=&\frac{1}{2}I\otimes\sigma_a,\non\\
\Gamma_i^a&=&-\frac{1}{2}\partial_iG^{-1}(p)I\otimes\sigma_a,
\eea
where $\partial_i$ and $\partial_\mu$ stand for $\partial/\partial p_i$,
$\partial/\partial p_\mu$ {}from here on, and the extra minus in the first
relation is consistent because implicitly $q_\mu\Gamma_\mu=q_0\Gamma_0
-q_i\Gamma_i$.

We now calculate $\Pi$ at small $q$. To zeroth order, use of the Ward identity
shows that the $J$-$J$ function gives zero, except when $\mu=\nu=i$.
In that case, it reduces to a constant that cancels the diamagnetic term
also present in just that case. Hence we require only the part first-order in
$q$. In the expression for $\Pi$ above, we first shift $p\to p-\frac{1}{2}q$,
so that $q$ no longer appears in any bare vertices, but does appear in the
Green's functions on both sides of the ladder, between the rungs which are the
interaction lines. Hence to first order, we obtain a factor
$\pm\frac{1}{2}\partial_\mu G=\mp\frac{1}{2} G\partial G^{-1} G$ in place of
$G$ in one position in the ladder. Since there may be any number of rungs
(including zero) between this and either of the vertices at the ends, the
terms can be summed up into a ladder dressing each vertex, evaluated at $q=0$.
Hence we obtain to first order
\bea
\Pi_{\mu\nu}^{ab}(q)&=&-\frac{i}{2}\int\frac{d^3p}{(2\pi)^3}{\rm
tr}\left[\Gamma_\mu^a(p,p) q_\lambda\partial_\lambda G
\Gamma_\nu^b(p,p)G(p)\right.\non\\
&&\left.\mbox{}-\Gamma_\mu^a(p,p)G(p) \Gamma_\nu^b(p,p) q_\lambda
\partial_\lambda G \right].
\eea
Using the Ward identity, this becomes
\bea
\Pi_{\mu\nu}^{ab}(q)&=&\frac{i}{8}q_\lambda\int\frac{d^3p}{(2\pi)^3}{\rm tr}
\left\{(I\otimes\sigma_a)(I\otimes\sigma_b)G\partial_\mu G^{-1}\right.\non\\
&&\left.\times[G\partial_\lambda G^{-1},G\partial_\nu G^{-1}]\right\}
\label{3form}
\eea
Since the $G$'s are independent of the spin-space indices, the explicit
$\sigma$'s factor off, and the result is $\delta^{ab}$ times a
spin-independent part. The latter can be simplified using the BCS-Hartree-Fock
form of $G$, by writing the latter as
\be
G(p)=\frac{p_0+\bE_{\bf p}\cdot\hbox{$\sigma$}\otimes I}
{p_0^2-E_{\bf p}^2}.
\ee
The spin-independent factor contains $\epsilon_{\mu\nu\lambda}$ since it is
antisymmetric in these labels. Keeping track of the signs, we find for the
quadratic term in the induced action
\be
\frac{1}{4\pi}\frac{\cal M}{4}\int d^3r A_\mu^a\frac{\partial A_\nu^a}{\partial
r_\lambda}\epsilon_{\mu\nu\lambda},
\ee
with $\cal M$ given by the topological invariant
\be
{\cal M}=\int \frac{d^2p}{8\pi}\epsilon_{ij}\bE_{\bf p}\cdot(\partial_i
\bE_{\bf p}\times\partial_j\bE_{\bf p})/E_{\bf p}^3.
\ee
The right hand side is exactly the Pontriagin winding number $m$, and is an integer as long as $\bE$ is a continuous,
differentiable function of $\bf p$; it is $2$ for the d-wave case.

To ensure SU(2) gauge invariance, the CS term should include also a term cubic
in $A$, with no derivatives. For this term we evaluate the triangle one-loop
diagrams with three insertions of $J$, with each vertex dressed by the ladder
series. Setting the external momenta to zero, the Ward identity can be used
for all three vertices, and the result can be seen to be
\bea
\label{triangle}
\Pi_{\mu\nu\lambda}^{abc}(0,0)&=&\mbox{}-\frac{1}{24}\int\frac{d^3p}{(2\pi)^3}
{\rm tr}\left[(I\otimes\sigma_a) G\partial_\mu G^{-1}\right.\non\\
&&\times\left.\{(I\otimes\sigma_b) G\partial_\nu G^{-1},(I\otimes\sigma_c)
G\partial_\lambda G^{-1}\}\right]\,.
\eea
The anticommutator $\{\,,\,\}$ arises since the result must be symmetric under
permutations of the index pairs $\mu$, $a$, etc.
The product $\sigma_a\sigma_b\sigma_c$, when traced over the spin-space
indices, yields a factor $2i\epsilon_{abc}$, which is antisymmetric, and so the
remainder must contain $\epsilon_{\mu\nu\lambda}$ to maintain symmetry; the
rest of the structure is the same as before.
Hence the full result is the SU(2) CS term, which we write in terms of the
$2\times2$ matrix vector potentials $A_\mu=\frac{1}{2}\sigma_aA_\mu^a$,
\be
\frac{k}{4\pi}\int d^3x\epsilon_{\mu\nu\lambda}{\rm tr}(A_\mu\partial_\nu
A_\lambda+\frac{2}{3}A_\mu A_\nu A_\lambda).
\ee
Here $k$ is the conventional notation for the coefficient of such a term, in
this same normalization; if we wished to quantize the theory by functionally
integrating over $A$, we would need $k=$ an integer. In our case
$k={\cal M}/2=1$ for d-wave.

For the spin-triplet case with an unbroken U(1) symmetry, we must use
the fact that $\Delta_{-\bk}=-\Delta_\bk$. For example, in
the two-dimensional A-phase, as occurs in the 331 state in the double-layer
FQHE system with zero tunneling, the pairs are
in the isospin $S_z=0$ triplet state $\up_i\down_j+\down_i\up_j$, and the U(1)
symmetry generated by $S_z$ is unbroken; we recall that the underlying
Hamiltonian is not assumed to have a full SU(2) symmetry. The effective
quasiparticle Hamiltonian becomes, in the Nambu-style notation,
\bea
K_{\rm eff}&=&\sum_\bk \Psi_\bk^\dagger\left[\xi_\bk(\sigma_z\otimes I)
   +{\rm Re}\,\Delta_\bk(\sigma_x\otimes \sigma_z)\right.\non\\
&&\left.\mbox{}-{\rm
Im}\,\Delta_\bk(\sigma_y\otimes \sigma_z)\right]\Psi_\bk.
\eea
The U(1) vector potential $A_\mu$ couples to $S_z$, and the vertex functions
contain $I\otimes\sigma_z$, which commutes with the BCS-Hartree-Fock Green's
function $G$. The tensors appearing in the three terms in $K_{\rm eff}$
obey the same algebra as the three in that for the spin-singlet case
(where they were
trivial in the second factor), and as in that case commute with
$I\otimes\sigma_z$. Consequently, the derivation for the induced action to
quadratic order in $A_\mu$ is similar to that for the SU(2) singlet case
above, and the traces in the Nambu indices can be carried out with the same
result as before, to obtain the abelian CS term
\be
\frac{1}{4\pi}{\cal M}\int d^3r A_\mu\frac{\partial A_\nu}{\partial
r_\lambda}\epsilon_{\mu\nu\lambda},
\ee
and no cubic term. In this case, $\cal M$ is again given by the winding
number $m$ which is $0$ or $\pm 1$ in the p-wave strong and weak-pairing phases
(respectively) discussed in this chapter.

%%%%%%%%%%%%%%%%%%%%%%%%%%%%%%%%%%%%%%%%%%%%%%%%%%%%%%%%%%%%%%%%%%%%
\section{Discussion and Generalizations}
\label{chp:Generalization}
%%%%%%%%%%%%%%%%%%%%%%%%%%%%%%%%%%%%%%%%%%%%%%%%%%%%%%%%%%%%%%%%%%%%

We note that the effect of the vertex corrections we included as
ladder series is to renormalize the $q=0$ vertices as shown in eq.\
(\ref{qzerovert}) for the spin-singlet case, and use these in one-loop
diagrams with no further corrections. This corresponds to the minimal
coupling $p \to p-A$ in the action, as one would expect by gauge
invariance. If we assume such a coupling, and treat the low-energy,
long-wavelength theory near the weak-strong transition as Dirac fermions
with relativistic dispersion and minimal coupling to the external gauge
field, then the expression for $\cal M$ as an integral over $\bf p$
covers only half the sphere in $\bf n$ space, and we would get $\pm 1$
(d-wave), $\pm 1/2$ (p-wave). The missing part results {}from the
ultraviolet regulator in the field theory version of the calculation
\cite{rahw}, or {}from a second fermion with a fixed mass in some lattice
models \cite{ludwig}. In our calculation, the remainder is provided by
the ultraviolet region, where $\Delta_\bk\to0$ as $\bk\to\infty$. At the
transition, $\mu=0$, the map is discontinuous and covers exactly half the
sphere in the p-wave case, so ${\cal M}=1/2$, as in other problems. In the
d-wave case with rotational symmetry, the value of $|v_\bk/u_\bk|$ as
$\bk\to0$ is nonuniversal, and hence so is
the value of $\sigma_{xy}^s$ at the transition. This is a consequence of
the non-relativistic form of the dispersion relation of the
low-energy fermions in this case. We may also note that for a paired system
on a lattice, as in models of high $T_c$ superconductors, a similar
calculation will give an integral over the Brillouin zone, which is a torus
$T^2$, instead of the $\bk$ plane which can be compactified to $S^2$. But maps
{}from $T^2$ to $S^2$ are again classified by the integers, and the
integer winding number is given by the same expression, so quantization
is unaffected.

We can also argue that the quantization result away {}from a transition is
exact in a translationally-invariant system, at least in all orders in
perturbation theory. For this we use the form in eq.\ (\ref{3form}) or
(\ref{triangle}), where the Ward identity for the vertex has been used.
Diagrammatically, it is clear that the exact expression can be similarly
written, using the exact (i.e., all orders in perturbation) Green's
function and vertex function.
(This is also true when the CS gauge field interaction is included.) The
Ward identity that relates them is exact, and the result for $\sigma_{xy}^s$
is of the same form as shown. The next step, the frequency integrals,
cannot be done explicitly in this case, because the precise form of the
Green's function is unknown, and the analogs of $\xi_\bk$, $\Delta_\bk$ (or of
$u_\bk$, $v_\bk$) do not exist. The latter do not exist because in general the
poles in the Green's function, which would represent the quasiparticles, are
broadened by scattering processes, except for the lowest energies for
kinematical reasons. However, the form in eq.\ (\ref{3form}) is itself a
topological invariant, as we will now argue. As long as there is a gap in the
support of the spectral function of $G$, $G(p)$ is continuous and
differentiable on the {\em imaginary} frequency axis, and tends to
$I\otimes I/p_0$ as $p_0\to\pm i\infty$. Thus $G^{-1}$ exists and never
vanishes. Considering the spin-singlet case for convenience, the spin-space
structure is trivial, so we may perform the corresponding traces, and then $G$
or $G^{-1}$ is a $2\times 2$ matrix, with the same reality properties on the
imaginary $p_0$ axis as in the BCS-Hartree-Fock approximation. (The
spin-triplet case should work out similarly, because of the
algebraic structure already mentioned.) It thus represents a real non-zero
4-component vector, in ${\bf R}^4-{0}$, which topologically is the same as
$S^3$. $S^3$ is obtained by dividing $G$ by its norm, $({\rm tr}G^\dagger G)
^{1/2}$, and the normalized $G$ is a $2\times 2$ unitary matrix with
determinant $-1$, so it lies in $S^3$. The $\bk$ space can be compactified
to $S^2$ as before, and the frequency variable can be viewed as an element of
the interval ${\cal I}=(-1,1)$, so the integral is over $S^2\times{\cal I}$.
However, since the limit of the Green's function as $p_0\to\pm i\infty$ for
fixed $\bk$ is independent of $\bk$, we can view this as simply $S^3$. Thus
we are dealing with maps {}from $S^3$ to $S^3$, the equivalence classes of
which are classified by the homotopy group $\pi_3(S^3)={\bf Z}$. The
integral we have obtained simply calculates the integer winding number or
Pontryagin index of the map, when properly normalized ($G$ can be
normalized to lie in SU(2) without affecting the integral). This
establishes the quantization of $\sigma_{xy}^s$ in a translationally-invariant
system with a gap, at least to all orders in perturbation theory, and
probably can be made fully non-perturbative (as the Ward identity is already).

\chapter{Adsorption on Carbon Nanotubes}
\label{chap:tube}

In this chapter we switch gears into one dimension and consider adsorption on nanotubes as discussed in the introductory chapter, Section \ref{sec:intro_tubes}.

%%%%%%%%%%%%%%%%%%%%%%%%%%%%%%%%%%%%%%%%%%%%%%%%%%%%%%%%%%%%%%%
\section{Nanotube geometry}
\label{eq:geom}
%%%%%%%%%%%%%%%%%%%%%%%%%%%%%%%%%%%%%%%%%%%%%%%%%%%%%%%%%%%%%%%

Fig. \ref{TubeCell} illustrates the way in which a nanotube is obtained by wrapping a graphite sheet.  The hexagons are at positions $R_{n,m}=na_++ma_-$, where $a_\pm$ are primitive lattice vectors of the honeycomb lattice.  The standard convention is to identify $R_{0,0}\equiv R_{N,M}$, and to simply label the tube $(N,M)$.   The case $(N,N)$ is
known as the armchair tube, $(N,0)$ is the zig-zag, and all others are
chiral.  There is a geometric frustration whenever the wrapping
destroys the tripartite nature of the infinite sheet, which occurs
when $(N-M) {\rm mod}\;3$ is non-zero (this criterion is familiar in
the context of electronic conductivity\cite{Dresselhaus}).  Thus, both
zig-zag and chiral tubes can be frustrated geometrically, whereas
armchair tubes cannot. 
\begin{figure}
\center
\epsfxsize=3.5 in
\epsfbox{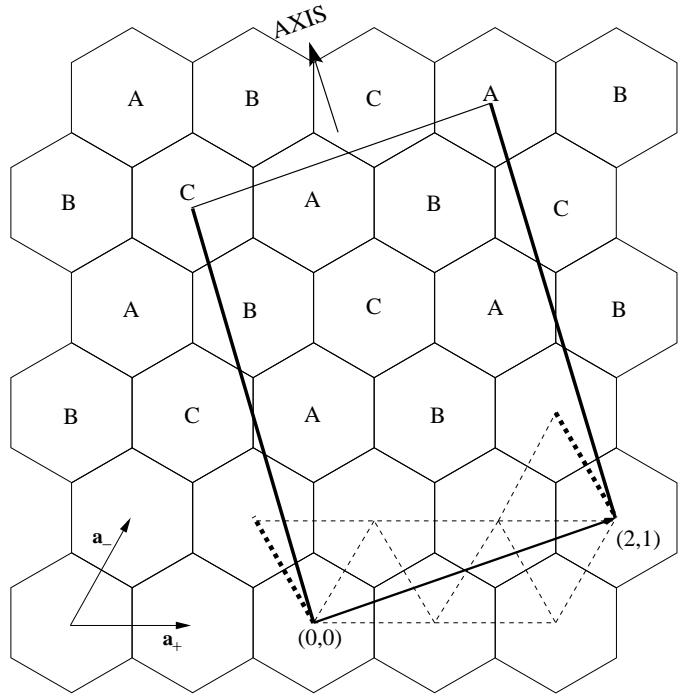}
\caption{\small An example of wrapping of the graphite sheet to make a $(2,1)$ tube. ${\bf a}_\pm$ are the primitive lattice vectors of the honecomb lattice.  The solid rectangle is the primitive cell of the tube.  The tube can also be built up by stacking the dotted region along the axis with the solid dotted lines identified.  Also shown is the tripartite lattice labeling A, B, C.}
\label{TubeCell}
\end{figure}
The adsorption sites form a triangular lattice wrapped on the cylinder, which is shown in Fig. \ref{Tube} for the $(7,0)$ zig-zag.  
\begin{figure}
\center
\epsfxsize=3.5 in
\epsfbox{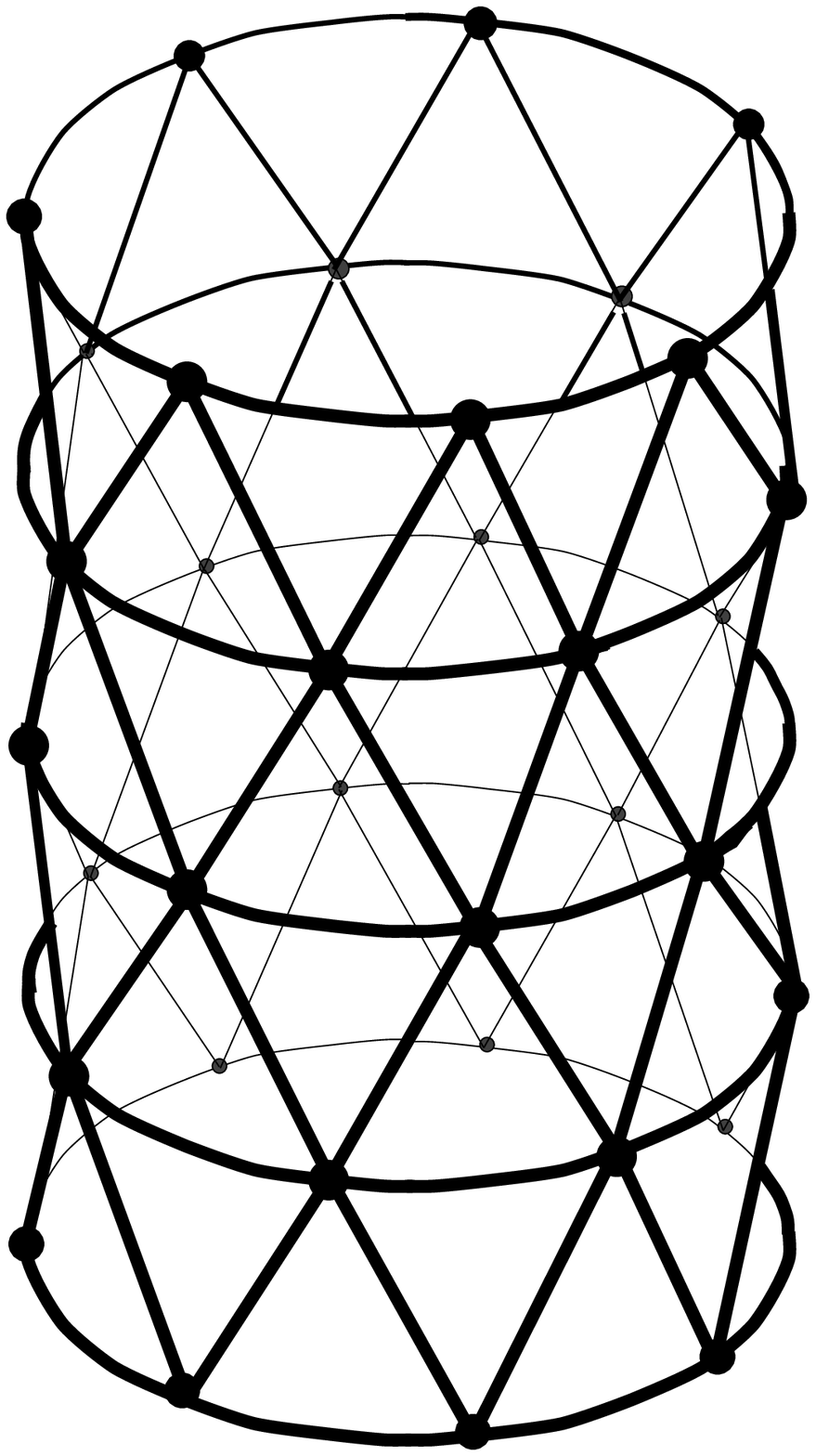}
\caption{\small Adsorption sites on a $(7,0)$ zig-zag nanotube}
\label{Tube}
\end{figure}
In the following section we specify the Hamiltonian and consider the simplest (classical) limit in which intersite tunneling of adatoms is prohibited.  We will then turn on the hopping perturbatively (the quantum case).

%%%%%%%%%%%%%%%%%%%%%%%%%%%%%%%%%%%%%%%%%%%%%%%%%%%%%%%%%%%%%%%%%
\section{The Hamiltonian and Classical Limit}
\label{sec:H_phase}
%%%%%%%%%%%%%%%%%%%%%%%%%%%%%%%%%%%%%%%%%%%%%%%%%%%%%%%%%%%%%%%%%%

When the adsorbed gas is a hard-core boson, the lattice gas is defined
by the Bose-Hubbard Hamiltonian\cite{Murthy,Auerbach}
\begin{equation}
 \label{eq:Hubbard}
{\cal H}\! =\! -t\sum_{\langle ij\rangle} b_i^\dagger b_j + b_j^\dagger
 b_i
 + V\sum_{\langle ij\rangle} n_i n_j - \mu\sum_i n_i~,
\end{equation}
where $n_i$ is the boson density at site $i$, $V$ is the nearest neighbor
repulsion and $t$ is the hopping amplitude. The occupation numbers $n_i$ are restricted to $0,1$ by the hard-core condition.  There is a familiar Heisenberg spin
representation \cite{Huang}, which identifies $S^z_i=n_i-1/2$, $S^+_i = b_i^\dagger$, and $S^-_i=b_i$.  The Hamiltonian is thus 
\begin{equation}
 \label{eq:Spin}
{\cal H}\! =\!  -2t  \sum_{\langle ij\rangle} S^x_i S^x_j\! +\! S^y_i S^y_j
 + V \sum_{\langle ij\rangle} S^z_i S^z_j \!-\! H\sum_i S^z_i ~,
\end{equation}
where $S^z_i=n_i-1/2$ and $H=\mu-3V$ is an effective external magnetic
field. Throughout the paper we will use the spin and density
representations interchangeably.  The spin models obtained in this way
are similar to recent examples of ``spin tubes''\cite{Andrei-Cabra}.

The Ising limit of the spin models, $t=0$ corresponds to the case when hopping is forbidden, and already contains
many interesting features.  We start the analysis in this regime,
obtaining the phase diagram as a function of the magnetic field, and
then consider quantum fluctuations perturbatively in $t/V$.  We
summarize our results first.

The phase diagram in the temperature-magnetic field plane of a typical
tube is shown in Fig.~\ref{PhaseDiagram}.
\begin{figure}
\vspace{0.25in}
\epsfxsize=3.5in 
\center
\epsfbox{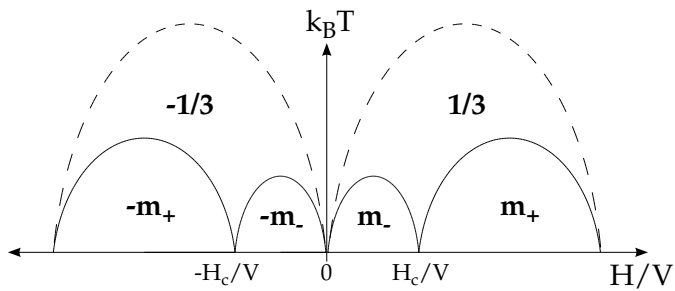}
\caption{\small Phase diagram of the $(N,M)$ tube}
\label{PhaseDiagram}
\end{figure}
When the index $q=(N-M){\rm mod}\;3$ is $1$ or $2$, we find four lobes
(solid lines), corresponding to two plateaus with magnetizations $m_-
< 1/3$ and $m_+ > 1/3$.  Here, we use the standard Ising notation in which
spin is $\pm1$.  Note that the plateaus are real phases only at
zero temperature because the tube is one-dimensional.  At finite
temperature, the boundaries should be interpreted as crossovers.
Nonetheless, deep within a lobe, at $k_BT\ll V$, the magnetizations
are well-defined.  Specifically, for $q=1$, we obtain the exact
expressions
\begin{eqnarray}
&\;& m_+=\frac{1}{3}\left(1+\frac{2}{2M+N}\right) \quad
m_-=\frac{1}{3}\left(1-\frac{2}{2N+M}\right) \nonumber\\
&\;&\hspace{2cm}
H_c=\left(4-\frac{2M}{N+M}\right)V
\label{eq:Exact}
\end{eqnarray}
The complementary case of $q=2$ is obtained by interchanging
$N\leftrightarrow M$.  On the other hand, those tubes without
geometric frustration ($q=0$) behave similarly to the flat sheet
(dotted lines) which has only two lobes with magnetizations $\pm 1/3$
\cite{Schick}.  In the flat sheet, the dotted lines are second order phase transitions in the universality class of the Potts-3 models \cite{Blote_Nightingale}. In our wrapped case, as the tube perimeter approaches the flat sheet limit, one
expects that the geometric frustration becomes irrelevant.  Indeed, as
$N$ or $M\rightarrow\infty$, $m_+$ and $m_-$ squeeze $1/3$ as the
inverse of the tube diameter and become indistinguishable.  Beyond the
lobes, where the field is strong enough to overcome all nearest
neighbor bonds ($|H|/V>6$ at $k_BT=0$), the tube is fully
polarized.  The filling fractions are
obtained from the magnetizations by $m=-2(n-1/2)$.  The phase diagram,
however, is more easily visualized in terms of spin since
spin reversal, $m\leftrightarrow-m$, corresponds to particle-hole
symmetry, $n\leftrightarrow1-n$.

We have verified this prediction numerically by transfer matrix
methods \cite{Huang} for zig-zag tubes up to $N=11$ and for the chiral tubes up to
$N+M=7$.\footnote{The programming used the {\sc Gnu} implementation of {\sc Fortran 77} on a {\sc PC}.  The main stumbling block is very large contributions to the partition function at low $T$; the {\sc Lapack} routine library was used to handle numbers outside ordinary machine range.}  The transfer matrix rows for a sample tube are delineated by dotted lines in Fig. \ref{TubeCell}.  It should be noted that similar transfer matrix calculations have been carried out for the special case of unfrustrated zig-zag tubes ($q=0$) \cite{Blote_Nightingale}.  The motivation in these earlier works was a finite size scaling analysis of the solid phases on flat graphite.  

Although $7$ is probably too small to be physical, we
believe that the arguments in this paper generalize to any tube.  In
Fig.~\ref{Magnetization} we display sample data for two zig-zag tubes
with different $q$: $(7,0)$ and $(8,0)$.  The magnetization curves
show clear plateaus whose values and transition fields match those
predicted by Eq.~(\ref{eq:Exact}).  By increasing the temperature and
following the evolution of the plateaus, we generate the phase diagram
above.

We find that a rather interesting feature of the zig-zag $(N,0)$ tubes
emerges, making them exceptional.  The insets in
Fig.~\ref{Magnetization} indicate an extensive entropy at zero
temperature, which has plateaus, too.  Upon enumerating the degenerate
space explicitly (Section \ref{sec:Quantum} below), we shall show that the entropy is exactly $s=({\rm
ln}2)/N$ and that it occurs in $m_+$ for $q=1$ and in $m_-$ for $q=2$.
In the presence of hopping, the non-degenerate plateaus retain their
gaps, whereas the degenerate ones become correlated states with a
unique ground state and {\it gapless} excitations.  More precisely,
conformal invariance develops and the effective theory has central
charge $c=1$ with a compactification radius, $R$, quantized by the
tube circumference, $R=N$.
\begin{figure}
\epsfxsize=3.5in
\center
\epsfbox{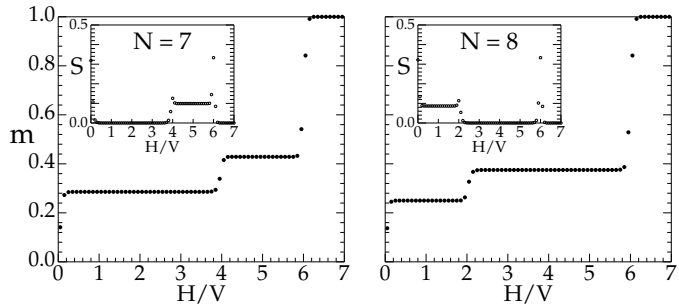}
\caption{\small Magnetization and entropy per site at $k_BT=0.05V$}
\label{Magnetization}
\end{figure}

In order to understand the magnetizations and nature of the geometric
frustration, it is more intuitive to use the original bosonic picture.
As a result of hard-core repulsion on the infinite graphite sheet, the
$m=1/3$ plateau corresponds to filling one of the three sublattices,
$A, B$ or $C$, of the triangular lattice.  This configuration
minimizes the repulsion, $Vn_i n_j$, while maximizing the filling, $\mu
n$. It is natural to try the same for nanotubes, as we illustrate in
Fig.~\ref{TubeFillReg} for $(5,0)$.
\begin{figure}
\epsfxsize=3.5in
\center
\epsfbox{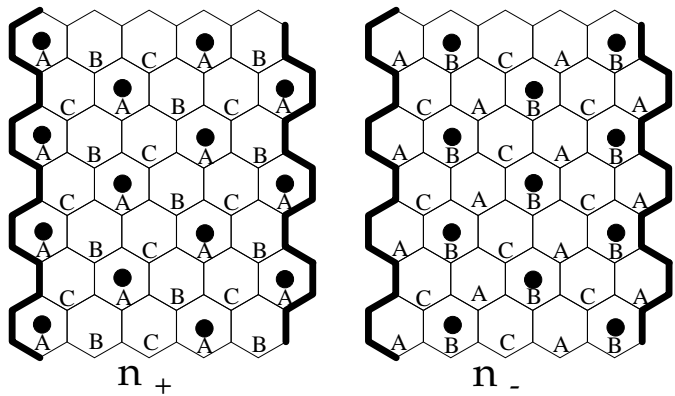}
\caption{\small Fillings and zipper of the $(5,0)$ zig-zag tube. $n_\pm$ corresponds to $m_\mp$}
\label{TubeFillReg}
\end{figure}
Upon wrapping, however, the thick vertical lines are identified and
the lattice is no longer tripartite.  In fact, the number of
sublattice sites is no longer equal, and there is a mismatch along the
thick line, which we term the ``zipper''.  On the left we fill the $A$
sublattice, obtaining the filling fraction $n_+=2/5$, and on the right
either $B$ or $C$ may be filled with the result that $n_-=3/10$.  For
general $(N,0)$ there are $2N$ hexagons in the unit cell, and the
filling fractions are $n_+=\lceil 2N/3\rceil/2N$ and $n_-=\lfloor
2N/3\rfloor/2N$, where $\lceil x\rceil$ and $\lfloor x\rfloor$ denote
the larger and smaller of the two bounding integers of $x$,
respectively.  The magnetizations in Eq.~(\ref{eq:Exact}) follow
directly by using the correspondence $m=-2(n-1/2)$.  Furthermore, due
to the sublattice mismatch, the number density of adjacent particles,
$n_b$, may be non-zero.  In the case of $(5,0)$, there are two broken
bonds per unit cell in $n_+$, and none in $n_-$.  This result
generalizes to any $q=2$ zig-zag tube: $n_{b+}=2/2N$ and $n_{b-}=0$.
For $q=1$, the argument goes through as before, except that
$n_{b+}=1/2N$.  We summarize this compactly by $n_{b+}=q/2N$.

Substituting these fillings into the Hamiltonian (\ref{eq:Hubbard})
yields two energies per site, $e_{\pm}(\mu)=Vn_{b\pm}\!-\!\mu
n_{\pm}$.  The transition occurs when these levels cross: $e_+=e_-$,
or
\begin{equation}
\frac{q}{2N}-\mu\frac{\lceil 2N/3\rceil}{2N}=-\mu\frac{\lfloor
2N/3\rfloor}{2N}
\label{eq:Transition}
\end{equation}
Solving for $\mu$ and using the correspondence
$H=\mu-3V$ gives precisely the critical field in Eq. (\ref{eq:Exact}).
In particular, this explains why there are exactly two independent
plateaus.  Note that, for the special case of the zig-zags, the
critical field depends only on $q$ and not on $N$ per se.

In the above analysis, we have made only one assumption, namely that
the zipper runs parallel to the tube axis. In general, the zipper may
wind helically around the tube or wiggle sideways. However, in all the
cases that we considered, the straight zipper has the lowest energy, and moreover, our transfer matrix
computations, which are blind to this assumption, are consistent with
our analysis.

The chiral tubes are different. Due to their geometry the zipper is
forced to wind, but, again, we find that the choice of the straightest
possible zipper reproduces our numerics for $N+M$ up to
$7$.  The determination of the fillings and level crossings is much
more involved than that of the zig-zag, and we leave it for a more
detailed paper.  In any case, our analysis reveals that the plateaus
in a chiral tube are not macroscopically degenerate, so that the
zig-zags are at a special degenerate point.

%%%%%%%%%%%%%%%%%%%%%%%%%%%%%%%%%%%%%%%%%%%%%%%%%%%%%%%%%%%%%%%%%%%%%
\section{Macroscopic Degeneracy and Quantum Fluctuations}
\label{sec:Quantum}
%%%%%%%%%%%%%%%%%%%%%%%%%%%%%%%%%%%%%%%%%%%%%%%%%%%%%%%%%%%%%%%%%%%%%

Having understood in detail the Ising limit, we now turn on a small
hopping, $t\ll V$, that introduces quantum fluctuations.  Deep within
a plateau, the substrate is maximally filled since adding a particle
increases $n_b$.  Consequently, all plateaus begin with a classical
gap of order $V$, and we work in the Hilbert space of the
classical ground states. Those plateaus which have only a
discrete symmetry must retain their gaps, but the macroscopically
degenerate plateaus are more complicated.

Let us reconsider the $n_+$ filling of the $(5,0)$ tube in
Fig.~\ref{TubeFillReg}.  Notice that a particle may hop laterally by
one site without changing $n_b$, as we illustrate in
Fig.~\ref{Hop-Lattice}, left.  Imagine building a typical $n_+$ state
layer-by-layer from top to bottom, with a total of $L$ layers.  Each
new layer must add exactly two filled sites and one nearest-neighbor
bond ($n_b=1/5$).  This constraint implies that no two adjacent sites
may be occupied within a layer; if they were, then, to conserve $n_b$,
two adjacent sites must be occupied in the next, and so on up the
tube.  However, this state is not connected to any other by a single
hop.  Similarly, the particles cannot hop from layer to layer because
this adds another intra-layer bond.
\begin{figure}
\epsfxsize=3.5in
\center
\epsfbox{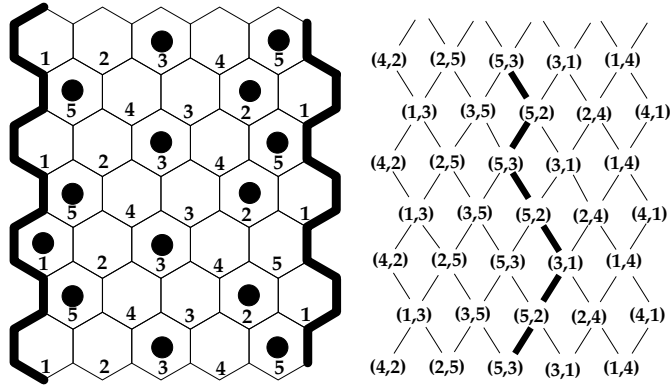}
\caption{\small LEFT: Typical configuration in $n_+$ (or $m_-$) of the
$(5,0)$ tube.  Alternating numbering within layers allows a symmetric
description from bottom-to-top or top-to-bottom.  RIGHT: Allowed
states as paths on a wrapped square lattice.  The vertex labels may be
dropped.}
\label{Hop-Lattice}
\end{figure}
An allowed state can be represented as a string of occupied sites,
$\sigma=\{\sigma_i\}$, $i=1,\ldots,L$, which in our example is
$\sigma=\{\cdots(5,3)(5,2)(5,3)\cdots\}$.  At each layer,
there are exactly two possibilities for the following one.  For
example, $(1,4)$ can be followed by $(1,4)$ or by $(2,4)$.  However,
the total number of possibilities at any given level is five.
Fig.~\ref{Hop-Lattice} (right) summarizes this structure succinctly as
a {\em square} lattice wrapped on the cylinder.  A typical state,
then, is a lattice path along the tube.  There is a recent Hubbard model considered by Henley and Zhang \cite{Henley} of spinless fermions on a square lattice in which the bookkeeping of states is very similar. 

Generalizing to $(N,0)$, we find $N$ possible states in each layer and
two in the succeeding one, and the structure of states is again that
of a wrapped square lattice with $N$ squares along the circumference.
The dimension of the Hilbert space is the number of lattice paths,
$N2^L$, so that in an infinitely long tube, the entropy per site is
exactly $({\rm ln}2)/N$, as claimed earlier.  Notice that constrained
paths introduce correlations along the length of the tube, despite the
absence of inter-layer hopping.

The matrix elements of the projected Hamiltonian connect only those
states that differ by a single hop:
\begin{equation}
\langle\tau|{\cal H}|\sigma\rangle = 
\left\{\begin{array}{l}
-2t\quad{\rm if}\quad\sum_i\delta_{\sigma_i \tau_i} = L-1\\
0\quad{\rm otherwise}
\end{array}\right.
\label{eq:Hopping}
\end{equation}
It turns out that this Hamiltonian is exactly solvable, being closely related to a class of solid-on-solid models that were introduced by Pasquier \cite{SOS}.  In the following section we derive the continuum limit of ${\cal H}$, and we confirm the result numerically in the succeeding section.
 
%%%%%%%%%%%%%%%%%%%%%%%%%%%%%%%%%%%%%%%%%%%%%%%%%%%%%%%%%%%%%%%%%%%%%%%
\subsection{Continuum Limit}
\label{sec:Analytic}
%%%%%%%%%%%%%%%%%%%%%%%%%%%%%%%%%%%%%%%%%%%%%%%%%%%%%%%%%%%%%%%%%%%%%%%

In the previous section, the paths $\sigma$ were labeled, for clarity, by a
string of occupied sites on the nanotube.  A simpler representation is to
work with the wrapped square lattice directly, where the path is uniquely
specified by an initial point and its direction in each layer.  We will label
the topmost layer by $i=1$ with $i$ increasing by $1$ with each downward
move.  There are $L$ layers of hexagons and we impose periodic boundary
conditions, $L+1\equiv1$.  In order for the layers to match, $L$ has to be even. 

To specify the initial point on $\sigma$, we chose an ``anchor'' $\alpha$ on
one of the $N$ sites in the $i=1$ layer ($\alpha=1,\ldots,N$).  Now,
represent a step to the right in layer $i$ by a fermion, $c^\dagger_i$, and a
step to the left by a hole, $c_i$.  A state $|\sigma\rangle$ in the Hilbert
space, $S$, is represented by 
\bea |\sigma\rangle=|\alpha\rangle\otimes
c^\dagger_{i_1}c^\dagger_{i_2}\cdots c^\dagger_{i_p}|0\rangle~,
\label{eq:state}
\eea 
where $\alpha$ is the anchor site in the first layer and $i_1\cdots i_p$
are the layers where the path steps to the right.  For instance, the portion
of the path in fig. \ref{Hop-Lattice} is $|\sigma\rangle=|3\rangle\otimes
c^\dagger_1c^\dagger_3c^\dagger_4\cdots |0\rangle$.   The
fermionic representation is convenient since there is exactly one step in
each layer, but hard-core bosons can also be used.  In any case, in one
dimension they are equivalent.   The number of particles
(steps to the right) and holes (steps to the left) must add up to $L$ in
order for the path to close on itself along the length of the tube. Each
path also has a topological character for the number of times that it winds
around the tube, which must be a multiple of $N$ for the path to close.
These two conditions may be written as
\bea
\label{eq:charge}
N_p+N_h&=&L\\
N_p-N_h&=&bN~,
\eea    
where $N_{p,h}$ is the number of particles or holes, and $b$ is an integer.
If the particles are assigned a charge, then $bN$ is the total charge.  Note that $b=0$ corresponds to half-filling, $N_p=N_h=L/2$. 

Whenever it is allowed within a layer, a single hop changes the step sequence
right-left to left-right and {\it vice versa}, which corresponds to
$c^\dagger_{i+1}c_i$ or $c^\dagger_ic_{i+1}$.  In a layer without a kink no
hops are possible, and the hopping terms vanish by fermionic statistics.
Since we are working in periodic boundary conditions, the boundary terms,
$c^\dagger_1c_L$ and $c^\dagger_Lc_1$, must be treated more carefully.  A hop
at this point is necessarily accompanied by a translation of the anchor point
by $|\alpha\rangle\mapsto|\alpha\pm 1\rangle$.  Let us represent this
operation by \bea R_\pm|\alpha\rangle=|\alpha\pm 1\rangle~ \eea with
$R^\dagger_-=R_+$.  Cylindrical wrapping requires a ${\bf Z}_N$ symmetry
because ${|\alpha\pm N\rangle\equiv|\alpha\rangle}$, i.e. $R_\pm^N=R_\pm$.
Putting the bulk and boundary hopping terms together, the Hamiltonian of eqn.
(\ref{eq:Hopping}) becomes \bea {\cal H} =
-2t\left[\sum_{i=1}^{L-1}c^\dagger_{i+1}c_i+R_-\otimes c^\dagger_1c_L\right]
+ h.c.~.
\label{H_hopping}
\eea 
We can think of ${\cal H}$ as describing free fermions on a periodic one
dimensional chain with a ${\bf Z}_N$ impurity on one of the bonds.

${\cal H}$ can be diagonalized exactly in momentum space.  Going around the
tube lengthwise contributes a phase $e^{ikL}$ while going around the
perimeter contributes $e^{i\phi}$, with $\phi = 2\pi a/N$ ($a=1,\ldots,N-1$).
Therefore toroidal boundary conditions require 
\bea 
\label{eq:phases}
e^{ikL}e^{i\phi} = 1~.
\eea 
Or, 
\bea k=\frac{2\pi v}{L}\left(n+\frac{a}{N}\right)~,
\label{eq:k}
\eea
where $v=2t$ is the velocity and $n$ is an integer.  The Hamiltonian contains the usual free particle dispersion, but with the allowed $k$ given by eqn. (\ref{eq:k}),
\bea
\label{eq:Ham_fermions}
{\cal H}=-4t\sum_k \cos k\,c^\dagger_k c_k~.
\eea
If the spectrum is linearized around the Fermi momentum, $|k_F|$ (at half-filling), then at small $k$, nonzero $a$ states cost an additional energy of $(2\pi v/L)(a/N)^2$.
 
The $a/N$ offset in $k$ can be thought of as a minimally coupled
vector potential such that the magnetic field is a $\delta$-flux tube through
the torus containing $(a/N)$ flux quanta.  In other words, one of the bonds
along the chain (the ``anchor'') had ${\bf Z}_N$ symmetry, whose effect is
equivalent to a flux tube.  Fig. \ref{fig:flux} illustrates this equivalence.
\begin{figure}
\epsfxsize=3.5in
\center
\epsfbox{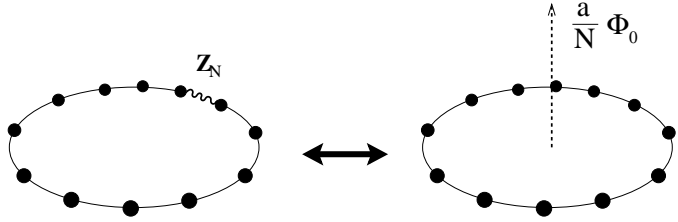}
%\vspace{.3 in}
\caption{\small The left ring shows the ${\bf Z}_N$ impurity on the anchor bond (wavy
  line).  The right ring shows the equivalent alternative, where the impurity
  is replaced by a flux tube through the torus.}
\label{fig:flux}
\end{figure}
The offset in $k$ is like a total current in the fermion system.

At this point, one can see two topological effects of the torus.  First is
the ${\bf Z}_N$ flux tube, or total current.  As we have seen, its
contribution to the energy near $|k_F|$ was $(2\pi v/L)(a/N)^2$.  Second
is the path winding along the length of the tube, or total charge,
eqn. (\ref{eq:charge}).  Its contribution to the energy near $|k_F|$ is
similar, $(2\pi v/L)(bN/2)^2$.  The total energy due to these topological
sectors is
\bea
\label{eq:E_top}
\Delta E_{a,b}=\frac{2\pi v}{L} \left(\frac{a^2}{N^2}+\frac{b^2 N^2}{4}\right)\,.
\eea 
This expression is familiar from the Luttinger liquid model of one-dimensional
spinless Fermions \cite{Haldane}.

We can now obtain the continuum limit of our model.  It is well known that
free fermions in one dimension are equivalent to free bosons.  The corresponding Lagrangian is 
\bea 
{\cal L} =\frac{1}{8\pi}[v^{-1}(\partial_t\varphi)^2 -
v(\partial_x\varphi)^2]~, 
\eea
where $\varphi$ is the bosonic field.  ${\cal L}$
is a conformally invariant theory with central charge $c=1$.  We conjecture
that the topological effects that we described above come from compactifying
$\varphi$ on a circle of radius $R$, 
\bea 
\phi\equiv\phi+2\pi R~.
\label{eq:compactification}
\eea 
By compactifying the boson, topological modes (or zero modes) appear.  In
field theory, they are conventionally obtained from electric and magnetic
monopoles.  The energy of the zero modes is
\bea
E^0_{a,b}=\frac{2\pi v}{L} \left(\frac{a^2}{R^2}+\frac{b^2 R^2}{4}\right) ,
\label{eq:ZeroModes}
\eea 
where $a$ and $b$ are integers labeling the fundamental cycles on the torus.
Comparing $E^0_{a,b}$ to $\Delta E_{a,b}$ (\ref{eq:E_top}), we find that $R=N$.  The ordinary phonon, (oscillator) modes exist on top of each topological sector and simply contribute the usual phonon energy, so that the complete dispersion is
\bea
E=E^0_{a,b}+\frac{2\pi}{L}|n|~.
\eea 

The overall picture of a compactified boson with central charge $c=1$ is
consistent with the solid-on-solid models of Pasquier \cite{SOS}.  
%%%%%%%%%%%%%%%%%%%%%%%%%%%%%%%%%%%%%%%%%%%%%%%%%%%%%%%%%%%%%%%%%%%%%%%
\subsection{Numerics} 
\label{sec:Numerics}
%%%%%%%%%%%%%%%%%%%%%%%%%%%%%%%%%%%%%%%%%%%%%%%%%%%%%%%%%%%%%%%%%%%%%%%
We have diagonalized the original Hamiltonian, eqn. (\ref{eq:Hopping}) numerically with periodic boundary conditions for system sizes up to $N=11$ and $L=10$.  Due to the sparseness of ${\cal H}$m we were also able to obtain the ground state energy up to $L=16$.  We will fix $2t=1$ in what follows. 

We find that the degeneracy is lifted and the ground state becomes
unique and uniform.  The ground state energy, $E_0(L)$, follows
${E_0\!\sim\!-0.61L\!-\!0.31\pi c/L}$.  The lowest
$N\!-\!1$ excited states are given by ${\Delta_a=a^2\Delta/(N^2L)}$,
with $\Delta=12.9\pm 0.5$, which is shown in Fig.~\ref{Gap} for
$a=1,2,3$.  All of these levels are doubly degenerate.
\begin{figure}
\epsfxsize=3.75in
\center
\epsfbox{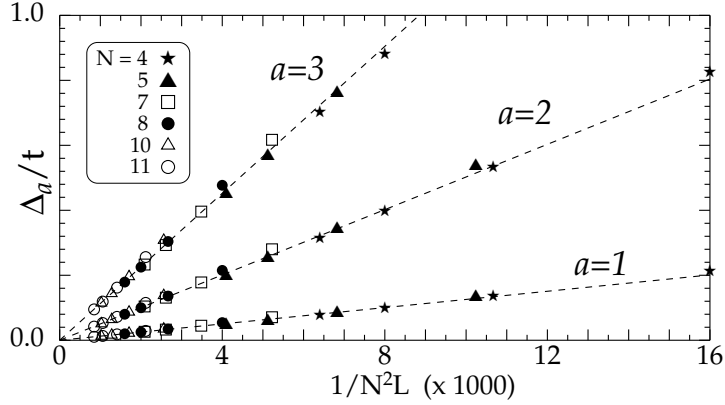}
\caption{\small The gap scales as $1/N^2L$.}
\label{Gap}
\end{figure}
This ground state energy and spectrum are in perfect agreement with
free bosons compactified on a radius $R\!=\!\zeta N$, as we described in the previous section. 

Right- and left-moving oscillator modes of energy
$\omega_n=vk_n$, where $k_n=2\pi n/L$, appear in the spectrum,
but for $a<N$ the zero modes are the lowest.  Our spectrum in
Fig.~\ref{Gap} corresponds to $E^0_{a,b}$ with $b=0$.  The modes with non-zero $b$ are very high in energy and are washed out by our small system size.  To fix $\zeta$,
we look at higher low-lying levels (which also scale like $1/L$). We
find that the $N$'th excitation energy is independent of $N$ and
quadruply degenerate.  This can happen only if the $N$'th zero mode,
$E^0_{\pm N,0}=2\pi v/\zeta^2L$, is degenerate with the lowest
oscillator mode, $\omega_{\pm 1}=2\pi v/L$, which fixes $\zeta=1$.
Thus, the compactification radius is $R=N$.  The velocity can be read
off from the slopes in Fig.~\ref{Gap} as $v=\Delta/2\pi$.  Within our accuracy, $v=2$.  The rest of
our spectrum is consistent with these parameters.  For instance, we find a
unique, zero-momentum state with $a=b=0$, which consists of one right-
and one left-moving oscillator mode with $n=2\pi/L$ at energy $E=2E_1$.  Note that $\zeta$ is fixed only by counting degneracies, not by fitting any parameters.

%%%%%%%%%%%%%%%%%%%%%%%%%%%%%%%%%%%%%%%%%%%%%%%%%%%%%%%
\subsection{Higher Order Corrections}
\label{sec:t2}
%%%%%%%%%%%%%%%%%%%%%%%%%%%%%%%%%%%%%%%%%%%%%%%%%%%%%%%

The preceeding discussion is valid to first order in $t/V$.  The next terms are of order $t^2/V$ and involve virtual transitions to adatom configurations that are not in the degenerate subspace $S$.  The generic form is 
\bea
-\frac{t^2}{V}\,{\cal P}_S\left[\sum_{{\langle ij\rangle}{\langle kl\rangle}}b^\dagger_ib_jb^\dagger_kb_l\right]{\cal P}_S~,  
\eea
where ${\cal P}_S$ is a projection operator into $S$.  Another way of writing the second order perturbation is the familiar form,
\bea
\langle\sigma^\prime|{\cal H}|\sigma\rangle \rightarrow 
\langle\sigma^\prime|{\cal H}|\sigma\rangle - \sum_\lambda\frac{\langle\sigma^\prime|{\cal H}|\lambda\rangle\langle\lambda|{\cal H}|\sigma\rangle}
{E_\lambda-E_\sigma}~,
\label{eq:virtual}
\eea
where $|\sigma\rangle, |\sigma^\prime\rangle\in S$ while $|\la\rangle\notin S$.  $E_\lambda$ is the energy of the virtual state.  $E_\sigma=E_{\sigma^\prime}$ are, of course, constant, and all energy differences are due to the nearest neighbor repulsion $Vn_in_j$. 

There are three types of virtual processes: (i) single particle hopping from $\sigma$ to $\sigma\neq\sigma^\prime$, (ii) two particle correlated hopping from $\sigma$ to $\sigma\neq\sigma^\prime$ and (iii) single particle diagonal hopping from  $\sigma$ back into $\sigma$.  For concreteness, consider process (iii) in the $(5,0)$ state in fig. \ref{Hop-Lattice}.  Whenever there is a kink in $\sigma$, such as in the third layer from the top, the contribution to eqn. (\ref{eq:virtual}) from all virtual hops is $-(35/6)4t^2/V$.  For example, the adatom on site $3$ can hop into any one of its six neighbors with the energy denominators $1/2+1/2+1/2+1/2+1/2+1/3$ (in units of $t^2/V$).  Similarly, the adatom on site $5$ contributes $1+1+1$, for a total of $35/6$ (it is forbidden to hop one site over to the right because the resulting state is in $S$).  On the other hand, if there is no kink, the contribution is $-8\cdot 4t^2/V$.  The criterion for a kink in layer $i$ is $2[1/4-(\w n_i-1/2)(\w n_{i+1}-1/2)]=1$, where $\w n_i=c^\dagger_ic_i$; otherwise this quantity vanishes.  Similarly, the absence of a kink is synonymous with the nonvanishing of $2[1/4+(\w n_i-1/2)(\w n_{i+1}-1/2)]$.  Thus, the total diagonal contribution to ${\cal H}$ can be written
\bea
\label{eq:4fermion}
\lefteqn{{\cal H}\rightarrow{\cal H}-}\non\\
&& \frac{8t^2}{V}\sum_\sigma\sum_i\frac{35}{6}
\left[\frac{1}{4}\!-\!\left(\w n_i\!-\!\frac{1}{2}\right)\left(\w n_{i+1}\!-\!\frac{1}{2}\right)\right]\!+\!
8\left[\left(\w n_i\!-\!\frac{1}{2}\right)\left(\w n_{i+1}\!-\!\frac{1}{2}\right)\!+\!\frac{1}{4}\right]|\sigma\rangle\langle\sigma|
\non\\
&&={\cal H}-\frac{4t^2}{V}\sum_\sigma\left[\frac{13}{3}\sum_i \w n_i \w n_{i+1}-\frac{13}{3}\sum_i \w n_i +8\right]|\sigma\rangle\langle\sigma|~.
\eea   
For general $N$, the correction scales like $N$.  The essential term in the last line of eqn. (\ref{eq:4fermion}) is the first one.  This four-fermion interaction renormalizes the radius $R$ by corrections of order $t/V$.

Let us return to processes (i) and (ii).  An example of (i) is the adatom in the third layer from the top, site $5$, hopping to the second layer, site $1$, and then back into the third layer, site $1$.  The intermediate state is not in $S$.  This process serves only to renormalize $t$ because its amplitude is the same for all kinks.  An example of (ii) is the particle in the fourth layer, site $5$, hopping to site $4$, {\it followed by} the particle in the third layer, site $5$, hopping to site $1$ in the same layer.  This correlated hopping occurs in a configuration containing the sequence particle-hole-hole or hole-particle-particle, which corresponds to a next-nearest neighbor interaction $c^\dagger_{i+2}c^\dagger_{i+1}c_{i+1}c_i+h.c.$.  We have not analyzed all such terms in detail, and it is possible that there is a delicate cancellation of the terms (ii) and (iii) when the fields are linearized around $|k_F|$, but we consider it more likely that they do not cancel so that $R$ is renormalized at order $t/V$.   

One can also consider the extreme limit in which $t\gg V$.  In this case, the XXZ Hamiltonian in eqn. (\ref{eq:Spin}) is simply the XY model on a cylinder.  Let us denote the spin angle relative to the cylindrical surface by $\varphi(x,\theta)$, where $x$ is the coordinate along the tube and $\theta$ is the coordinate around the perimeter.  Uniqueness of the wavefunction requires that $\varphi$ has the periodicity $\varphi(x,\theta+2\pi)=\varphi(x,\theta)+2\pi m$, where $m$ is an integer.  The low energy excitations are purely along the length of the tube; excitations around the perimeter will cost an energy on the order of $1/N$, which is large compared to $1/L$.  Thus we can freeze the $\theta$ coordinate, and the energy density is proportional to $|\partial_x\varphi|^2$.  Since the periodicity is still $\varphi\equiv \varphi+2\pi m$, we end up with a free boson compactified on radius $R_{XY}=1$.  The question is how the adsorption regime $t\ll V$, which is also a compactified boson but on radius $R=N$, is reached.  

%%%%%%%%%%%%%%%%%%%%%%%%%%%%%%%%%%%%%%%%%%%%%%%%%%%%%%%
\section{Special Case: $N=2$}
%%%%%%%%%%%%%%%%%%%%%%%%%%%%%%%%%%%%%%%%%%%%%%%%%%%%%%%

Before concluding with the effective theory, we should point out that the geometry of the $(2,0)$ tube
is special; all sites in adjacent layers are interconnected.  As a result,
all of its plateaus have an extensive entropy, and we find that hopping opens
a gap in both plateaus.  Fig. \ref{fig:N=2} illustrates this exception.  At either filling, the adatom in each layer is free to hop to either site---both configurations are iso-energetic because each site is contiguous to all sites in the neighboring layers.  Hence both plateaus are macroscopically degenerate.  In the presence of hopping, each adatom lives in a double well potential, which has a gap of order $t$.  
\begin{figure}
\epsfxsize=3.5in
\center
\epsfbox{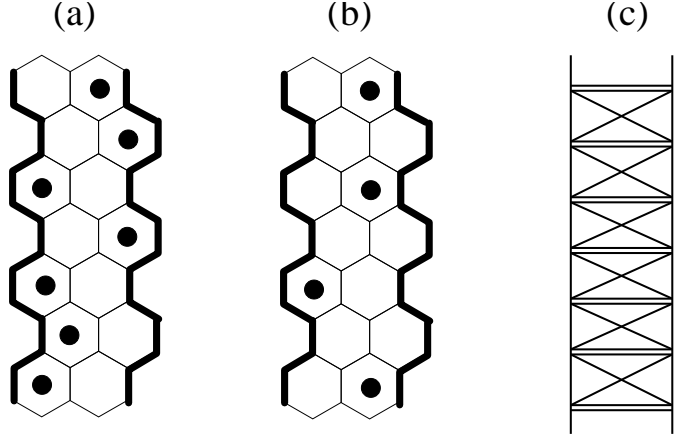}
\caption{\small The fillings for $N=2$. $n_+=1/2$ (a) and $n_-=1/4$ (b).  Each adatom can live at either site in its layer because each site is connected to every site in the neighboring layers.  (c) shows the triangular lattice in the plane; the horizontal double bond is due to the periodicity around a cylinder.   (c) is exactly the geometry of the spin ladder studied by other authors (albeit with different coupling).}
\label{fig:N=2}
\end{figure}
In fact, this tube can be written as a spin chain that has been studied
at isotropic coupling\cite{Sakai}, $-2t=V$.  Two plateaus were found in
this case, and it is tempting to speculate whether the two regimes are
connected adiabatically.

%%%%%%%%%%%%%%%%%%%%%%%%%%%%%%%%%%%%%%%%%%%%%%%%%%%%%%%%%%%%
\section{Discussion and Conclusion}
%%%%%%%%%%%%%%%%%%%%%%%%%%%%%%%%%%%%%%%%%%%%%%%%%%%%%%%%%%%%

One observable consequence of conformal symmetry is that the low temperature
heat capacity is fixed by $c$\cite{CFT}: 
\begin{equation}
C=c\frac{\pi vk_B^2}{3}T=\frac{\pi vk_B^2}{3}T
\nonumber
\end{equation}

It is noteworthy that, even though the dispersion of the oscillator
modes is independent of $N$, the spectrum remembers, via the
zero-modes, the finite radius of the nanotube. Furthermore, $R$ is
quantized by $N$; in the language of Luttinger liquids, this means
that the Luttinger parameter is fixed by topology, similarly to the
case of edge states in a fractional quantum Hall fluid\cite{Wen}, and
in contrast to quantum wires (where the Luttinger parameter can vary
continuously). Because there is no inter-layer hopping, $\phi$ is tied
to transverse, rather than to longitudinal, density fluctuations along
the tube.  

Finally, let us briefly view the spin tube as a quantum spin ladder to see if
it yields a gapless state in the degenerate plateaus.  A standard approach is to use a
Lieb-Schultz-Mattis (LSM) argument, in which the spins are deformed
slowly along the length\cite{Affleck}.  Applying it to our tube, we
find that a plateau is gapless if $S-M$ {\it is not} an integer, where
$S$ and $M$ are the total spin and magnetization, respectively, per layer  
(a layer being the $N$ sites around the perimeter).  Using
$S=N/2$ and the magnetizations from Eqn.~(\ref{eq:Exact}), we find
that $S-M$ {\it is} an integer in the macroscopically degenerate
plateaus, so that the LSM argument is insufficient in this case. A
conclusive argument must take the geometric frustration into account, which is further evidence that our state is strongly correlated.

In conclusion, we have studied the problem of monolayer adsorption on
carbon nanotubes and identified several interesting filling fraction
plateaus. Since the difference between the plateaus decreases slowly,
as the inverse of the tube diameter, experimental measurement should
be feasible for large enough tubes. We have identified the zig-zag
tubes as exceptional, in which the geometric frustration together with
quantum fluctuations lead to conformal symmetry.  The effective theory is free compactified boson, which has a quantized radius to first order in the hopping.  The only other such theory in nature that we are aware of are the chiral edge states in a quantum Hall fluid, where the radius is quantized by the bulk filling fraction.  There are interesting questions related to the large hopping limit.

\chapter{Summary}
\label{sec:summary}

It is widely believed by both theorists and experimentalists that the composite particle construct is required for a full understanding of the fractional quantum Hall effect (FQHE).  In this thesis we take the point of view that a great deal of the FQHE physics can be understood by projecting to the lowest Landau level (LLL) at the outset, which allows us to develop a composite particle formalism at various filling fractions, $\nu$. Initially the composites are abstract operators building up the many-particle Fock space, but later analysis of physical operators, many-particle wavefunctions, and response functions reveals an interpretation in terms of vortices bound to an underlying particle.  Our theory departs from most other theoretical work in this area, which uses singular flux attachment to map the problem into a Chern-Simons action.  We treat both fermionic and bosonic statistics of the underlying particles.  

When the underlying particles (or particles, for short) are bosons, the appropriate formulation is in terms of composite fermions.  The original Hamiltonian for the bosons is mapped exactly into a Hamiltonian for composite fermions with an infinite number of constraints.  To preserve the constraints, we use a conserving approximation to derive an effective theory microscopically.  In addition to a self-consistent theory of incompressible quantum liquids in the LLL, our approach provides a method to calculate the effective mass (gap) and the single particle spectrum of the composite fermions, which arise solely from the interactions between the particles.  This calculation extends previous work in the special case $\nu=1$ to arbitrary $\nu$.  A recent proposal raises the intriguing prospect of observing a FQHE of bosons in rotating atomic Bose-Einstein condensates.

The complementary case is composite bosons.  The ground state for a perturbation expansion is macroscopically degenerate, which precludes a microscopic derivation of an effective theory.  We follow an alternate approach by constructing a phenomenological Landau-Ginzburg action based on a symmetry analysis.  A crucial ingredient in the action is the internal structure of the composite particle, which, at $\nu=1/p$, consists of $p$ vortices bound to the particle.  For $p=\mbox{even}$ the particles are bosons while for $p=\mbox{odd}$ they are fermions.  Our model incorporates this structure through gauge potentials that couple to the internal degrees of freedom.  The spectrum of the effective theory contains the so-called magnetoroton excitation, which seems to be the first analytic observation of this phenomenon.

The next portion of the thesis is an examination of paired states of bosons or fermions in two dimensions.  The class of pairings that we investigate includes non-zero relative angular momentum, which is a state that breaks both parity and time reversal invariance.  Part of our method is based on Bardeen-Cooper-Schreiffer (BCS) theory and is largely independent of the FQHE.  However, in the context of the FQHE, we use a mean field approximation that maps particles in a net magnetic field into composite particles in zero field.  The BCS approach then applies to the composite particles.  

For bosonic particles, we consider p-wave pairing of spin-$1/2$ bosons and d-wave pairing of spinless bosons.  In the FQHE, the former case is a singlet known as the ``permanent'' state and the latter is the ``Haffnian''.  Each state can be written as a trial wavefunction that is the unique ground state of its corresponding Hamiltonian.  By analyzing the spectrum of the Hamiltonian directly or by applying the mean field BCS theory, we find that the permanent sits on the transition between the polarized Laughlin state (a Bose condensate of composite particles) and a Bose condensate with helical order.  The spin order of the permanent is that of an anti-Skyrmion.  Similarly, the Haffnian is on the transition between a Laughlin state and a strong coupling paired state.  Of course, these conclusions can stand alone without reference to the FQHE because they can be derived solely within the BCS framework.  

For fermionic particles, we carry through a conserving approximation for the spin conductivity.  We show that the induced action for an external gauge field that couples to spin is a Chern-Simons (CS) term.  In the d-wave case we obtain the non-abelian SU(2) CS term because the system is spin-rotationally invariant, whereas the p-wave case gives an abelian U(1) term since it has only U(1) symmetry.  In both cases, the Hall spin conductivity is a topological invariant, which characterizes the winding of the order parameter in momentum space.  This is a microscopic proof of quantization that was proposed in earlier works by other authors.  

The last part of the thesis deals with adsorption on carbon nanotubes, which is a one-dimensional problem.  The hexagon centers serve as adsorption sites for hard core atoms, allowing the system to be treated as a lattice gas on a triangular lattice wrapped on a cylinder.  This model is equivalent to a type of quantum spin tube.  The wrapping introduces geometric frustration on top of the frustration of the triangular lattice, leading to interesting physics. In the spin language, we find magnetization plateaus in all tubes in the Ising limit, which is confirmed both analytically and numerically.  However, the zig-zag tubes are exceptional and contain plateaus that are macroscopically degenerate.  When quantum hopping is allowed, the special plateaus become gapless phases that are described by a $c=1$ conformal theory of a compactified boson.  The theory is derived analytically and confirmed numerically.  Perhaps the most remarkable feature is that the radius of compactification is quantized by the tube diameter.  This brings us back to the FQHE, where the theory of edge states is also characterized by a (chiral) conformal boson compactified on a quantized radius.

%\addtocontents{toc}{\contentsline{chapter}{Appendices:}{}}
\addcontentsline{toc}{chapter}{Appendix: Non-Commutative Fourier Transform} 
%%%%%%%%%%%%%%%%%%%%%%%%%%%%%%%%%%%%%%%%%%%%%%%%%%%%%%%%%%%%%%%%%%%%%%%%%%
\chapter*{Appendix: Non-Commutative Fourier Transform}
\label{sec:HP}
%%%%%%%%%%%%%%%%%%%%%%%%%%%%%%%%%%%%%%%%%%%%%%%%%%%%%%%%%%%%%%%%%%%%%%%%%%

Our ultimate aim in this subsection is to introduce constructs that will allow us to take the thermodynamic limit, $N\rightarrow\infty$, and to introduce the momentum $\bk$ so as to take advantage of translational invariance.  As discussed by several authors \cite{Read94,SM,ReadHalf}, ${\bk}$ is a good quantum number and $\hat{\bf z}\times{\bf k}$ will turn out to be the dipole moment.  The formalism in this appendix has been discussed in more detail by \cite{ReadHalf}, and we include it here for completeness.

Consider the special case of one attached vortex with $B_1=-B_2$, as introduced in Sections \ref{sec:TwoParticles} and \ref{sec:Fock_Physical}.  We first introduce real space wavefunctions by analogy to the usual matter field:
\bea
  c(z,\ol\eta) &=& \sum_{mn}u_m(z)\ol{u_n(\eta)}~c_{mn}\non\\
  c^\dagger(\eta,\ol z) &=& \sum_{mn}u_n(\eta)\ol{u_m(z)}~c^\dagger_{nm}~.
\label{eq:MatterField}
\eea   
Our convention is to use $z$ for the left coordinate and $\ol\eta$ for the right.  Complex conjugation reflects the two opposite charges.  There is only one magnetic length $\ell_B$ in this problem because the magnitude of the charges is equal.  This is a rather singular limit of our two-particle construction in Section \ref{sec:TwoParticles}; in this limit the effective magnetic field is $B=B_1+B_2=0$, and the pseudomomentum and translation operators, $\pi$ and $K$, are identical (eqn. (\ref{eq:PiK})).

In the $z$, $\eta$ basis, the densities become
\bea
\rho^R(\eta,\ol\eta^\prime)&=&\int d^2z~c^\dagger(\eta,\ol z)c(z,\ol\eta^\prime)
\non\\
\rho^L(z,\ol z^\prime)&=&\int d^2\eta~c^\dagger(\eta,\ol z^\prime)c(z,\ol\eta) 
\label{eq:RhoZ}
\eea
Thus integration has replaced summation over indices.  It is convenient to introduce a binary operation $*$ to represent integration over one set of coordinates,
\bea
(\hat a*\hat b)(z,\ol z^\prime)=\int d^2z_1~a(z,\ol z_1)b(z_1,\ol z^\prime)~,
\label{eq:Star}
\eea
of two operators, $\hat a$ and $\hat b$, which is just matrix multiplication.  It is also convenient to define the $*$-commutator by 
\bea
[\hat a\stackrel{*}{,}\hat b] = \hat a*\hat b-\hat b*\hat a~.
\label{eq:StarComm}
\eea
This allows us to write
\bea
\hat\rho^R &=&\hat c^\dagger~*~\hat c \non\\
\hat\rho^L &=&:\hat c~*~\hat c^\dagger:
\label{eq:RhoStar}
\eea 
Note the normal ordering in $\hat\rho^L$ necessary to avoid sign ambiguities.

Consider the plane wave, $e^{i\bk\cdot\bf r}$, projected to the LLL.  Following the previous discussion of two particles in Section \ref{sec:TwoParticles}, eqn. (\ref{eq:rEff}), ${\bf r}=({\bf R}_1+{\bf R}_2)/2$ in zero effective field.  The plane wave operator now becomes
\bea
e^{i\bk\cdot\bf r} = e^{i\bk\cdot({\bf R}_1+{\bf R}_2)/2}~.
\label{eq:PlaneWave}
\eea
Its representation in the $z$, $\eta$ coordinates is obtained by acting on the lowest weight eigenfunction $\psi_{0,0}$.  Recall that, for $B_1=-B_2$, $\psi_{0,0}(z,\ol\eta)= \frac{1}{2\pi}e^{-\frac{1}{4}|z|^2-\frac{1}{4}|\eta|^2+\frac{1}{2}z\ol\eta}$,
where the magnetic length has been set to unity.  It is not difficult to show that $\psi_{0,0}(z,\ol\eta)$ is identical to the delta function in the LLL.
\bea
\delta(z,\ol\eta) = \sum_m u_m(z)\ol{u_m(\eta)}~.
\label{eq:Delta}
\eea
In the LLL, $\delta$ acts as expected: $\hat\delta*\hat a=\hat a*\hat\delta=\hat a$.  The constraint is thus $\hat\rho^R=\hat\delta$, or
\bea
\rho^R(\eta,\ol\eta^\prime)=\delta(\eta,\ol\eta^\prime)~.
\non
\eea

The differential representation of ${\bf R}_{1,2}$ in the $z$, $\eta$ coordinates was constructed in Section (\ref{sec:TwoParticles}).  According to the prescription, the plane wave acting on $\hat\delta$ yields the LLL representation
\bea
\tau_\bk(z,\ol\eta)=\delta(z,\ol\eta)e^{\frac{1}{2}i(\ol kz+k\ol\eta)-\frac{1}{4}|k|^2}~.
\label{eq:tau_k}
\eea
By either straightforward integration or by using the commutator of ${\bf R}_{1,2}$, eqn. (\ref{eq:RCommutator2}), we find that the $\hat\tau_\bk$ obey
\bea
\hat\tau_\bk*\hat\tau_{\bk^\prime}=\hat\tau_{\bk+\bk^\prime}~
e^{\frac{1}{2}i\bk\wedge\bk^\prime}~,
\label{eq:TauComm}
\eea
where we have introduced the shorthand notation, $\wedge{\bf k} = -\hat z\times{\bf k}$ and $\bk\wedge\bk^\prime\equiv\bk\cdot\wedge\bk^\prime$.  The phase factor is the area of a triangle formed by $\bk$ and $\bk^\prime$ so that the phase counts the flux enclosed by the triangle.  Hence we interpret $\hat\tau_\bk$ as a magnetic translation in the plane by $\wedge\bk$ \cite{ReadHalf}.  The connection to plane waves extends to completeness and orthonormality properties, which defines a ``noncommutative Fourier transform''.  In particular,
\bea
\mbox{Tr}~\hat\tau_\bk*\hat\tau_{\bk^\prime}&=&2\pi\delta(\bk+\bk^\prime)~,\\
\int\frac{d^2\bk}{2\pi}~\tau_\bk(z,\ol z^\prime)\,\tau_{-\bk}(\eta,\ol\eta^\prime)&=& \delta(z,\ol\eta^\prime)\delta(\eta,\ol z^\prime)~.
\label{eq:Orthonormality}
\eea
The $\mbox{Tr}$ stands for a trace defined by $\mbox{Tr}~\hat a=\int d^2z~a(z,\ol z)$.  We can now define the Fourier transform and its inverse:
\bea
c(z,\ol\eta)&=&\int\frac{d^2\bk}{(2\pi)^{3/2}}~c_\bk~\tau_\bk(z,\ol\eta)~,\\
c_\bk &=& (2\pi)^{1/2}~\mbox{Tr}~\hat c*\hat\tau_{-\bk}~.
\label{eq:FT}
\eea
The extra $\sqrt{2\pi}$ factors are not used in general; they are specific to the Fourier transform of the $\hat c$'s in order for the commutators to retain their conventional form
\bea
[c_\bk,\,c^\dagger_{\bk^\prime}]_{_\pm}=(2\pi)^2\delta(\bk-\bk^\prime)~.
\label{eq:cComm}
\eea
Therefore, we have constructed composite particles with momentum $\bk$ and dipole moment $\wedge\bk$, which emerged from magnetic translations perpendicular to $\bk$.  When the underlying particles are bosons, anticommuators are required in eqn. (\ref{eq:cComm}).  This defines the composite fermion.  The Fock space of fermions at $\nu=1$ is trivial, consisting of exactly one function, so composite bosons are not useful in this case.  

Next we would like to apply the Fourier transform to the left and right densities in eqn. (\ref{eq:RhoZ}), which become
\bea
\hat\rho^R_\bq&=&\int\frac{d^2\bk}{(2\pi)^2}~e^{-\frac{1}{2}i\bk\wedge\bq}~ c^\dagger_{\bk-\frac{1}{2}\bq}c_{\bk+\frac{1}{2}\bq}~\\
\hat\rho^L_\bq&=&\int\frac{d^2\bk}{(2\pi)^2}~e^{\frac{1}{2}i\bk\wedge\bq}~ c^\dagger_{\bk-\frac{1}{2}\bq}c_{\bk+\frac{1}{2}\bq}~,
\label{eq:RhoK}
\eea
%and obey
%\bea
%[\rho^R_\bq,~\rho^L_{\bq^\prime}] = 0\non
%\eea
%\bea
%[\rho^R_\bq,~\rho^R_{\bq^\prime}] =     -2i\sin\left(\frac{1}{2}\bq\wedge\bq^\prime\right)\rho^R_{\bq+\bq^\prime}
%\eea
%\bea
%[\rho^L_\bq,~\rho^L_{\bq^\prime}] = 2i\sin %\left(\frac{1}{2}\bq\wedge\bq^\prime\right) \rho^L_{\bq+\bq^\prime}\non
%\label{eq:RhoComm}
%\eea
The commutators of $\hat\rho_\bq$ are familiar in the quantum Hall effect \cite{GMP,WInfinity,NonCommutativeSpace}, defining an infinite Lie algebra known as $W_\infty$.  The constraints appear particularly simple in momentum space:
\bea
\hat\rho^R_\bq-2\pi\overline\rho~\delta(\bq)=0~,
\label{eq:ConstraintsMomentum}
\eea
which enforces a uniform vortex density, $\overline\rho$.

\end{document}